\documentclass[fleqn,usenatbib]{mnras}

\usepackage{newtxtext,newtxmath}

\usepackage[T1]{fontenc}
\usepackage{ae,aecompl}

\usepackage{graphicx}	
\usepackage{amsmath}	
\usepackage{amssymb}	

\usepackage{float}
\usepackage{natbib}
\usepackage{color,soul}
\usepackage{adjustbox,lipsum}
\usepackage{array}
\usepackage{subfigure}
\usepackage{cleveref}
\usepackage{booktabs}
\usepackage[flushleft]{threeparttable}
\usepackage{caption}

\title[The AGN-SF Connection at Cosmic Noon]{Exploring AGN and Star Formation Activity of Massive Galaxies at Cosmic Noon}

\author[Florez et al.]{
Jonathan Florez,$^{1}$\thanks{E-mail: jflorez06@utexas.edu}
Shardha Jogee,$^{1}$ 
Sydney Sherman,$^{1}$ 
Matthew L. Stevans,$^{1}$ 
\newauthor 
Steven L. Finkelstein,$^{1}$ 
Casey Papovich,$^{2}$ 
Lalitwadee Kawinwanichakij,$^{2,3}$ 
\newauthor 
Robin Ciardullo,$^{4,5}$
Caryl Gronwall,$^{4,5}$ 
C. Megan Urry,$^{6,7}$ 
Allison Kirkpatrick,$^{8}$ 
\newauthor 
Stephanie M. LaMassa,$^{9}$ 
Tonima Tasnim Ananna$^{10}$
and Isak Wold$^{11}$
\\
$^{1}$Department of Astronomy, University of Texas at Austin, Austin, TX 78712, USA\\
$^{2}$Department of Physics and Astronomy, Texas A\&M University, College Station, TX 77843, USA\\
$^{3}$Kavli Institute for the Physics and Mathematics of the Universe, The University of Tokyo, Kashiwa, Japan 277-8583 (Kavli IPMU, WPI)\\
$^{4}$Department of Astronomy and Astrophysics, The Pennsylvania State University, University Park, PA 16802, USA\\
$^{5}$The Institute for Gravitation and the Cosmos, The Pennsylvania State University, University Park, PA 16802, USA\\
$^{6}$Yale Center for Astronomy \& Astrophysics, New Haven, CT 06520, USA\\
$^{7}$Department of Physics, Yale University, PO BOX 201820,New Haven, CT 06520, USA\\
$^{8}$Department of Physics \& Astronomy, University of Kansas, Lawrence, KS 66045, USA\\
$^{9}$Space Telescope Science Institute, 3700 San Martin Dr, Baltimore, MD 21218, USA\\
$^{10}$Department of Physics \& Astronomy, Dartmouth College, 6127 Wilder Laboratory, Hanover, NH 03755, USA\\
$^{11}$NASA Goddard Space Flight Center, Greenbelt, MD 20771
}

\date{Accepted XXX. Received YYY; in original form ZZZ}

\pubyear{2020}

\begin{document}
\label{firstpage}
\pagerange{\pageref{firstpage}--\pageref{lastpage}}
\maketitle

\begin{abstract}
We investigate the relation between AGN and star formation (SF) activity at $0.5 < z < 3$ by analyzing 898 galaxies with X-ray luminous AGN ($L_X > 10^{44}$ erg s$^{-1}$) and a large comparison sample of $\sim 320,000$ galaxies without X-ray luminous AGN{}. Our samples are selected from a large (11.8 deg$^2$) area in Stripe 82 that has multi-wavelength (X-ray to far-IR) data. The enormous comoving volume ($\sim 0.3$ Gpc$^3$) at $0.5 < z < 3$ minimizes the effects of cosmic variance and captures a large number of massive galaxies ($\sim 30,000$ galaxies with $M_* > 10^{11} \ M_{\odot}$) and X-ray luminous AGN{}. While many galaxy studies discard AGN hosts, we fit the SED of galaxies with and without X-ray luminous AGN with Code Investigating GALaxy Emission (CIGALE) and include AGN emission templates. We find that without this inclusion, stellar masses and star formation rates (SFRs) in AGN host galaxies can be overestimated, on average, by factors of up to $\sim 5$ and $\sim 10$, respectively. The average SFR of galaxies with X-ray luminous AGN is higher by a factor of $\sim 3$ to $10$ compared to galaxies without X-ray luminous AGN at fixed stellar mass and redshift, suggesting that high SFRs and high AGN X-ray luminosities may be fueled by common mechanisms. The vast majority ($> 95 \%$) of galaxies with X-ray luminous AGN at $z=0.5-3$ do not show quenched SF: this suggests that if AGN feedback quenches SF, the associated quenching process takes a significant time to act and the quenched phase sets in after the highly luminous phases of AGN activity.
\end{abstract}




\section{Introduction} \label{introduction}
The epoch of $z \sim 1-3$ (when the universe was only $\sim 2-6$ Gyr old, corresponding to $\sim 15\%$ to $50\%$ of cosmic history) is one of the most important and active epochs of galaxy formation. During this period, star formation (SF) and active galactic nuclei (AGN) activity in galaxies peaked, massive clusters collapsed into existence, and galaxies underwent significant growth. Although SF and AGN activity both peaked during this epoch, it is unclear how the two processes are related. Observations show that the cosmic star formation rate (SFR) density and black hole accretion rate density peak at around $z \sim 2$ and decrease rapidly down to $z \sim 0$ \citep{2003ApJ...587...25D, 2007A&A...474..755B, 2007ApJ...654..731H, 2008MNRAS.385..687W, 2009ApJ...697.1971J, 2011MNRAS.416.1900R, 2014ARA&A..52..415M, 2014MNRAS.439.2736D}. Furthermore, it is well known that the central black hole mass of galaxies correlates with the host galaxy bulge mass \citep{1998AJ....115.2285M, 2002MNRAS.331..795M}  and bulge velocity dispersion \citep{2000ApJ...539L...9F, 2000ApJ...539L..13G, 2013ARA&A..51..511K}. These relationships have led to suggestions that the growth of black holes and galaxies may be closely intertwined, but the issue of coevolution is the subject of debate \citep[e.g.,][]{2013ARA&A..51..511K, 2011ApJ...734...92J}.

AGN activity, which is a direct result of gas accreting onto a host galaxy's central black hole, has often been proposed as a mechanism that can reduce or suppress SF activity, as radiation, winds, and jets expel gas or heat it enough to prevent it from forming stars. Different forms of AGN feedback impact their host galaxies in different ways. AGN feedback from radiation and winds from the accretion disk can heat or expel galactic gas on different physical scales \citep[e.g.,][]{2011ApJ...738...16H, 2012ARA&A..50..455F, 2013MNRAS.436.3031V, 2015MNRAS.449.4105C, 2015ApJ...800...19R, 2016MNRAS.458..816H, 2017MNRAS.464.1854B}. Jets from AGNs are thought to play a large role in heating the intracluster medium (ICM) in clusters of galaxies, thus preventing gas from cooling and accreting onto galaxies and ultimately halting future episodes of star formation \citep[e.g.,][]{2008MNRAS.391..481S, 2006PhR...427....1P, 2007ARA&A..45..117M, 2009Natur.460..213C, 2012ARA&A..50..455F, 2014ARA&A..52..589H, 2019MNRAS.486.2827D}. In numerical simulations, some source of heating, such as AGN feedback, is thought to be crucial in solving the "overcooling" problem in galaxy formation, where in absence of feedback, the gas inside dark matter halos cools to form galaxies with mass functions resembling those of the dark matter halos \citep{1978MNRAS.183..341W, 2017ARA&A..55...59N, 2015ARA&A..53...51S}.

While AGN activity has been linked to a potential suppression of SF as described above, there are phases of galaxy evolution where AGN and SF activity coexist. A so-called AGN-SF connection has been claimed at $z < 0.1$ \citep{1988ApJ...325...74S, 2017MNRAS.471.3226M}. This connection, however, is currently a topic of debate as other studies \citep[e.g.,][]{2016MNRAS.455L..82L} claim  to find AGN activity associated with depressed SF activity at $z < 0.1$. Any potential connection between AGN and SF activity may be due, at least in part, to gas fueling both the circumnuclear SF activity and AGN activity when the angular momentum problem can be overcome \citep[e.g.,][and references therein]{2006LNP...693..143J}, as is the case in gas-rich mergers \citep{2008ApJS..175..356H}. At higher redshifts, the  AGN-SF connection is less well-studied. Some studies have claimed $1 < z < 3$ galaxies with X-ray luminous AGN, where X-ray emission is measured in the hard band (typically at 2-10 keV) or ultra-hard band (at 14-195 keV), have enhanced SFRs compared to galaxies without X-ray luminous AGN and/or that X-ray luminous AGN are preferentially found in star-forming galaxies out to $z \sim 2$ \citep{2018A&A...618A..31M, 2012A&A...540A.109S, 2013ApJ...771...63R, 2017MNRAS.466.3161S}, while other studies \citep[e.g.,][]{2015MNRAS.452.1841S} have claimed that galaxies with X-ray luminous AGN have decreased SFRs compared to galaxies without luminous AGN{}.

The goal of this work is to explore the general connections, if any, between AGN and SF activity  in galaxies at early cosmic times. As outlined above, the relationship between AGN and SF activity might be expected to depend on the evolutionary phase of a galaxy and the AGN activity cycle. It is possible that early-on, X-ray luminous AGN and SF activity coexist when large gas inflows on different physical scales fuel both types of activity. At later times, the AGN feedback phase may start if AGN-driven radiation, winds, and jets eject or heat galactic gas on different scales. Our study aims to explore the relation between AGN and SF activity, and to constrain the relative importance and timescales of these different evolutionary phases.

The processes that regulate SF and AGN activity are important for our understanding of galaxy evolution as they are intimately tied to the growth of galaxies. Observations have confirmed the existence of massive quenched galaxies up to $z \sim 4$ \citep[][Stevans et al. submitted]{2006ApJ...649L..71K, 2012MNRAS.421..621B, 2020MNRAS.491.3318S, 2018ApJ...858..100F, 2013ApJ...777...18M, 2013ApJ...768...92S, 2017Natur.544...71G}, suggesting that quenching mechanisms can efficiently suppress SF in galaxies at early cosmic times. It is unclear, however, what role (if any) AGN feedback plays in the quenching of massive galaxies. Simulations, such as Illustris \citep{2015MNRAS.449..361W} and IllustrisTNG \citep{2018MNRAS.477.1206N, 2018MNRAS.475..676S, 2018MNRAS.475..624N}, largely implement AGN and stellar feedback as the main quenching mechanisms for massive galaxies in their simulations. However, these simulations struggle to produce massive quenched galaxies by $z \sim 3$. We aim to explore the properties of massive, quiescent galaxies with and without X-ray luminous AGN in order to further constrain the processes that quench star formation.

Studying the SF activity of galaxies with X-ray luminous AGN can be rather complicated, as the emission from an AGN can dominate the galaxy spectral energy distribution (SED) at UV, optical, mid-IR, and far-IR wavelengths, thus making it difficult to disentangle the emission between AGN and star forming processes in SED fitting. For this reason, many studies that explore the global stellar mass-SF main sequence of galaxies explicitly remove or ignore the contribution from galaxies with X-ray luminous AGN \citep{2014ApJS..214...15S}. Although difficult, several studies have attempted to compare the star forming properties of galaxies with X-ray luminous AGN to galaxies without X-ray luminous AGN, with X-ray emission measured in the hard band (at 2-10 keV) \citep{2018A&A...618A..31M} and the ultra-hard band (at 14-195 keV) \citep{2017MNRAS.466.3161S}, and find that galaxies with X-ray luminous AGN tend to have enhanced SFRs with respect to galaxies without X-ray luminous AGN. The work presented in this paper will utilize a control sample of galaxies without X-ray luminous AGN that is larger than the control samples presented in the aforementioned studies and will use the same SED fitting code to measure galaxy properties for galaxies with and without X-ray luminous AGN.

For this study, we create a sample of galaxies with X-ray luminous AGN, where X-ray emission is measured in the full band of Stripe 82X at 0.5-10 keV, and a sample of galaxies without X-ray luminous AGN using the same photometric data and perform SED fits of both samples using Code Investigating GALaxy Emission (CIGALE) \citep{2019A&A...622A.103B,2015A&A...576A..10C,2009A&A...507.1793N} which is capable of disentangling the emission from AGN and star forming processes. This allows us to measure and compare the star forming properties of both samples in the same self-consistent way, unlike many other studies. We select our samples from a very large 11.8 deg$^2$ field where the Stripe 82X X-ray survey \citep{2016ApJ...817..172L} and the Spitzer-HETDEX Exploratory Large Area (SHELA) IRAC survey overlap \cite{2016ApJS..224...28P}. Our samples have extensive multi-wavelength coverage (e.g., X-ray, UV, optical, near-to-mid-IR, and some far-IR/submillimeter) over the 11.8 deg$^2$ field, which corresponds to a very large comoving volume of $\sim 0.3$ Gpc$^3$ at $z=0.5-3$. Such a large comoving volume minimizes the effects of cosmic variance and captures a large sample of rare massive galaxies ($\sim 30,000$ galaxies with $M_* > 10^{11} \ M_{\odot}$) and X-ray luminous AGN ($\sim 700$ objects with $L_X > 10^{44}$ erg s$^{-1}$), allowing us to provide some of the strongest constraints to-date on the relation between AGN and SF activity
at $z \sim 1-3$.

This paper is organized as follows. In Section \ref{dataset}, we discuss the data we use to select and create our samples of galaxies with and without X-ray luminous AGN{}. In Section \ref{sed_fitting}, we describe the SED fitting method and tests we use to obtain galaxy properties, such as stellar mass and SFR{}. In Section \ref{properties}, we discuss the stellar mass and SFR distributions of our samples. In Section \ref{results}, we compare the stellar mass-SFR relation of galaxies with and without X-ray luminous AGN{}. In Section \ref{quenched_galaxies}, we present the fraction and properties of quenched galaxies as a function of mass and redshift. We discuss our findings in Section \ref{discussion} and in Section \ref{summary} we summarize our results. Throughout this paper we assume $H_0 = 70$ km s$^{-1}$ Mpc$^{-1}$, $\Omega_M = 0.3$, and $\Omega_{\Lambda}=0.7$.

\section{Data \& Sample Selection} \label{dataset} 

The goal of this work is to estimate and compare the star-forming properties of galaxies with and without X-ray luminous AGN at $z=0.5-3$. To accomplish this, we utilize the large-area, multi-wavelength data available in the SHELA/HETDEX footprint, which consists of five photometric data sets: Dark Energy Camera (DECam) $u,g,r,i,z$ \citep{2019ApJS..240....5W}, NEWFIRM $K_S$ (Stevans et al. submitted), $Spitzer$-IRAC 3.6 and 4.5 $\mu$m \citep{2016ApJS..224...28P}, $Herschel$-SPIRE far-IR/submillimeter \citep[HerS,][]{2014ApJS..210...22V} and Stripe 82X X-ray \citep{2016ApJ...817..172L}. We also utilize available $J$ and $K_S$-band data from the VISTA-CFHT Stripe 82 (VICS82) Near-Infrared Survey \citep{2017ApJS..231....7G} and mid-IR photometry from the WISE survey \citep{2010AJ....140.1868W} to supplement this work. In the near future, optical integral-field spectroscopy between 3500 and 5000 \AA{} of this region will be available from the Hobby Eberly Telescope Dark Energy Experiment \citep{2016ASPC..507..393H}.

\subsection{NEWFIRM K-band Selected Catalog} \label{shela}
For the analysis presented in this paper, we use photometry from a NEWFIRM $K_S$-band selected catalog (Stevans et al. submitted) that includes DECam $u,g,r,i,z$ as well as IRAC 3.6 and 4.5 $\mu$m photometry spanning an area of 17.5 deg$^2$ across the SHELA field. The catalog is created using an approach similar to that in \cite{2019ApJS..240....5W} and described in detail in Stevans et al. (submitted). In summary, they use Source Extractor \citep[SExtractor;][]{1996A&AS..117..393B} to identify sources in their $K_s$-band images and report a $5 \sigma$ depth of $\sim 22.4$ AB mag inside fixed 2 arcsecond diameter apertures. The $K_S$-band fluxes that we use for the analysis presented here are the Kron aperture fluxes from SExtractor (e.g., {\tt FLUX\_AUTO}). Quality flags in the catalog allow us to exclude objects with saturated, truncated, or corrupted pixels as well as regions in close proximity to saturated stars from our analysis. The DECam and IRAC magnitudes are derived through forced photometry of NEWFIRM $K_S$ selected sources using the $Tractor$ image modeling code \citep[see][for description of code]{2016ascl.soft04008L}. The $Tractor$ code uses the source positions and surface brightness profiles of higher resolution bands to model and derive fluxes for remaining lower resolution bands. This allows the catalog to include individual fluxes for potentially blended sources in IRAC{}. These deblended DECam and IRAC fluxes are used to create the $K_S$-selected DECam-NEWFIRM-IRAC catalog that we use for this work.

\subsection{Stripe 82X} \label{stripe_82x}
Stripe 82X is an X-ray survey that covers 31.3 deg$^2$ of the SDSS Stripe 82 Legacy Survey \citep{2013MNRAS.436.3581L, 2013MNRAS.432.1351L, 2016ApJ...817..172L, 2017ApJ...850...66A}. The original catalog described in \cite{2016ApJ...817..172L} introduced the release of the XMM-Newton Announcement Opportunity 13 (“AO13”) data and is combined with archival XMM-$Newton$ and $Chandra$ X-ray Observatory X-ray data \citep{2013MNRAS.436.3581L, 2013MNRAS.432.1351L}. We utilize two updated versions of the catalog described in \cite{2017ApJ...850...66A} (hereafter A17) and \cite{2019ApJ...876...50L} (hereafter LM19). The XMM AO13 footprint largely overlaps ($\sim 11.8$ deg$^2$ overlap) with the SHELA field, whereas the archival $Chandra$ and XMM-$Newton$ data footprints hardly overlap with the SHELA field. For this reason, for our analysis we only consider XMM AO13 data from the Stripe 82X catalog, which has a spatial resolution of $\sim 6$ arcseconds \citep{2001A&A...365L..18S}.

 The A17 Stripe 82X catalog provides multiwavelength counterpart matches to the X-ray sources by crossmatching to the SDSS coadded catalogs of \cite{2016MNRAS.456.1359F} and includes IRAC photometry \citep{2016ApJS..224...28P,2016ApJS..225....1T} in the crossmatching as well. The A17 catalog also provides photometric redshifts for every source in the catalog as well as quality flags that indicate the reliability of the match. A description of how photometric redshifts are obtained for the sample can be found in A17. The photometric redshifts ($z_{\rm phot}$) when compared to the available spectroscopic redshifts ($z_{\rm spec}$) have a scatter of $|\Delta(z)| / (1+z) \sim 0.06$ where $\Delta(z) = (z_{\rm phot} - z_{\rm spec})$ and a $\sim 14 \%$ outlier fraction, where outliers are defined as objects with $|\Delta(z)| / (1 + z_{\rm spec}) > 0.15$. We use the A17 catalog to obtain full X-ray fluxes ($0.5-10$ keV), WISE counterpart fluxes, and redshifts for our sample of galaxies with X-ray luminous AGN{}. We utilize the aforementioned LM19 catalog  to add additional spectroscopic redshifts to the sources in XMM AO13, which brings the spectroscopic redshift completeness for our sample of galaxies with X-ray luminous AGN to $\sim 75 \%$.

\subsection{VICS82 and WISE Supplemental Data} \label{supplemental}
A significant portion of the SHELA footprint overlaps with the VICS82 survey \citep{2017ApJS..231....7G} and the WISE \citep{2010AJ....140.1868W} mid-IR survey. The DECam-NEWFIRM-IRAC catalog we utilize includes matches to VICS82 $J \& K_S$-band photometry. We make use of the VICS82 $J \& K_S$-band photometry in our analysis in order to better constrain the SED fit at near-IR wavelengths. We crossmatch sources in the DECam-NEWFIRM-IRAC catalog to the AllWISE source catalog \cite{2013yCat.2328....0C} by considering any objects within 1.5 arcseconds of the NEWFIRM source a match. We utilize the magnitude values from the profile-fitting photometry and only consider sources with S/N $> 2$ for this work. Due to the lower resolution of WISE photometry ($\sim 6-7"$ for WISE-1-3, $\sim 12"$ for WISE-4), we only consider objects a match for this work if the WISE source has no more than one NEWFIRM match within a 5 arcsecond radius, thus avoiding blended sources. In addition, we apply flag quality cuts to make sure none of our WISE sources have image pixels in the measurement aperture that are confused with nearby sources, and/or contaminated by saturated or otherwise unusable pixels. WISE photometry, particularly in the WISE-3 (12 $\mu\rm{m}$) and WISE-4 (22 $\mu\rm{m}$) bands, is particularly useful for constraining the SED in the mid-IR wavelengths where emission from an AGN can contaminate the emitted light at $3-20$ $\mu\rm{m}$ and dust emission from SF radiates at $8-1000$ $\mu\rm{m}$. As discussed in the following sections and illustrated in Figure \ref{sample_flowchart} and Table \ref{tab1}, we unfortunately find that only a small fraction of our sample has WISE-3 or WISE-4 photometry. Therefore, we mainly use WISE data to determine whether and how our results might change if mid-IR photometry is included (see Appendix).

\subsection{Sample Selection} \label{matching}
\begin{figure*}
\begin{center}
\includegraphics[scale=0.8]{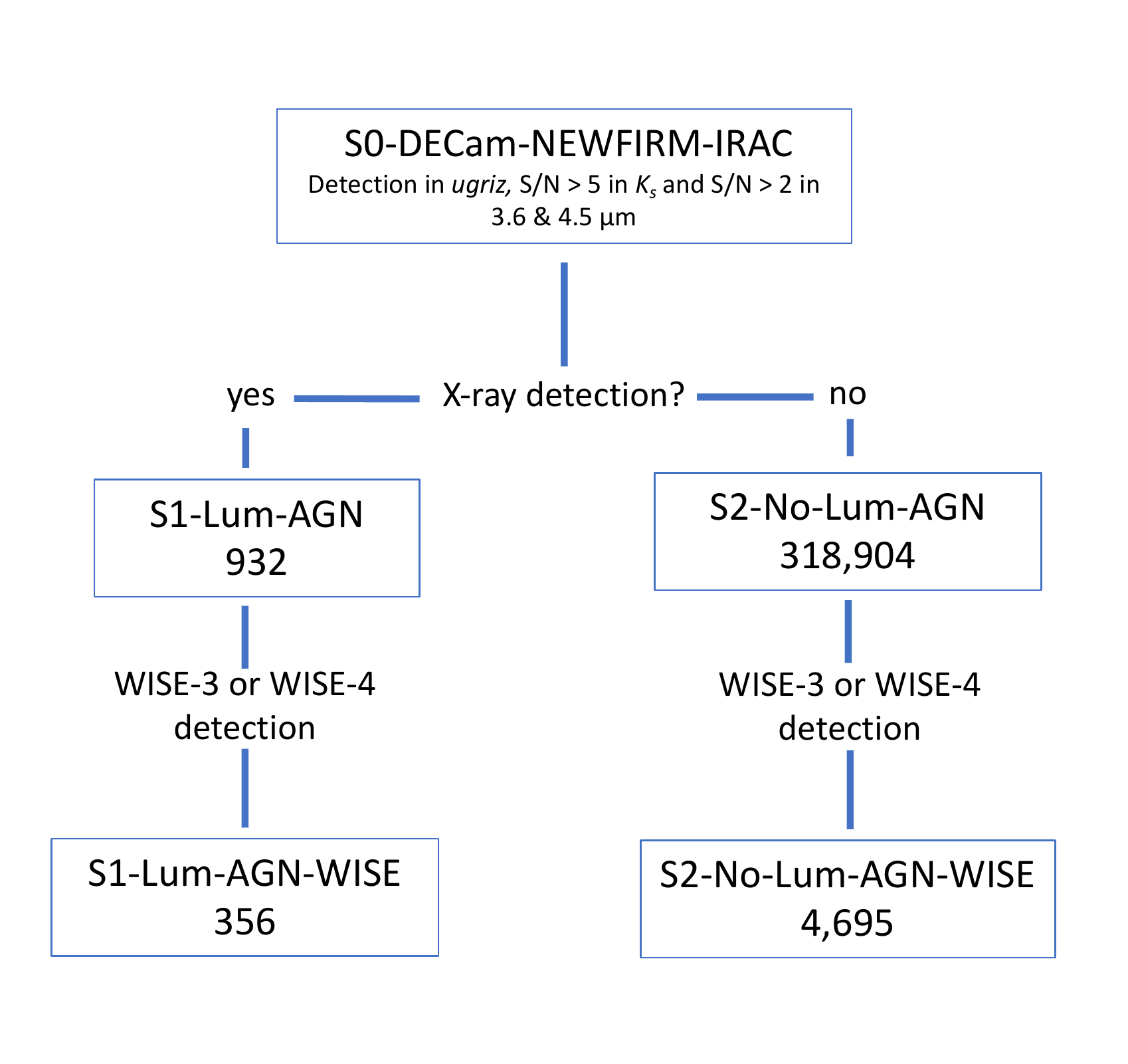}
\caption{Flowchart that demonstrates how our samples are selected. We start with the DECam-NEWFIRM-IRAC catalog and create a sample of sources that fit our selection criteria, which we refer to as S0-DECam-NEWFIRM-IRAC (see Sections \ref{shela} and \ref{matching}). We then crossmatch S0-DECam-NEWFIRM-IRAC with the Stripe 82X catalog via the MLE method and create samples with and without an X-ray match, which we refer to as S1-Lum-AGN (see Section \ref{gal_xray_agn}) and S2-No-Lum-AGN (see Section \ref{galaxy_faint_no_agn}), respectively. We then search for WISE detections in both S1-Lum-AGN and S2-No-Lum-AGN and create two more samples, which we refer to as S1-Lum-AGN-WISE and S2-No-Lum-AGN-WISE. The reader can refer to Table \ref{tab1} for number of objects in each sample complete in X-ray luminosity and stellar mass. We note that although we have 932 sources with X-ray luminous AGN in sample S1-Lum-AGN, only 898 of these are have a good SED fit with a reduced $\chi^2$ of less than 5. We therefore only analyze these 898 sources in S1-Lum-AGN for this work.}
\label{sample_flowchart}
\end{center}
\end{figure*}

In this section we describe how we obtain our samples of galaxies with and without X-ray luminous AGN{}. We start by selecting all sources in the $K_S$-selected DECam-NEWFIRM-IRAC catalog that overlap the Stripe 82X AO13 footprint on the sky. From these sources, we create a sample of objects, which we refer to as S0-DECam-NEWFIRM-IRAC, that have a detection in the DECam $u,g,r,i,z$ bands, a signal-to-noise (S/N) $>5$ in the NEWFIRM $K_S$-band, and S/N $>2$ in the IRAC 3.6 and 4.5 $\mu$m bands. The reader can refer to the flowchart in Figure \ref{sample_flowchart} for an illustration of how the samples with and without X-ray luminous AGN are selected from S0-DECam-NEWFIRM-IRAC (described in further detail below).

\subsubsection{Galaxies with X-ray Luminous AGN} \label{gal_xray_agn}
 In order to create a sample of X-ray selected AGN, we crossmatch between XMM AO13 sources from the Stripe 82X catalog and sources in S0-DECam-NEWFIRM-IRAC using the maximum likelihood estimator (MLE) method of \cite{1992MNRAS.259..413S}. The MLE method has been widely used to perform crossmatching between X-ray data and multiwavelength counterparts \citep[see][]{2017ApJ...850...66A, 2016ApJ...817..172L, 2013MNRAS.436.3581L, 2010ApJ...716..348B}. We implement the same MLE methodology used in \cite{2016ApJ...817..172L}. That is, we set our search radius to 7 arcseconds and the background search radius to be between 10 to 45 arcseconds. We use X-ray positional error values given in the \cite{2016ApJ...817..172L} catalog and assume that the positional errors in the DECam-NEWFIRM-IRAC catalog are negligible compared to those of the X-ray. The only difference in our method to that of \cite{2016ApJ...817..172L} is that we consider objects that have a reliability value ($R$ in equation (2) of \cite{2016ApJ...817..172L}) greater than 0.5 to be true matches, as any object with $R > 0.5$ is the most likely source inside the search radius to be the true counterpart.  We only crossmatch to the $K_S$-band photometry in the DECam-NEWFIRM-IRAC catalog as we are using a $K_S$-band selected catalog for the analysis presented here.
 
 \begin{figure*}
\includegraphics[scale=0.65]{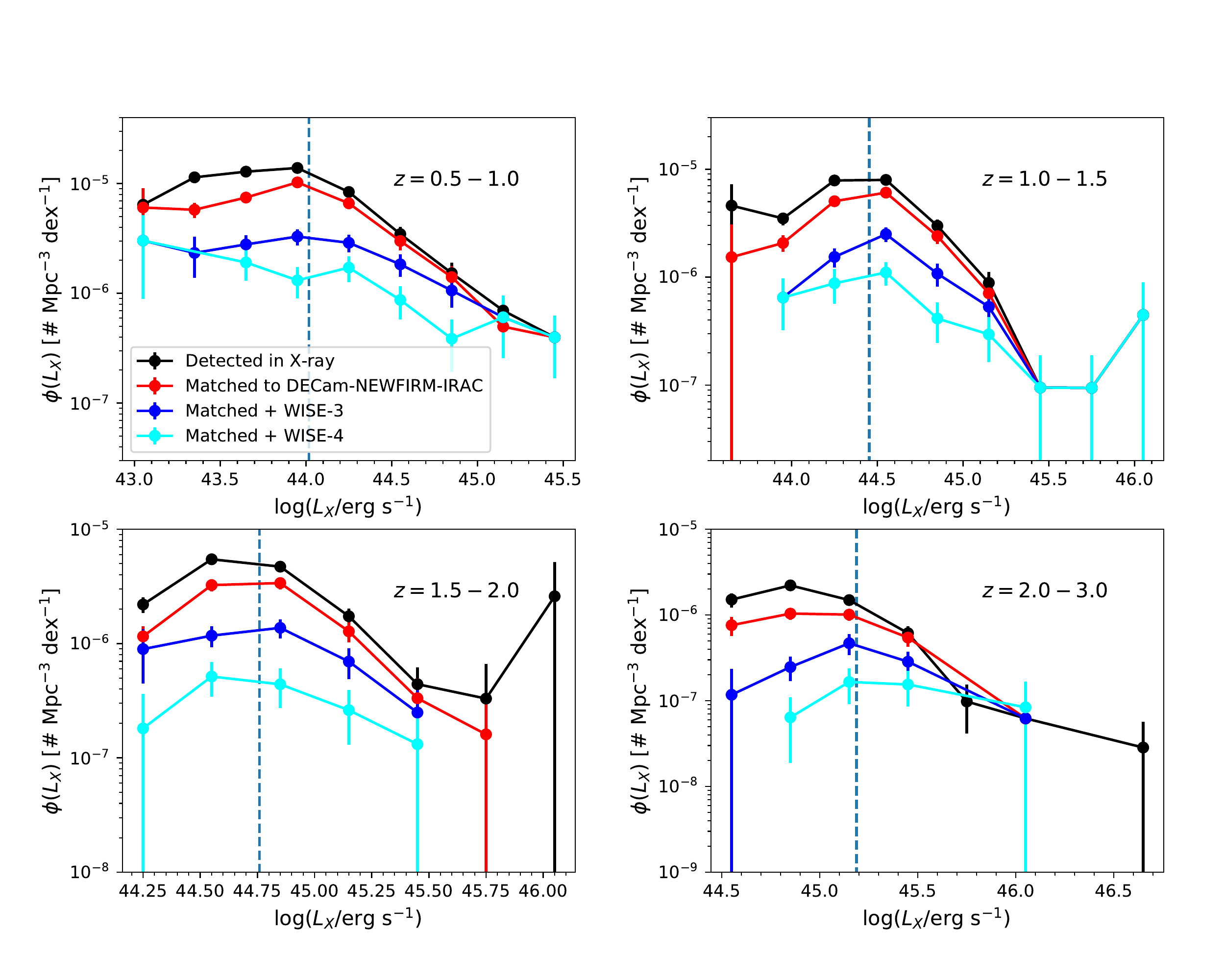}
\caption{The full (0.5-10 keV) X-ray luminosity function of all XMM AO13 sources that fall in SHELA (black), those that have a reliable counterpart in the DECam-NEWFIRM-IRAC catalog through crossmatching with the MLE method (red), matched sources with a WISE-3 detection (dark blue) and matched sources with a WISE-4 detection (light blue) in four different bins of redshift. The luminosity function here is computed using the $V_{\rm max}$ method (see Section 2.4.1). The dashed line in each panel indicates the $80 \%$ X-ray luminosity completeness limit, computed from the $80 \%$ flux limit at the upper edge of each redshift bin.}
\label{xray_LF}
\end{figure*}

 Once we find a match between S0-DECam-NEWFIRM-IRAC and XMM AO13, we assign it a photometric or spectroscopic (if available) redshift from A17. We create the sample S1-Lum-AGN from sources in S0-DECam-NEWFIRM-IRAC that have an X-ray match and $z=0.5-3$. From this sample, we search in A17 for any sources that have a WISE detection and include those in our subsample S1-Lum-AGN-WISE. Out of 1,356 unique XMM AO13 X-ray sources that fall in the SHELA footprint at $z=0.5-3$, we find a total of 932 reliable matches in the DECam-NEWFIRM-IRAC catalog. Of these, only 356 sources have a detection in either the WISE-3 or WISE-4 bands.
 
In Figure \ref{xray_LF}, we show the full (0.5-10 keV) X-ray luminosity function of all sources in XMM AO13, those matched with S0-DECam-NEWFIRM-IRAC, and those that are matched with a WISE-3 or WISE-4 detection. Although we calculate the X-ray luminosity function here to show the effects of requiring photometric completeness and WISE photometry on our sample, we refer the reader to \cite{2019ApJ...871..240A} for the latest evolving X-ray luminosity function which includes the effects of X-ray absorption and also includes data from multiple surveys in order to overcome the luminosity-redshift correlation in any one flux-limited survey. The X-ray luminosity function here is calculated using the $1/V_{\rm{max}}$ method described in \cite{1968ApJ...151..393S}. In the $1/V_{\rm{max}}$ method, the luminosity (or mass) function of a sample of galaxies is calculated by dividing the number of galaxies at a given luminosity (or mass) bin by the bin width times the differential comoving volume $\Delta V_C$, where $\Delta V_C$ is the difference between the comoving volume ($V_{\rm max}$) at a maximum redshift  that a source of given luminosity (or mass) can probe and the comoving volume at the low edge of the redshift bin. This method tries to correct the luminosity (or mass) function for the fact that flux-limited observational surveys increasingly fail to detect faint sources at higher redshifts, and are therefore biased to produce an artificial drop in the number density of faint sources at higher redshifts. We note that the X-ray luminosity function drops by almost an order of magnitude for sources with WISE-3 counterparts above the X-ray completeness limit, and even more for sources with WISE-4 counterparts. This means that requiring a WISE-3 or WISE-4 detection would cause us to incompletely sample the X-ray luminosity function at $z=0.5-3$. Furthermore, as shown in Figure \ref{sample_flowchart} and Table \ref{tab1}, our sample of X-ray luminous AGN (S1-Lum-AGN) would be reduced from 932 to 356 sources if we require WISE detections. Therefore, in order to better sample the X-ray luminosity function and prevent a drastic reduction in sample
size, we do not use S1-Lum-AGN-WISE in our main analysis. However, in the Appendix, we perform tests on this sample to verify that the inclusion of WISE mid-IR data would not change the results of this work.
 
Table \ref{tab1} shows the total number of sources ($N_{\rm tot}$) in the full sample based on the DECam-NEWFIRM-IRAC catalog (S0-DECam-NEWFIRM-IRAC), the sample of X-ray luminous AGN (S1-Lum-AGN), and the subset of the latter sample with WISE detections (S1-Lum-AGN-WISE), as well as the number of sources above the X-ray completeness limit and stellar mass completeness limit for each sample. The X-ray luminosity completeness limit we use here is computed from the $80 \%$ completeness full flux band limit of XMM AO13 ($F_X = 2.0 \times 10^{-14}$ erg s$^{-1}$ cm$^{-2}$), estimated from the flux area curves given in \cite{2016ApJ...817..172L}. We take this flux limit and convert to a luminosity at the high end of each given redshift range to generate an $80 \%$ luminosity completeness limit of $L_x = 10^{44.03}$ erg s$^{-1}$ at $z=0.5-1$, $L_x = 10^{44.47}$ erg s$^{-1}$ at $z=1-1.5$, $L_x = 10^{44.78}$ erg s$^{-1}$ at $z=1.5-2$, and $L_x = 10^{45.21}$ erg s$^{-1}$ at $z=2-3$. The source of X-ray emission for all galaxies in this sample should be entirely dominated by the respective AGN component of each galaxy, given that we only analyze sources with $L_X > 10^{44}$ erg s$^{-1}$. AGN activity is the most likely mechanism in a galaxy capable of producing such high X-ray luminosities \citep{2015A&ARv..23....1B}.

As an additional test, we explore whether our selection of X-ray luminous AGN using the X-ray energy band at 0.5-10 keV might lead us to pick low-luminosity AGN with enhanced SF by comparing our study with other studies that use the more traditional X-ray hard band luminosities, at 2-10 keV, for selecting AGN \citep{2019ApJ...876...50L,2018A&A...618A..31M, 2015A&ARv..23....1B, 2005ARA&A..43..827B}. These studies classify sources with hard X-ray band luminosities greater than $L$(2-10 keV) $\sim 10^{42}$ erg $^{-1}$ as sources hosting X-ray luminous AGN, while galaxies without AGN typically have hard X-ray emission below this threshold and an extremely small number of starburst galaxies may exceed this threshold, but not by much \citep{2008ApJ...681.1163L}. We compute the rest-frame hard X-ray luminosity for each source as described in \cite{2019ApJ...876...50L} and find that all sources in our sample with X-ray full band (0.5-10 keV) luminosities above $10^{44}$ erg s$^{-1}$ have rest-frame X-ray hard band (2-10 keV) luminosities greater than $10^{43.5}$ erg s$^{-1}$, which is more than a magnitude greater than the $L$(2-10 keV) $= 10^{42}$ erg s$^{-1}$ luminosity threshold used in the aforementioned studies. Therefore, we conclude it is unlikely that many of our X-ray luminous AGN are actually low-luminosity AGN with enhanced SF.

It is important to note here that our sample likely contains a large number of Type I AGN (i.e., AGN whose broad-line region is visible with respect to the observer) and a very small number of Type II AGN (i.e., AGN whose broad-line region is obscured with respect to the observer) as several studies \citep{2016ApJ...817..172L, 2011ApJ...728...58B} find very few obscured AGN with X-ray luminosities above $10^{44}$ erg s$^{-1}$. These obscured sources in Stripe 82X tend to be optically classified as "normal" galaxies but are considered AGN because their X-ray luminosities exceed the $10^{42}$ erg s$^{-1}$ luminosity threshold, while quasars and broad-line sources can be found at all X-ray luminosities and make up the majority of sources with X-ray luminosities greater than $10^{44}$ erg s$^{-1}$. We therefore note that the results of the analysis presented in this paper do not extend to all populations of AGN types, and likely apply mostly to Type I AGN (broad-line sources and quasars). While it is possible that we could also be missing obscured sources in our sample that are intrinsically X-ray luminous AGN, we note that \cite{2020ApJ...891...41P} estimate how many additional X-ray luminous AGN would be added to their high luminosity sample ($L_X > 10^{44.5}$ erg s$^{-1}$, where $L_X$ is the X-ray full band luminosity at 0.5-10 keV) if they corrected their luminosities for dust obscuration assuming a column density distribution matching that of XMM-XXL AGN \citep{2016MNRAS.459.1602L} and found that their sample would only increase by $\lesssim 4\%$. We therefore conclude that our sample of X-ray luminous AGN is unlikely to be missing many luminous AGN due to dust obscuration.

\begin{table*}
\begin{center}
\begin{threeparttable}
\caption{Total number of galaxies in each sample at a given redshift range.}
\begin{tabular}{l c c c c c}
\hline
\hline
Sample & All $z$ bins ($z=0.5-3.0$) & $z = 0.5-1.0$ & $z = 1.0-1.5$ & $z = 1.5-2.0$ & $z = 2.0-3.0$ \\
(a) & (b) & (c) & (d) & (e) & (f) \\
\hline
\hline
(1) S0-DECam-NEWFIRM-IRAC & 319,836 & 156,978 & 91,193 & 22,753 & 48,912\\
\hline
(2) S1-Lum-AGN\\
i.) Total Number $N_{\rm{tot}}$ & 932 & 352 & 273 & 184 & 123\\
ii.) $L_X > L_{X,\rm{lim}}$ & 386 & 147 & 136 & 71 & 32\\
iii.) $L_X > L_{X,\rm{lim}}$ \& $M_* > M_{*,95\% \rm{lim}}$ & 258 & 131 & 83 & 30 & 14\\
\hline
(3) S1-Lum-AGN-WISE\\
i.) Total Number $N_{\rm{tot}}$ & 356 & 133 & 112 & 72 & 39\\
ii.) $L_X > L_{X,\rm{lim}}$ & 193 & 75 & 64 & 35 & 19\\
iii.) $L_X > L_{X,\rm{lim}}$ \& $M_* > M_{*,95\%\rm{lim}}$ & 130 & 71 & 37 & 12 & 10\\
\hline
(4) S2-No-Lum-AGN \\
i.) Total Number $N_{\rm{tot}}$ & 318,904 & 156,626 & 90,920 & 22,569 & 48,789\\
ii.) $M_* > M_{*,95\% \rm{lim}}$ & 153,765 & 95,048 & 42,155 & 7,991 & 8,571\\
\hline
(5) S2-No-Lum-AGN-WISE\\
i.) Total Number $N_{\rm{tot}}$ & 4,695 & 3,487 & 903 & 116 & 189\\
ii.) $M_* > M_{*,95\% \rm{lim}}$ & 4,040 & 3,353 & 836 & 98 & 132\\
\hline 
\end{tabular}
\begin{tablenotes}
\small
\item *Note: (1) The sample S0-DECam-NEWFIRM-IRAC contains galaxies which have a detection in the $u,g,r,i,z$ bands, $\rm S/N > 5$ in the K-band, and $\rm S/N > 2$ in the two IRAC bands, and overlap with the Stripe 82X survey (see flowchart in Figure \ref{sample_flowchart} for an illustration of how samples are selected). (2) The sample S1-Lum-AGN contains a subset of galaxies in S0-DECam-NEWFIRM-IRAC that have an X-ray luminous AGN{}. The total number of galaxies in the sample, as well as the number of galaxies with X-ray luminosities above the completeness limit ($L_X > L_{X,\rm{lim}}$) and stellar masses above the $95 \%$ stellar mass completeness limit ($M_* >M_{*,95 \%\rm{lim}}$) are shown. (3) The sample S1-Lum-AGN-WISE contains the much smaller subset of galaxies in S1-Lum-AGN that have matching WISE photometry. (4) The sample S2-No-Lum-AGN contains the subset of galaxies in S0-DECam-NEWFIRM-IRAC that do not contain an X-ray luminous AGN{}.
(5) The sample S2-No-Lum-AGN-WISE contains the smaller subset of sources in S2-No-Lum-AGN that also have WISE photometry.
\end{tablenotes}
\label{tab1}
\end{threeparttable}
\end{center}
\end{table*}

\subsubsection{Galaxies without X-ray luminous AGN} \label{galaxy_faint_no_agn}
We create a sample of galaxies without X-ray luminous AGN, which we refer to as S2-No-Lum-AGN, from sources in S0-DECam-NEWFIRM-IRAC that do not have an X-ray counterpart. We derive photometric redshifts for this sample using the EAZY-py photometric redshift SED fitting code \citep[description of original EAZY code in][]{2008ApJ...686.1503B} and keep all sources with $z_{\rm phot} = 0.5-3$. EAZY-py fits a set of Flexible Stellar Population Synthesis (FSPS) templates \citep{2009ApJ...696..620C, 2009ApJ...699..486C} that span a wide range of galaxy types (e.g., star-forming, quiescent, dusty, etc.) in non-negative linear combination. The photometric redshift, $z_{\rm phot}$, is determined from the combination of templates that have the lowest $\chi^2$ value. To ensure we have a sample of galaxies that are all well fit by the EAZY-py templates, we implement a cut of $\chi^2 < 10$ on the entire sample. Comparison with the available SDSS spectroscopic redshifts reveal a $1 \sigma$ scatter of $\Delta z / (1 + z_{\rm spec}) = 0.037$. Although this comparison is only done for bright, low redshift ($z < 1$) sources, \cite{2020MNRAS.491.3318S} find a $1 \sigma$ scatter of $\Delta z / (1 + z_{\rm spec}) = 0.168$ for a sample of 16 bright $1.5 < z < 3.5$ galaxies with HETDEX spectroscopic redshifts using EAZY-py and the same photometry we use in this work, indicating fair agreement between the EAZY-py redshifts and spectroscopic redshifts. We explore the impact of uncertainty in redshifts on the SFRs of our sample of galaxies without X-ray luminous AGN by shuffling the EAZY-py redshifts by $1 \sigma$. For the aforementioned case of $\sigma=0.037$ (applicable to bright sources with $z < 1$), there is a scatter of $\sim 2-3$ when comparing the reshuffled SFRs to the original SFRs, but the average SFR remains the same. For the case of $1 \sigma = 0.168$ (which applies for galaxies at $1.5 < z < 3.0$), we find that after shifting the photometric redshift of our galaxies at $z > 1.5$ by $1 \sigma$, the majority ($> 90 \%$) of galaxies have reshuffled SFRs within 0.5 dex of the original SFR, while a small percentage ($< 2 \%$) have SFRs that shift by more than 1 dex. We therefore conclude that this photometric redshift uncertainty would not change the main results of this paper.

We create a subsample of objects in S2-No-Lum-AGN with a WISE-3 or WISE-4 counterpart, which we refer to as S2-No-Lum-AGN-WISE by crossmatching with the WISE catalog. Similar to what is described for S1-Lum-AGN-WISE in the previous section, we do not include S2-No-Lum-AGN-WISE in our main analysis. Rather, in the Appendix, we perform tests on this sample S2-No-Lum-AGN-WISE to verify that the inclusion of WISE mid-IR data would not change the results of this work.

Our sample S2-No-Lum-AGN has a total of 318,904 sources. Of these, only 4,695 have a WISE-3 or WISE-4 detection. The number of sources in S2-No-Lum-AGN and S2-No-Lum-AGN-WISE is also shown in Table \ref{tab1}, along with the number of sources above the stellar mass completeness limit at a given redshift range.


\begin{figure*}
\centering
\subfigure{\includegraphics[scale=0.5]{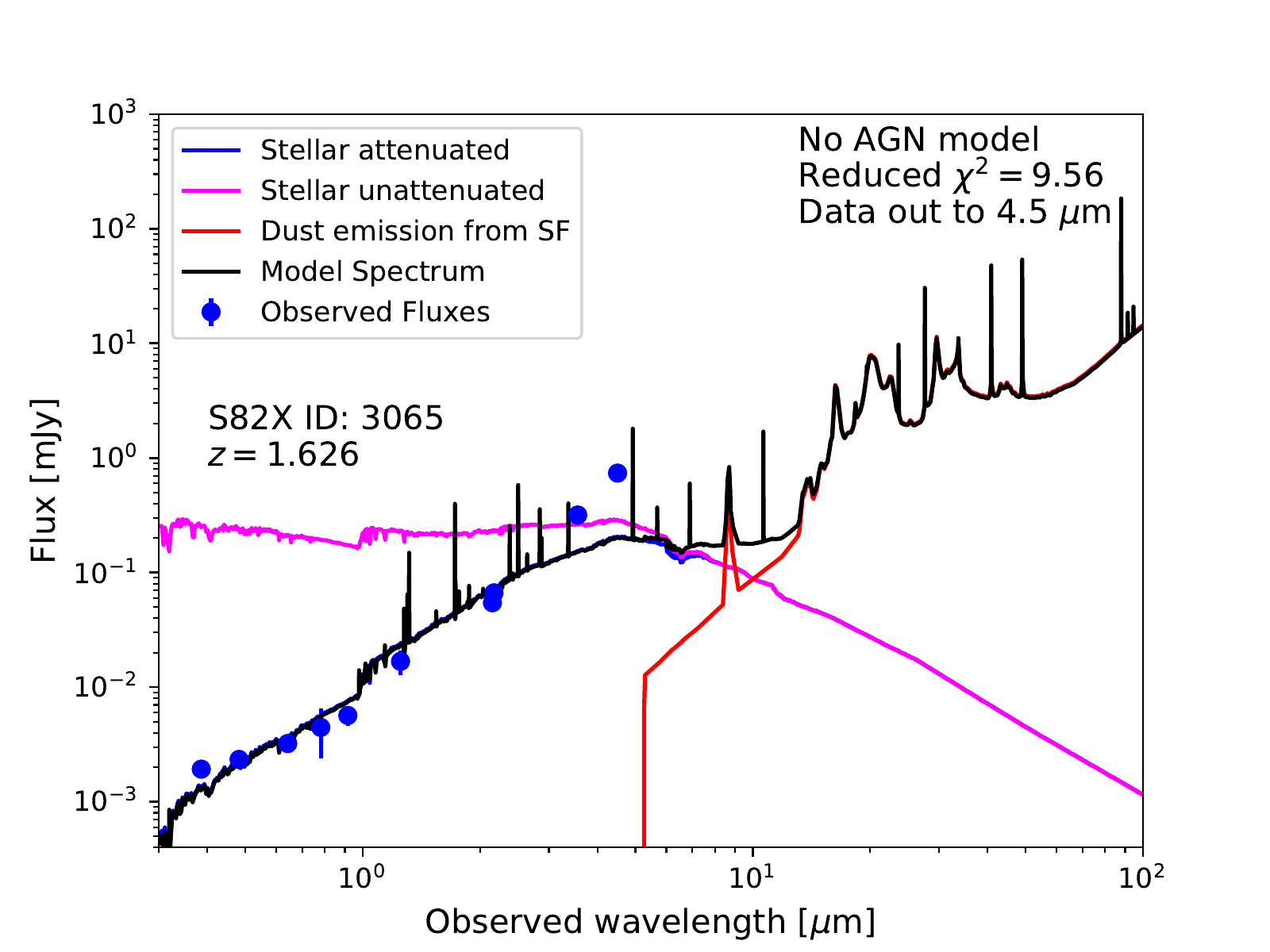}}
\subfigure{\includegraphics[scale=0.5]{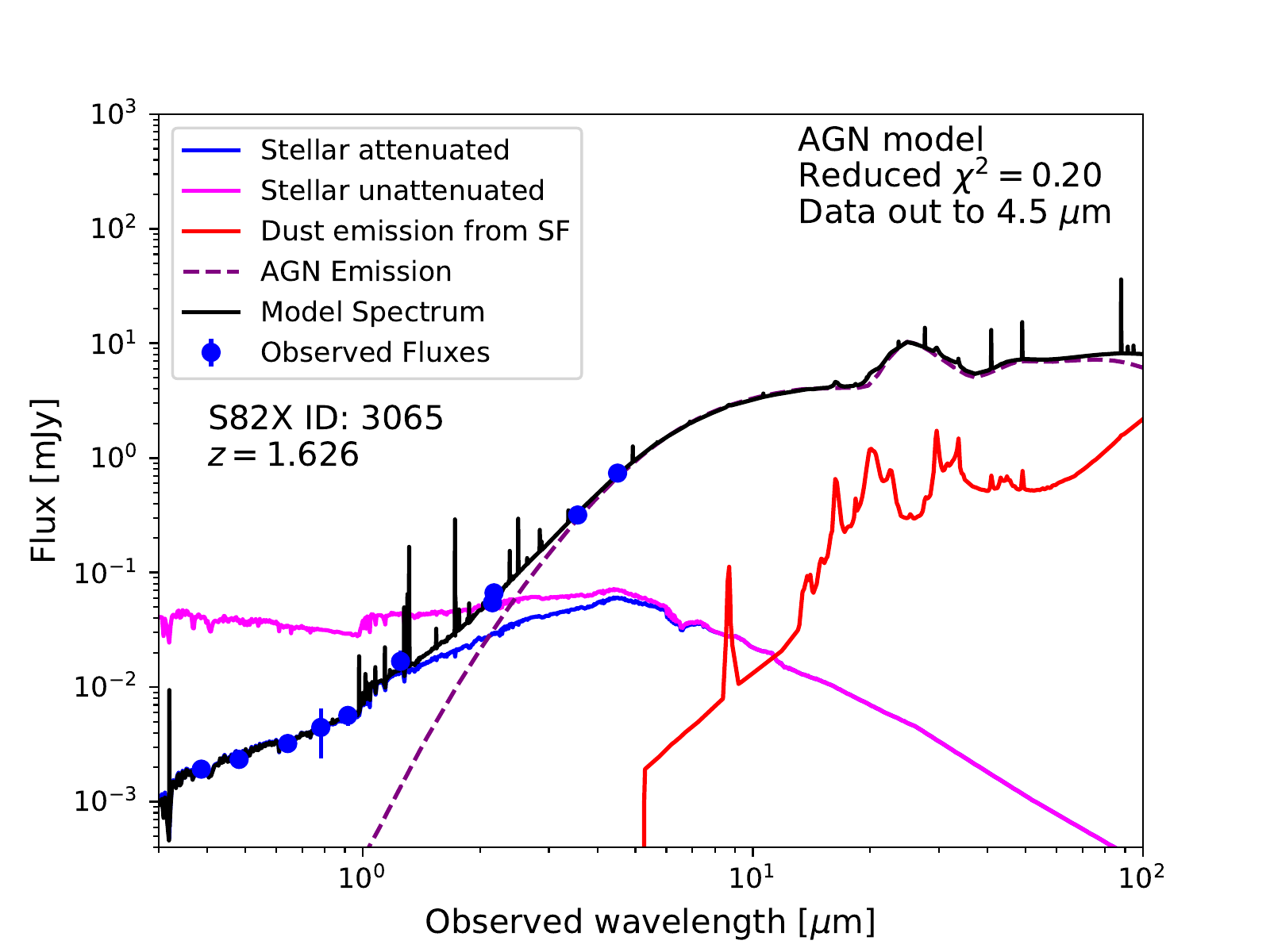}}
\subfigure{\includegraphics[scale=0.5]{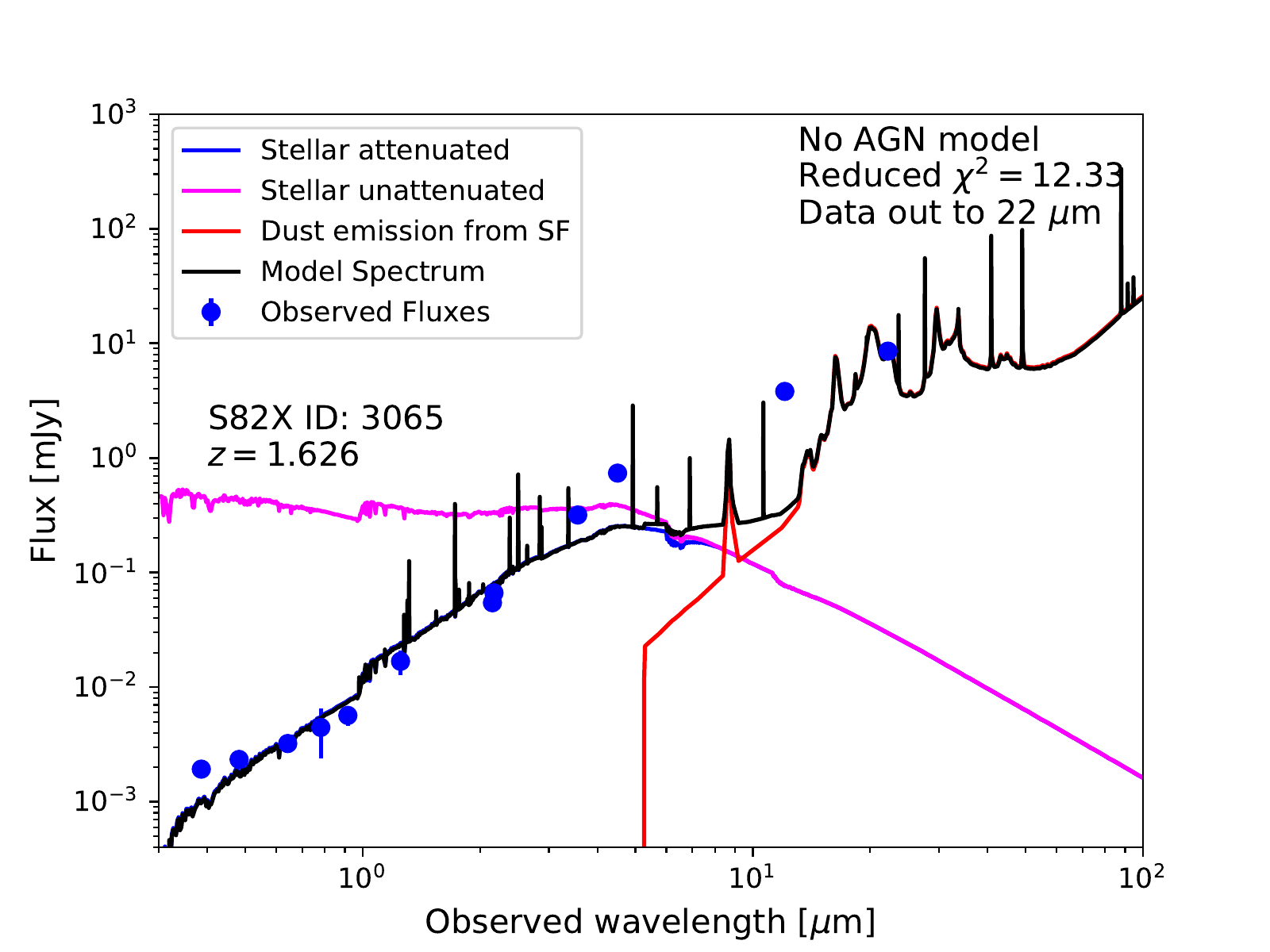}}
\subfigure{\includegraphics[scale=0.5]{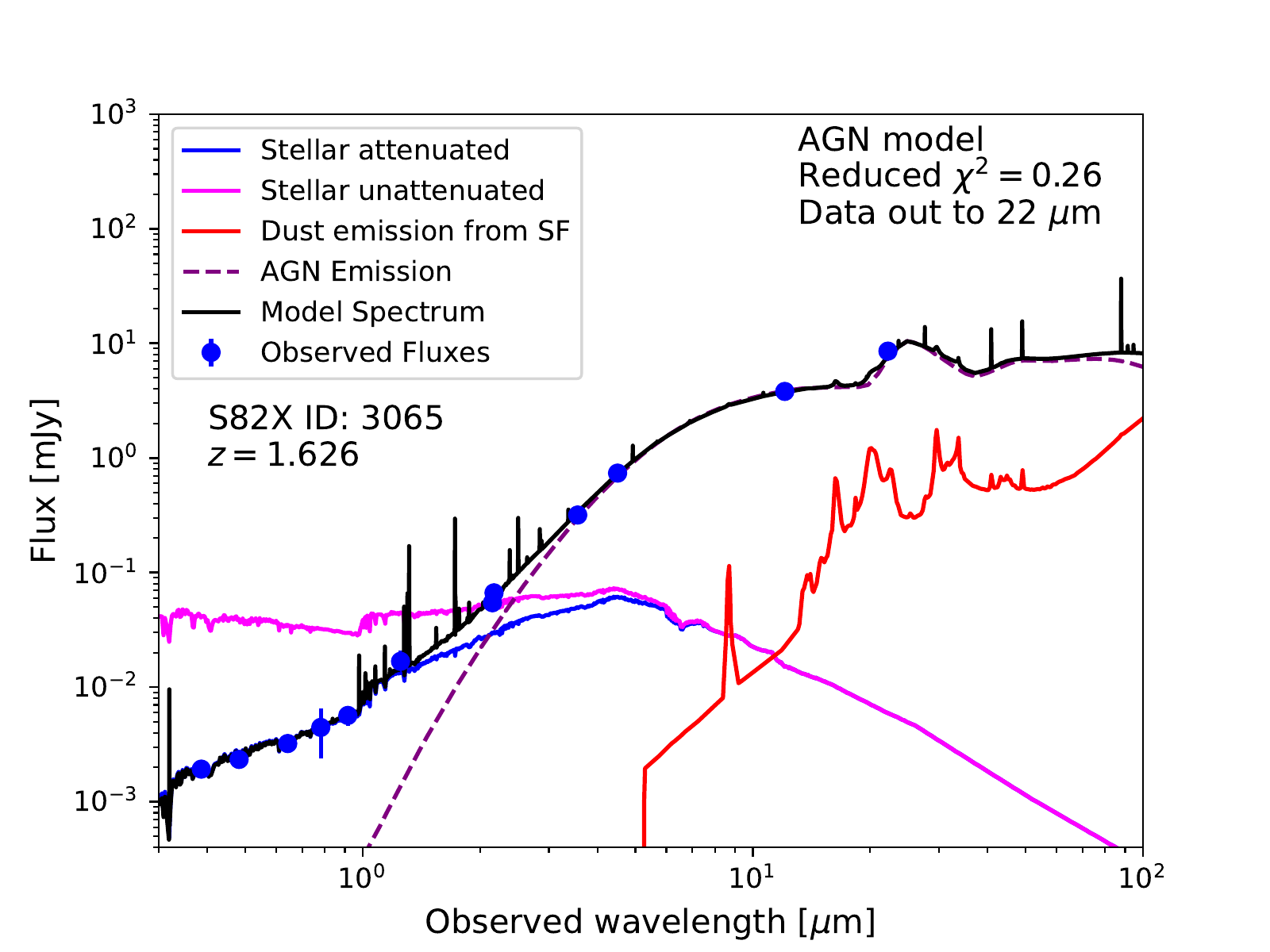}}
\caption{This figure compares model SED fits for a galaxy, whose redshift and Stripe 82X ID is displayed in each panel, with a type II X-ray luminous AGN in our samples S1-Lum-AGN (top two panels, with data coverage out to 4.5 $\mu$m and S1-Lum-AGN-WISE (bottom two panels with data coverage out to 22 $\mu$m) for SED fits that do not include AGN emission (left) and include AGN emission (right). The final model SED fit (solid black line) with AGN emission (right) is made up of the attenuated stellar emission (blue; which is inferred from the unattenuated stellar emission (magenta)), the dust emission from dust heated by massive stars from recent SF (red), the combined AGN emission (purple) from the accretion disk (particularly important at UV+optical wavelengths) and the dusty torus (particularly important at the $3-1000$ $\mu$m wavelength range). The best-fit model SED without AGN emission (left) clearly cannot provide a good fit to the observed fluxes at wavelengths past 1 $\mu$m, therefore, the AGN emission templates are needed in order to constrain all emission above 1 $\mu$m. While WISE data at 12 and 22 $\mu$m can provide important constraints on the SED at longer wavelengths, a comparison of the top and bottom right panels of Figure \ref{SEDs} shows that for the galaxy fitted here, the IRAC 3.6 and 4.5 $\mu$m photometry alone, without any WISE photometry, can provide important constraints on the SED fits with AGN emission templates.} 
\label{SEDs}
\end{figure*}

\begin{figure*}
\begin{center}
\includegraphics[scale=0.6]{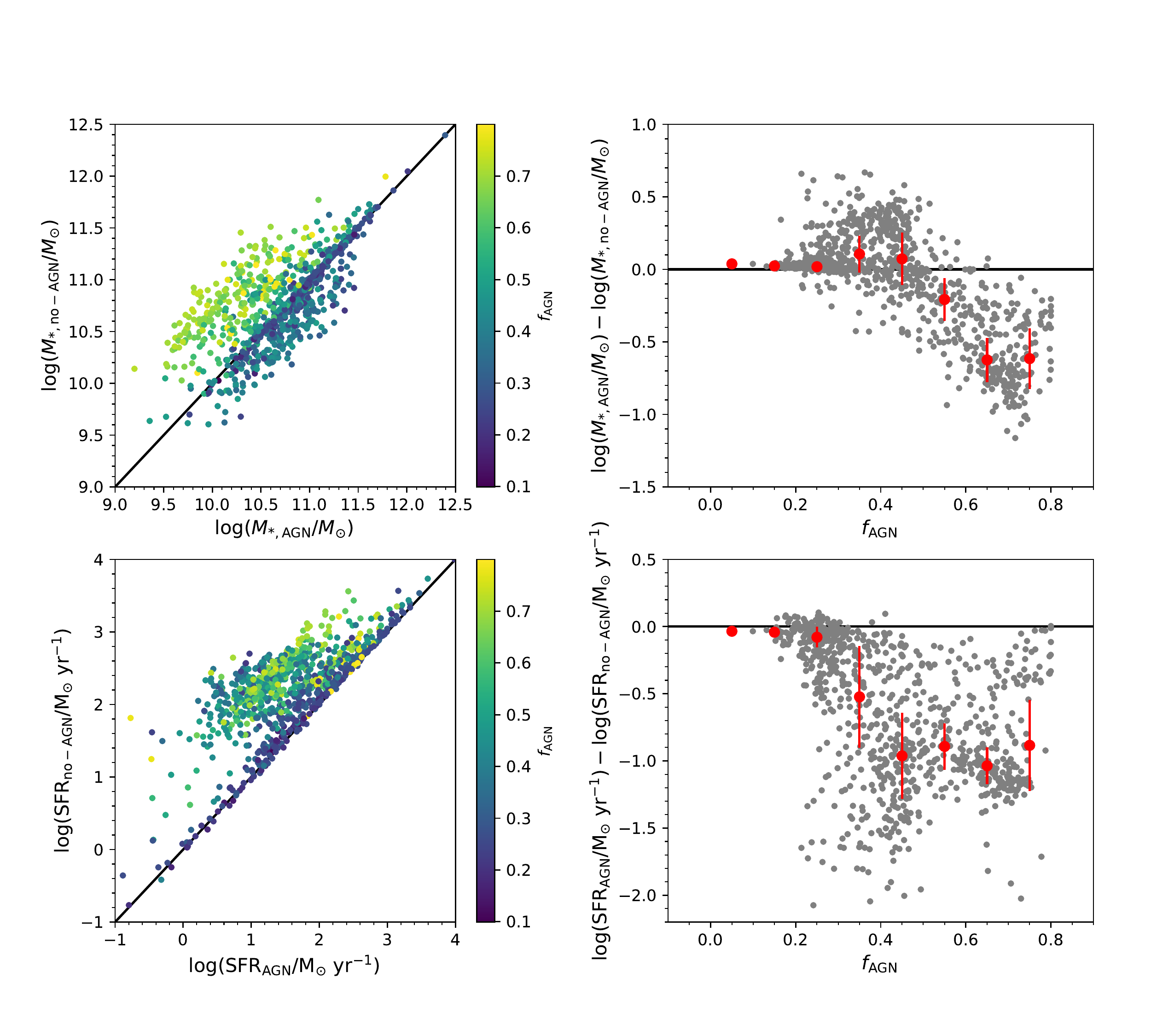}
\caption{Left: Stellar mass and SFR estimates for our sample of galaxies with X-ray luminous AGN (S1-Lum-AGN) when AGN emission is included in the SED fit (x-axis) versus when AGN emission is not included (y-axis). Points are colored according to their fractional AGN contamination ($f_{\rm{AGN}}$), defined as the fraction of light in the $8-1000$ $\mu$m wavelength range that is contributed by the AGN{}. Right: Difference in log stellar mass and SFR as a function of the fractional AGN contamination. Also shown is the median (red circles) log difference of stellar mass and SFR with and without the AGN emission in the SED fit in four bins of $f_{\rm{AGN}}$ with the median absolute deviation shown as error bars. Note that for $f_{\rm{AGN}} > 0.4$ stellar masses and star formation rates (SFRs) can be overestimated on average, by a factor of up to $\sim 5$ and $\sim 10$, respectively, if AGN emission templates are not included in the SED fit.}
\label{SED_comp}
\end{center}
\end{figure*}

\section{SED Fitting} \label{sed_fitting}
In this section we describe the SED fitting process that allows us to derive stellar masses and SFRs for our samples. One challenge that has persisted in the study of AGN host galaxies is determining the impact of AGN emission on the host galaxy's SED and having the ability to accurately decompose the galaxy SED into stellar, dust, and AGN components. An AGN can have a significant effect on the light of the galaxy SED across a wide range of wavelengths, and the magnitude of the effect depends on the strength and phase of the AGN, and the orientation of the AGN with respect to the observer. Another challenge that persists in the study of AGN is due to the wide variety of SED fitting codes. Many SED fitting codes do not include AGN emission templates in their code (e.g., EAZY-py \citep{2008ApJ...686.1503B}, MAGPHYS \citep{2008MNRAS.388.1595D}, iSEDfit \citep{2013ApJ...767...50M}, GalMc \citep{2011ApJ...737...47A}, etc.),  while others that do include AGN emission (e.g., AGNfitter \citep{2016ApJ...833...98C}, SED3FIT \citep{2013A&A...551A.100B}, etc.) will often try to include an AGN component where none may exist, thus making such codes unsuitable for fitting samples of galaxies without X-ray luminous AGN{}. This creates a further issue: many studies that measure and compare the properties of galaxies with luminous AGN to galaxies without luminous AGN do not estimate the properties of both populations in a self-consistent manner \citep{2015MNRAS.452.1841S, 2017MNRAS.466.3161S, 2018A&A...618A..31M, 2012A&A...540A.109S}. 

In order to accurately and consistently estimate properties of galaxies with and without X-ray luminous AGN, we perform SED fitting of both samples using the CIGALE \citep[version 2018.0;][]{2009A&A...507.1793N, 2015A&A...576A..10C} SED fitting code. The CIGALE code offers several advantages over other SED fitting codes. First, CIGALE allows one to optionally include AGN emission templates in the SED fitting. This means one can quantify how the derived galaxy properties are affected by the inclusion or exclusion of AGN emission templates in the fit. More significantly, however, it allows us to derive stellar mass and SFR for galaxies with and without X-ray luminous AGN accurately and consistently using the exact same code.

We fit a total of 9 free parameters to the SED fit for our samples of galaxies with X-ray luminous AGN: 4 parameters for the star formation history (SFH), 1 parameter for the dust attenuation, and 4 parameters for the AGN emission templates. For galaxies without X-ray luminous AGN, we only fit 5 free parameters as we omit fitting the AGN emission templates for these galaxies. In our SED fitting of galaxies with X-ray luminous AGN, we include the AGN emission models of \cite{2006MNRAS.366..767F} for emission from the AGN accretion disk and dusty torus in the fit. For these AGN emission templates, we fit the AGN fraction contamination, $f_{\rm AGN}$, which is the total amount of light emitted at $8-1000 \ \mu$m that is attributed to the AGN, the orientational angle of the AGN ($\Psi = 0^{\circ}$ for a type II AGN, $\Psi = 90^{\circ}$ for a type I AGN), the optical depth at 9.7 $\mu$m, and the ratio of the maximum to minimum radii of the dusty torus. For both samples of galaxies with and without X-ray luminous AGN, we include models of dust emission attributed to SF \citep{2014ApJ...784...83D}, and the stellar populations of \cite{2003MNRAS.344.1000B}. We assume attenuation of the galaxy SED by dust as described in \cite{2000ApJ...533..682C}, a Chabrier initial mass function \citep[IMF;][]{2003PASP..115..763C}, and a delayed exponential SFH with a constant burst/quench term described by the following equation:

\begin{equation}
\begin{aligned}
    &\rm{SFR}(t) \propto t \times \exp{(-t / \tau)}, \rm{when}\; t \leq t_{\rm{trunc}}
    \\
    &\rm{SFR}(t) \propto r \times \rm{SFR}(t=t_{\rm{trunc}}), \rm{when}\; t > t_{\rm{trunc}}
\end{aligned}
\end{equation}

\noindent
where $r$ is the constant burst/quench term, defined as the ratio between $\rm{SFR}(t)$ and $\rm{SFR}(t= t_{\rm trunc})$ at $t> t_{\rm trunc}$, and $t_{\rm{trunc}}$ is the time at which the SFR experiences an instantaneous increase or decrease given by the burst/quench term $r$ \citep[see][for description of SFH]{2018A&A...615A..61C, 2016A&A...585A..43C}. Previous studies \citep[e.g.,][]{2011A&A...528A..46F, 2006ApJ...651..811B} have proposed using a delayed SFH which could undergo a strong decrease in SFR{}. Such an SFH would allow for more flexibility in modeling the recent SFH of quenched galaxies or starbursts \citep{2018A&A...615A..61C, 2016A&A...585A..43C}. \cite{2017A&A...608A..41C} showed that this SFH provides a good estimate of the SFR on main-sequence galaxies, starbursts, and rapidly quenched systems at all redshifts. The SFRs we report here are the instantaneous SFRs given by this SFH, and thus, are obtained from equation (1) at $t=t_{\rm{age}}$, where $t_{\rm{age}}$ is the presently observed age of the galaxy. 

We make note here of the various factors that contribute to the dispersion of estimated SFRs. Work done by \cite{2014A&A...561A..39B} assesses the reliability of estimated SFRs from CIGALE for galaxies at $1 < z < 3$. They report full consistency between the instantaneous SFRs output by CIGALE, assuming different SFHs, to total SFRs estimated by empirical recipes using UV and FIR luminosities, suggesting the choice of SFH in CIGALE does not have a strong impact on the estimated SFR. Additionally, it is important to note that differences in the IMF and metallicity can affect the estimate of SFR. \cite{2014A&A...561A..39B} find that variation in the IMF changes the derived SFRs by a factor of up to $\sim 0.17$ dex, and variation in metallicity can change the derived SFRs by a factor of up to $\sim 0.2$. All of these factors are important to consider when estimating and reporting SFRs for a given sample, however, we believe the choice of SFH, IMF, and metallicity for this study should not have a large impact on our results.

We perform our analysis with the \cite{2006MNRAS.366..767F} smooth dusty torus AGN emission templates because they are by far the most flexible AGN emission models available in CIGALE and cover a large range of parameters. However, we discuss here the implications of selecting the smooth dusty torus models of \cite{2006MNRAS.366..767F} over the clumpy dusty torus models that several other studies have investigated \citep{2011MNRAS.414.1082M, 2014ApJ...794..152M}.  Observations have provided evidence in favor of both clumpy \citep{2007A&A...474..837T} and smooth \citep{2007A&A...466..531I} dusty toroidal distributions which are both often used in modeling the SED of X-ray luminous AGN. \cite{2012MNRAS.426..120F} perform a comparison of both smooth and clumpy dust torus distributions widely used in the literature, comparing the \cite{2006MNRAS.366..767F} models, for a smooth dusty torus, to the \cite{2008ApJ...685..147N} models, for a clumpy dusty torus. They find that models with matched parameters between smooth and clumpy distributions do not produce similar SEDs and only a very limited number of random parameter combinations can produce seemingly identical SEDs for both distributions. Interestingly, they find that most of the differences in the SEDs between these two published models are due to different dust chemical composition rather than dust morphology.

In terms of differences caused by dust morphology, \cite{2012MNRAS.426..120F} find that the clumpy AGN emission templates peak at slightly longer wavelengths, tend to have wider IR bumps, and steeper mid-IR slopes than the smooth dust models. It is possible that mean SFRs would be slightly lower when clumpy emission templates are applied as more emission at $8-1000$ $\mu$m would be attributed to AGN activity, however, this would not be universal as the smooth templates could have wider IR bumps (within a matched parameter space than the clumpy templates).

\subsection{Impact of AGN Emission on SED Fit and Derived Physical Properties} \label{impact}
It is especially important to include AGN emission in the SED fit of galaxies with X-ray luminous AGN as it can have a drastic impact on the derived properties of the host galaxy, such as stellar mass and SFR{}. In the UV+optical wavelength regime, a large portion of the emitted light of a luminous type I AGN (i.e., an AGN with the broad line region visible to the observer) is attributed to the AGN accretion disk. In the case of type I and possibly type II AGN (i.e., an AGN whose broad line region is obscured with respect to the observer), at rest-frame wavelengths greater than $1 \mu$m, the SED of a galaxy with a luminous AGN is impacted by the AGN's dusty torus as the AGN's power-law flux drowns the polycyclic aromatic hydrocarbon (PAH) features in the SED{}. The CIGALE code includes the AGN emission of the accretion disk, as well as emission from the dusty torus in the SED fit, then carefully removes this emission from the derived SED when estimating the host galaxy properties. 

In the top row of Figure \ref{SEDs}, we show as an example the SED of a galaxy with an X-ray luminous AGN before and after AGN emission is included in the fit for galaxies with data out to $4.5$ $\mu$m. Note that the AGN emission affects all emission above 1 $\mu$m. The bottom row of Figure \ref{SEDs} shows the same comparison, but this time also including WISE 12 and 22 $\mu$m photometry. While WISE data at 12 and 22 $\mu$m can provide important constraints on the SED at long wavelengths, a comparison of the top and bottom right panels of Figure \ref{SEDs} shows that in the galaxy fitted here, the IRAC 3.6 and 4.5 $\mu$m photometry alone, without any WISE photometry, can provide important constraints on the SED fits with AGN emission templates.

The inclusion of the AGN component can have a strong impact on the derived stellar masses and SFRs as the former will depend on whether the entire SED is dominated by a luminous AGN and the latter will depend on whether the UV and the near-to-far-IR light is attributed solely to stars and dust associated with SF or to a mix of stars, dust associated with SF, and AGN{}. When including AGN emission in the model SED fit, the CIGALE code will estimate the fraction ($f_{\rm{AGN}}$) of light in the $8-1000$ $\mu$m wavelength range that is contributed by the AGN{}. In Figure \ref{SED_comp}, we quantify the effect of not including an AGN component in the SED fitting using our sample of galaxies with X-ray luminous AGN (S1-Lum-AGN), which have data from the UV band out to IRAC $4.5 \mu$m, but no WISE data. Galaxies with $f_{\rm{AGN}} < 0.4$ can have SFRs that are overestimated by a factor of up to 2, on average, and those with $f_{\rm{AGN}} > 0.4$ can have SFRs overestimated by a factor of up to 10, on average, when AGN emission templates are not included in the SED fitting. In a few cases, the SFR can be overestimated by a factor $\sim 100$ when AGN emission templates are excluded from the SED fit. For this reason, we emphasize that SED fitting of galaxies with X-ray luminous AGN should require AGN emission templates.

Figure \ref{SED_comp} also shows the impact on derived stellar masses. For $f_{\rm{AGN}} > 0.4$, stellar masses can, on average, be overestimated by a factor of up to 5 when AGN emission templates are not included in the SED fit. This happens because without the AGN emission template, all of the of the light emission of the galaxy is assumed to come from stellar sources and dust as opposed to the AGN central engine. For galaxies with $f_{\rm AGN} < 0.4$, stellar masses can be underestimated by a factor of up to 3 if AGN emission templates are not included in the SED fit. The underestimate of stellar masses in some of the galaxies with $f_{\rm AGN} < 0.4$ happens because without the AGN emission templates, the AGN-boosted mid-IR luminosities of these galaxies will be fit by dusty stellar population templates instead of the AGN dusty torus templates. This in turn causes CIGALE to assume that the galaxy SED has a younger, dustier stellar population than it really does, thereby lowering the stellar masses estimates.

As mentioned earlier, the results shown in Figure \ref{SED_comp} are based on the sample of galaxies with X-ray luminous AGN (S1-Lum-AGN) that have data from the UV band out to 4.5 $\mu$m, but no WISE data at 12 and 22 $\mu$m.  We made the decision not to limit our analysis to only sources with WISE data as our sample size would be drastically reduced (see Figure \ref{sample_flowchart}). Instead, we confirmed that not including the WISE data does not change the results in this paper by performing additional tests in the Appendix. Figures \ref{SED_comp-WISE} and \ref{WISE-no_WISE_comparison} in the Appendix show that the exclusion of WISE data does not change the results of Figure \ref{SED_comp} and does not have a systematic effect on the derived stellar mass and SFR{}.  This can in part be understood by the fact that in many galaxies, such as the one shown in Figure \ref{SEDs}, the SED fit is already constrained by the data just below 5 micron and remains unchanged with or without WISE photometry included in the fit.

We include AGN emission templates when we fit the samples (S1-Lum-AGN and S1-Lum-AGN-WISE) of galaxies hosting X-ray luminous AGN{}. However, we do not include AGN emission templates in the SED fit of the sample (S2-No-Lum-AGN and S2-No-Lum-AGN-WISE) of galaxies without X-ray luminous AGN because (a) it is computationally expensive to fit the AGN emission models to $> 100,000$ galaxies and (b) when we do include the AGN emission in the SED fit of a small subsample of galaxies in S2-No-Lum-AGN-WISE we find that $f_{\rm AGN}$ is extremely low ($< 0.05$) for $90 \%$ of the sample, moderately low ($< 0.2$) for $8 \%$ of the sample, and that there is no systematic change in the derived SFRs.

\section{Derived Stellar Mass and SFR of Sample Galaxies} \label{properties}
We discuss in the following subsections the stellar mass completeness, stellar mass function (SMF) (Section \ref{mass-distrib}) and SFR distributions (Section \ref{sfr-distrib}) of our sample of galaxies with (S1-Lum-AGN) and without (S2-No-Lum-AGN) X-ray luminous AGN{}.

\begin{figure*}
    \includegraphics[scale=0.7]{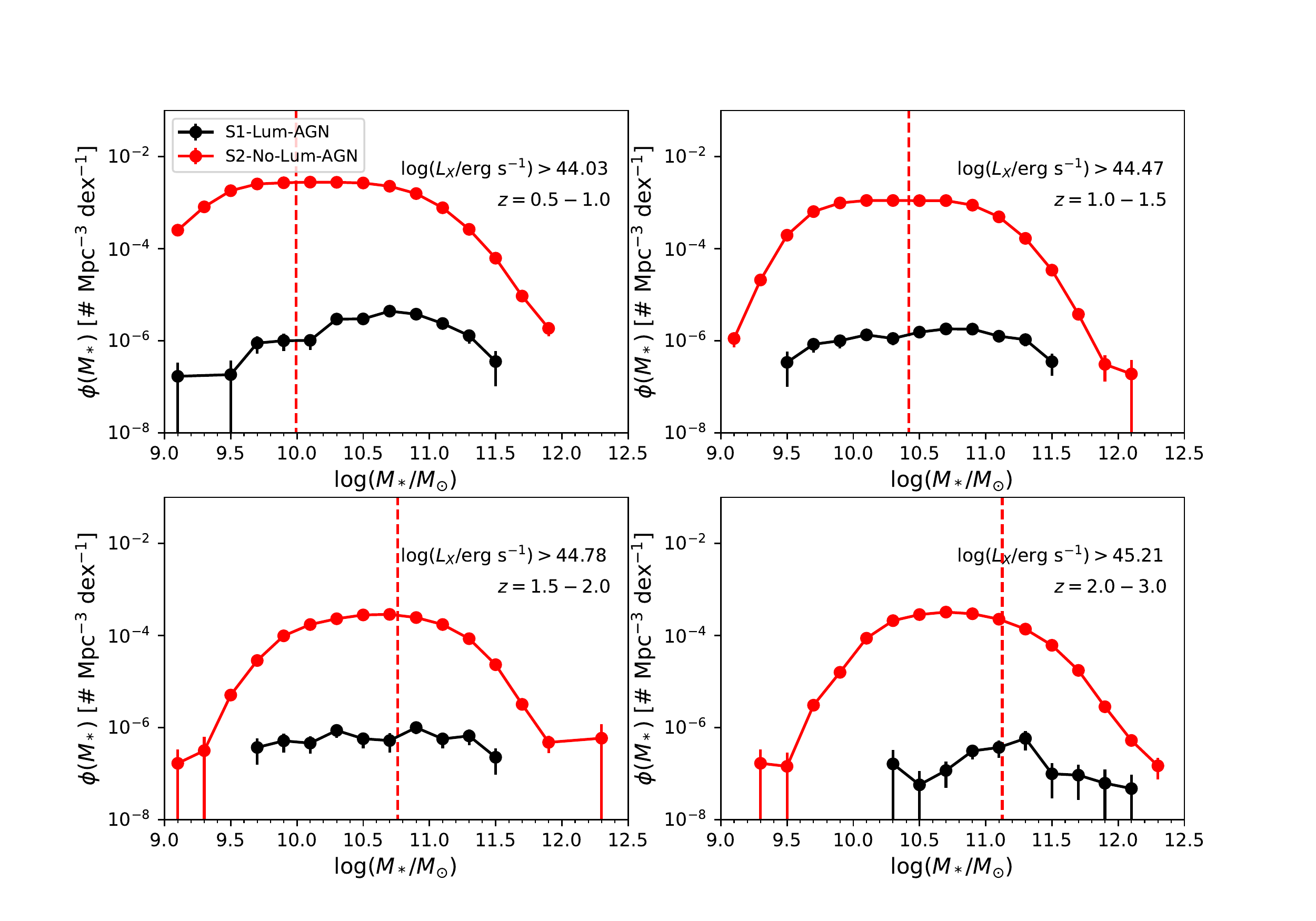}
    \caption{Galaxy stellar mass function (SMF) for our samples of galaxies with (S1-Lum-AGN, black) and without (S2-No-Lum-AGN, red) X-ray luminous AGN at four different redshift ranges. All galaxies with X-ray luminous AGN (S1-Lum-AGN) are complete in X-ray luminosity at their respective redshift bin; we show the X-ray completeness limit in each panel. The dashed vertical line indicates the stellar mass completeness of the sample of galaxies without X-ray luminous AGN (S2-No-Lum-AGN) at each bin. The error bars at each stellar mass bin are Poisson errors. The SMF is calculated using the $1/V_{\rm max}$ correction described in Section \ref{gal_xray_agn} for the X-ray luminosity function. We find that the SMF of galaxies with X-ray luminous AGN is much lower than the SMF of galaxies without X-ray luminous AGN by $\sim 2$ orders of magnitude at $M_* > 10^{11} \ M_{\odot}$ and by $\sim 3$ orders of magnitude at $M_* < 10^{11} \ M_{\odot}$.}
    \label{SMF}
\end{figure*}

\subsection{Distribution of Stellar Masses} \label{mass-distrib}
To estimate stellar mass completeness for each sample, we follow the procedure described in \cite{2010A&A...523A..13P} and \cite{2013A&A...558A..23D}. As per this method, we assume that the mass completeness limit of a survey can be estimated from the mass of the least massive galaxy that can be detected in a given bandpass with a magnitude equal to the magnitude limit of the survey in that bandpass. At each redshift, we select a representative sample from the faintest $20 \%$ of galaxies and scale their stellar mass, $\log(M_*)$, using the following equation:

\begin{figure*}
    \includegraphics[scale=0.7]{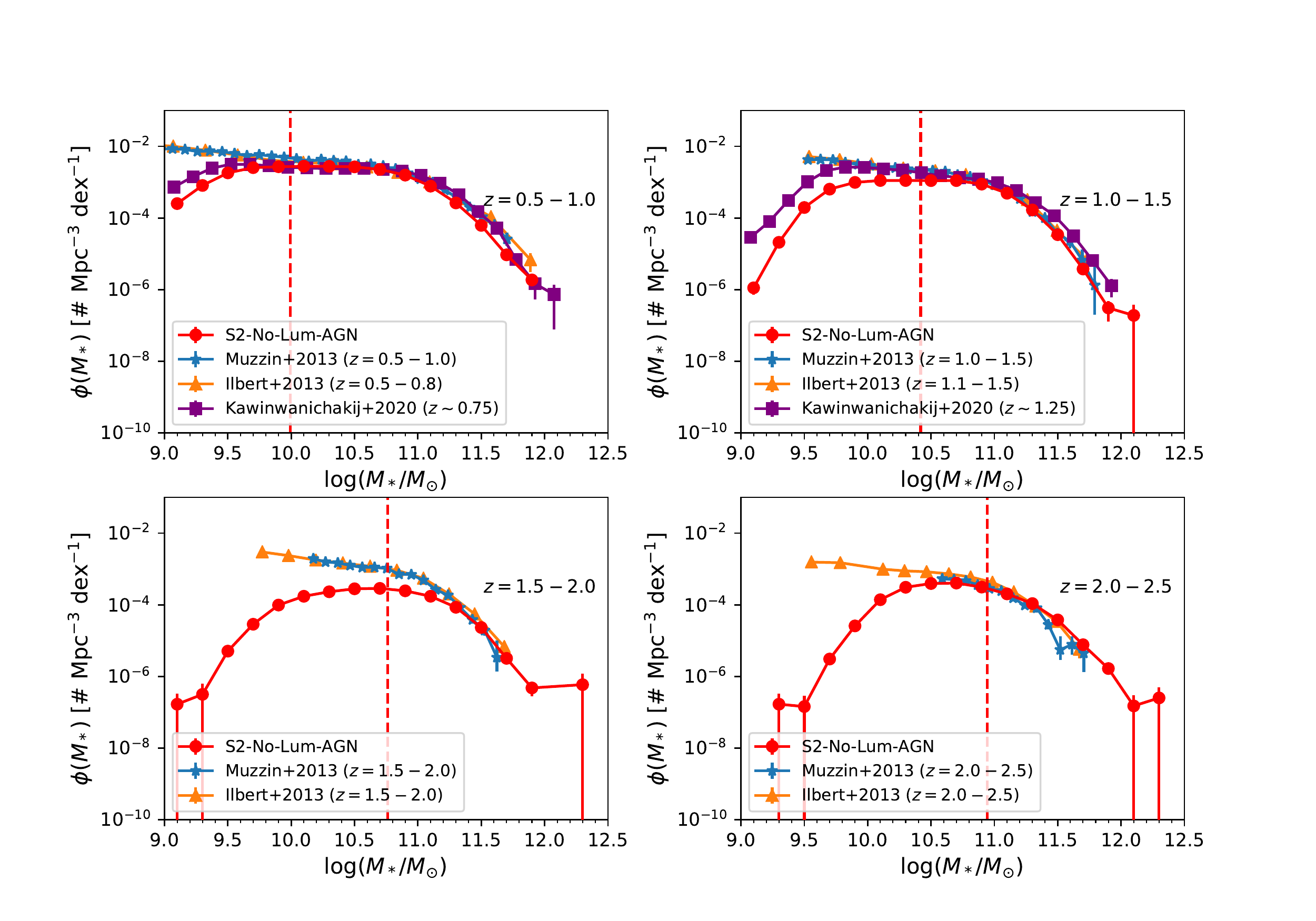}
    \caption{he observed galaxy SMF of our sample of galaxies without X-ray luminous AGN (S2-No-Lum-AGN, red) in four redshift bins with observed literature values at the corresponding redshifts plotted from \protect\cite{2013ApJ...777...18M} at $z=0.5-2.5$ (blue, stars), \protect\cite{2013A&A...556A..55I} at $z=0.5-2.5$ (orange, triangles), and \protect\cite{2020ApJ...892....7K} at $z \sim 0.75$ and $z \sim 1.25$ (purple, squares). As in Figure \ref{SMF} for our sample, the red dashed line represents the stellar completeness limit of our sample, error bars represent Poisson errors, and the SMF is calculated using the $1/V_{\rm max}$ method. We end the last redshift bin here at $z=2.5$, instead of $z=3.0$, for comparison purposes. We find good agreement with other observed SMFs at all redshifts for galaxies with $\log(M_* / M_{\odot}) > 11$ and at all stellar masses above our completeness limit for all redshift ranges except $z=1.5-2.0$ (see text).}
    \label{SMF_comparison}
\end{figure*}

\begin{equation}
    \log(M_{*,m=m_{\rm lim}}) = \log(M_*) + 0.4(m - m_{\rm lim})
\end{equation}

Here, $m$ is the measured AB magnitude, and $m_{\rm lim}$ is the AB magnitude limit of the survey in a given bandpass. After scaling the stellar masses of the faintest $20 \%$ of objects, we take the 95th percentile of the scaled mass distribution to be the mass completeness limit at each redshift bin. For our $K_S$-band selected sample, the $K_S$-band limiting magnitude is the $5 \sigma$ depth magnitude (22.4) measured in NEWFIRM. If we use this as $m_{\rm lim}$ in equation (2) to estimate the stellar mass completeness for sample S2-No-Lum-AGN, we find $\log(M_{*,95\% \rm  lim}/M_{\odot}) = 9.99, \ 10.42, \ 10.76, \ 11.12$ at $z=0.5-1.0, \ z=1.0-1.5, \ z=1.5-2.0$, \ $z=2.0-3.0$, respectively, for the mass completeness limit.

Although different bandpasses, such as the IRAC 3.6 \& 4.5 $\mu$m bandpasses, are better suited to tracing the stellar mass buildup of galaxies at $z \sim 2$, the \cite{2010A&A...523A..13P} method for estimating stellar mass completeness takes into account the range in mass-to-light ($M/L$) ratios at different redshifts using a bandpass that is not close to rest-frame $K$. We find that the stellar mass completeness limits do not change by more than $\sim 0.1$ dex at all redshifts if we use IRAC 3.6 $\mu$m-band photometry instead of $K_S$-band photometry to estimate stellar mass completeness. We also find similar stellar mass completeness limits when using 4.5 $\mu$m-band photometry to those estimated from $K_S$-band photometry (i.e., do not vary by more than $\sim 0.1$ dex) at $z < 2$. At $z=2.0-3.0$, the stellar mass completeness limits estimated from IRAC 4.5 $\mu$m-band photometry is 0.3 dex lower than when estimated from $K_S$ band photometry. Because our sample is a $K_S$-band selected sample, and stellar mass completeness limits do not vary much when using $K_S$, 3.6, or 4.5 $\mu$m-band photometry, we decide to use the $K_S$ limiting magnitude and photometry to estimate stellar mass completeness. 

Lastly, we apply this method to our sample of galaxies with X-ray luminous AGN (S1-Lum-AGN) to estimate stellar mass completeness limits for that sample and find that the stellar mass completeness limits of S1-Lum-AGN are lower by 0.1-0.2 dex at $z < 1.5$, and by 0.3-0.5 dex at $z > 1.5$, when compared to those of our sample of galaxies without X-ray luminous AGN (S2-No-Lum-AGN). We use the more stringent completeness limit found from S2-No-Lum-AGN as the stellar mass completeness limits for both samples in this analysis.

In Figure \ref{SMF}, we plot the observed SMF of galaxies with (S1-Lum-AGN) and without (S2-No-Lum-AGN) X-ray luminous AGN in four different bins of redshift across $z=0.5-3$. The SMF is estimated using the $1/V_{\rm{max}}$ method described in Section \ref{gal_xray_agn} for the X-ray luminosity function. The dashed line in each panel indicates the completeness limit of the sample S2-No-Lum-AGN{}. \textit{The SMF for the sample (S1-Lum-AGN) of galaxies with X-ray luminous AGN shows that the number density of such systems is about three orders of magnitude lower than galaxies without X-ray luminous AGN at $\log(M_*/M_{\odot}) < 11$ and about two orders of magnitude lower at $\log(M_*/M_{\odot}) > 11$, at all redshifts}. All galaxies in S1-Lum-AGN are complete in X-ray luminosity at each given redshift range, and AGN-host galaxies with $L_X > 10^{43}$ erg s$^{-1}$ have been shown to exhibit a turnover with decreasing number densities towards lower masses at $\log(M_*/M_{\odot}) < 11$ \citep{2016A&A...588A..78B}, which is consistent with what we observe in our SMF of galaxies with X-ray luminous AGN{}. 

We compare the observed SMF of our sample of galaxies without X-ray luminous AGN (S2-No-Lum-AGN) to the observed SMF from other studies in Figure \ref{SMF_comparison}. We plot the observed values of \cite{2013ApJ...777...18M} and \cite{2013A&A...556A..55I} at $z=0.5-2.5$ and the observed values of \cite{2020ApJ...892....7K} at $z \sim 0.75$ and $z \sim 1.25$. Our observed SMF for S2-No-Lum-AGN agrees well with those from other studies at all redshifts for galaxies with $\log(M_* / M_{\odot}) > 11$ and at all redshifts, except $z=1.5-2.0$, at stellar masses above the 95$\%$ completeness limit. The discrepancy between our SMF values and those of \cite{2013ApJ...777...18M} and \cite{2013A&A...556A..55I} at $z=1.5-2.0$ could arise from a number of factors, such as cosmic variance, since both \cite{2013ApJ...777...18M} and \cite{2013A&A...556A..55I} probe the same small area ($< 1.6$ deg$^2$) on the sky, and/or a lack of photometric filters in our data that allow us to probe certain features in model spectra during SED fitting that can break degeneracies between different redshifts and cause us to miss certain galaxies at $z=1.5-2.0$. Although we do not believe the latter to be the full cause of this discrepancy, we do not have other SMFs to compare to at this redshift range and our SMF agrees with that of Sherman et al. (submitted) at $z=1.5-2.0$.

\subsection{Distribution of Star Formation Rates} \label{sfr-distrib}

\begin{figure*}
    \centering
    \includegraphics[scale=0.6]{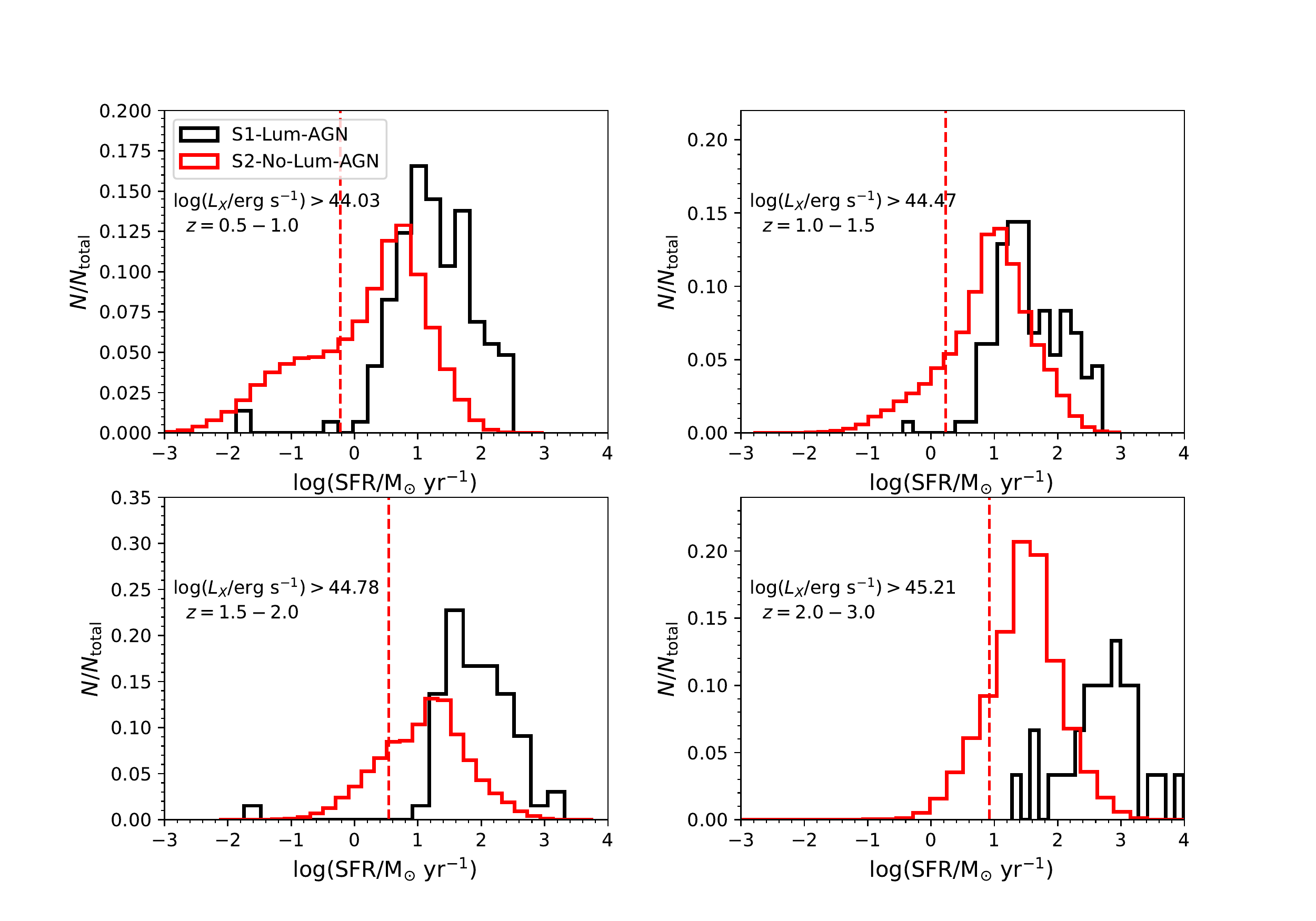}
    \caption{We plot the normalized histograms of the log of the measured SFR for our sample of galaxies with (S1-Lum-AGN, black) and without (S2-No-Lum-AGN, red) X-ray luminous AGN across four different redshift bins. The measured SFR refers to the intrinsic dust-corrected SFR derived from the SFH produced by SED fitting with CIGALE of the observed photometry from the UV to the IRAC 4.5 $\mu$m band.  Note that galaxies with X-ray luminous AGN have a distribution of intrinsic SFRs skewed towards higher values than galaxies without X-ray luminous AGN{}.  We also plot as dotted lines our estimated $5 \sigma$ completeness limit for the observed dust-extincted FUV-based SFR, which is estimated from the  $5 \sigma$ detection limit in the $u$-band and $g$-band filters, which most closely trace the rest-frame 1500 \AA{} luminosity at  $z=0.5-1.5$  and $z=1.5-3.0$, respectively. See text in section 4.2 for details.}
    \label{SFR_hist}
\end{figure*}

In Figure \ref{SFR_hist}, we show the normalized SFR distribution for galaxies with (S1-Lum-AGN) and without (S2-No-Lum-AGN) X-ray luminous AGN{}. The SFR histograms are normalized by dividing the number of galaxies in each bin by the total number of objects in the redshift bin, such that the sum of all the histogram bins is equal to 1. We remind the reader that the SFRs reported here are the instantaneous SFRs given by equation (1) in Section \ref{sed_fitting} from the SFH of the SED fit. The SFR from an SED fit refers to the intrinsic extinction-corrected SFR and is typically constrained by fitting observed photometry ranging from the UV to the IRAC 4.5 $\mu$m band.

Figure \ref{SFR_hist} shows that the SFRs for galaxies with X-ray luminous AGN (S1-Lum-AGN) have a distribution skewed towards higher values, and therefore, have higher SFRs than galaxies without X-ray luminous AGN (S2-No-Lum-AGN). The median SFR values for both samples at each redshift range is $\log(\rm{SFR}/M_{\odot}\rm{yr}^{-1})$ = $1.18,\ 1.41,\ 1.77,\ 2.68$ for S1-Lum-AGN at $z=0.5-1,\ z=1-1.5,\ z=1.5-2$, and $z=2-3$, respectively, and $\log(\rm{SFR}/M_{\odot}\rm{yr}^{-1})$ = $0.41,\ 0.96,\ 1.15,\ 1.49$ for S2-No-Lum-AGN, respectively. The median SFR values are roughly a factor of 5-10 higher for galaxies with X-ray luminous AGN than galaxies without X-ray luminous AGN{}.

It is often useful in studies of galaxy evolution to estimate the completeness limit down to which one can measure the SFR{}. This is not so easy to do for our analysis as our measured SFRs refer to the intrinsic extinction-corrected SFR derived by fitting observed photometry over a wide range of wavelengths from the UV to the IRAC 4.5 $\mu$m band using a large set of  SED templates with many free parameters. Getting a true completeness limit for this measured SFR would be complicated and requires the analysis of many observed bands and SED templates. Instead, we attempt here to make a simpler estimate of the observed dust-extincted FUV-based SFR by using the detection limits in just a few bands (DECam $u$ and $g$ bands) that trace the rest-frame UV luminosity from massive SF across our redshift range of interest.

For this estimate of the detection limit for the observed dust-extincted SFR, we start by converting the observed magnitude closest to the rest-frame FUV wavelength into a luminosity (i.e., DECam $u$-band at $z < 1.5$, DECam $g$-band at $z > 1.5$). We then compute the 100 Myr FUV-based dust-extincted SFR for our sample of galaxies without X-ray luminous AGN (S2-No-Lum-AGN), using the SFR calibrator from \cite{2011ApJ...741..124H} and assume a Kroupa IMF \citep{2001MNRAS.322..231K}.  The fact that SFRs computed in CIGALE assume a \cite{2003PASP..115..763C} IMF is not a problem as  the shifts in SFR calculated from either IMF are essentially negligible \citep{2014ApJS..214...15S}.

We estimate the $5 \sigma$ completeness limit for the observed dust-extincted FUV-based SFR similarly to how we estimate stellar mass completeness. That is, we take the 20\% faintest galaxies in each redshift bin, using the $u$-band at $z < 1.5$ and the $g$-band at $z > 1.5$, and scale their FUV SFRs to the value they would have if their magnitude was equal to the limiting magnitudes of the survey (25.0 AB magnitude for $u$-band, 24.8 AB magnitude for $g$-band), which are the $5 \sigma$ magnitude depths reported in \cite{2019ApJS..240....5W} using the following equation:

\begin{equation}
    \log (\rm{SFR}_{\rm{FUV,\rm{lim}}}) = \log (\rm{SFR}_{\rm FUV}) + 0.4 (m - m_{\rm{lim}})
\end{equation}

\noindent
From the resulting distribution of values, we take the 95th percentile value as a lower limit to the $5 \sigma$ completeness limit for the observed dust-extincted FUV-based SFR{}. The values are $\log (\rm{SFR}/ M_{\odot}\rm{yr}^{-1}) = -0.22$, $0.23$, $0.54$, and $0.92$ at $z=0.5-1.0$, $z=1.0-1.5$, $z=1.5-2.0$, and $z=2.0-3.0$, respectively.  We thus expect our survey to detect dust-extincted FUV-based SFR above these values at the $5 \sigma$ level.  In practice, we do not use a S/N cut for the $u$ or $g$-band fluxes in our analysis, so we should be able to measure dust-extincted FUV-based SFRs below these $5 \sigma$ limit values.

We plot  our estimated $5 \sigma$ completeness limit for the observed dust-extincted FUV-based SFR as dotted lines on Figure \ref{SFR_hist}, which also  shows the distribution of  measured SFRs for the samples of galaxies with (S1-Lum-AGN) and without (S2-No-Lum-AGN) X-ray luminous AGN{}.  It is important to bear in mind that the measured SFRs from the SED fit and the observed dust-extincted FUV-based SFRs are two very different quantities. The measured SFRs from the SED fit refer to the intrinsic dust-corrected SFR of a galaxy based on the SED fit and observed photometry from the UV to the IRAC 4.5 $\mu$m band and are likely higher than the observed dust-extincted FUV-based SFRs. The fact that we see some measured intrinsic SFR values lower than the 5 $\sigma$ completeness limit for the observed dust-extincted FUV-based SFR is likely due to the fact that we do not use a S/N cut for the $u$ or $g$-band fluxes in our analysis.

\subsection{Testing CIGALE SFRs Against Previously Published Empirical SFRs}

In this section and the following section, we explore whether the SFRs derived by CIGALE for galaxies hosting X-ray luminous AGN are reliable by performing two separate tests which are based, respectively, on previously published SFRs of real galaxies with X-ray luminous AGN and a set of synthetic mock galaxy SEDs. 

For the first test, we compare SFRs derived from CIGALE with SFRs from the CANDELS catalog \citep{2011ApJS..197...35G, 2011ApJS..197...36K}, where SFRs are derived from SED fitting using near-UV to near-IR photometry \citep[see][for description of SFR estimates]{2019MNRAS.485.3721Y,2017ApJ...842...72Y}. The SFRs presented in \cite{2019MNRAS.485.3721Y} are median values of SFRs obtained from separate teams who perform SED fitting of the same sample and take the total SFR values from the SFH of the SED fit. We do not use the FIR-derived SFRs presented in \cite{2019MNRAS.485.3721Y,2017ApJ...842...72Y} as they are not corrected for AGN emission and thus would overestimate the SFR when converting between FIR luminosity and SFR (see Section 3.1 and Fig. 3). We select 38 sources in the CANDELS catalog with a Chandra X-ray detection with $L_X > 10^{43}$ erg s$^{-1}$ in order to obtain SFRs, using CIGALE, for a sample of galaxies with X-ray luminous AGN{}. For these sources, we select available photometry from the COSMOS and Ultra-deep Survey (UDS) fields in CANDELS (i.e., CFHT MegaCam $u,g,r,i,z$-bands, NEWFIRM $K_S$-band \citep{2009PASP..121....2V}, IRAC 3.6 and 4.5 $\mu$m) to perform SED fitting using photometric bands that cover the same wavelength ranges that we use to perform SED fitting of our S1-Lum-AGN and S2-No-Lum-AGN samples. We also adopt the redshifts from the CANDELS catalog when we perform the SED fit in CIGALE.

\begin{figure}
    \centering
    \includegraphics[scale=0.5]{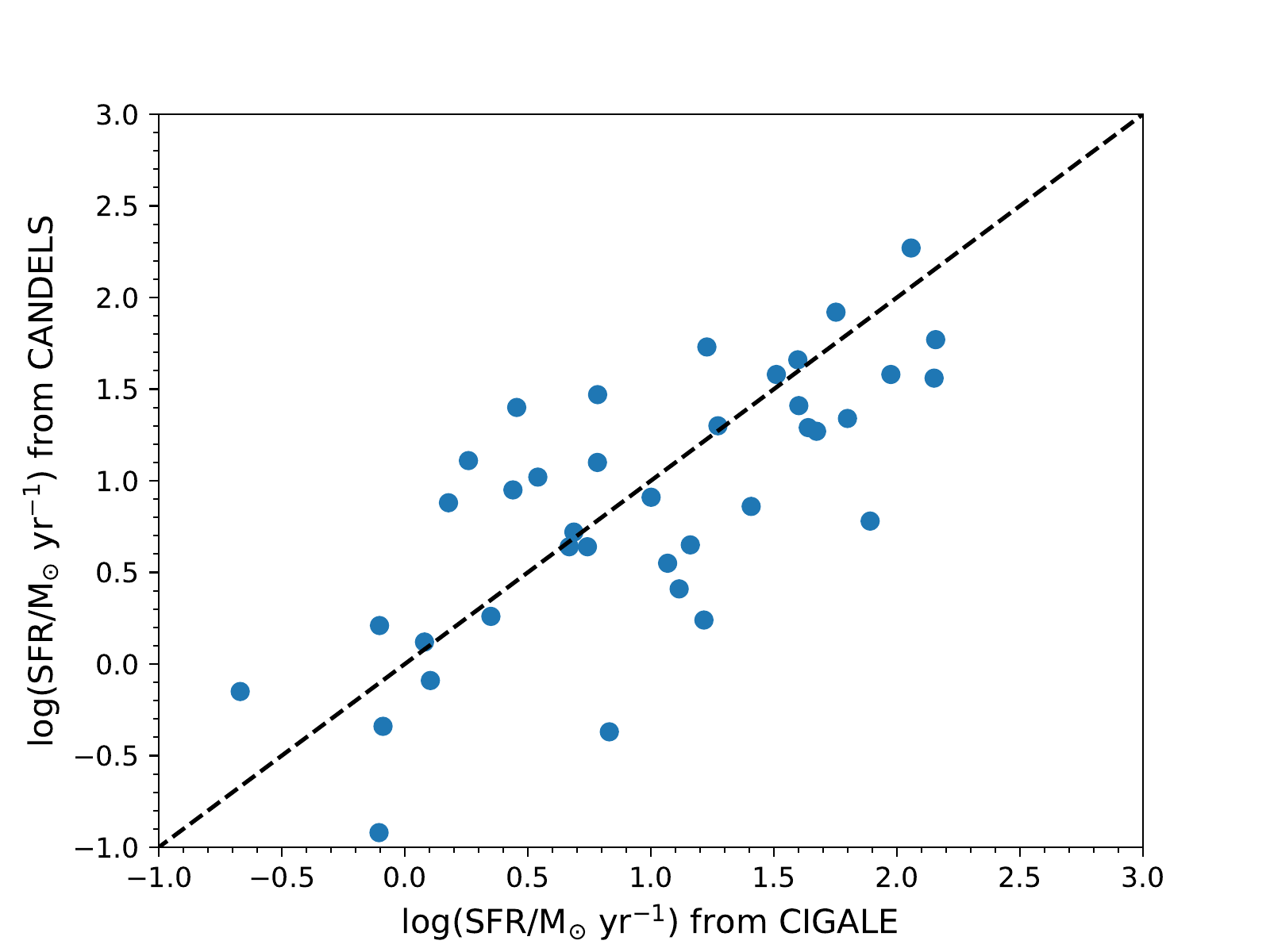}
    \caption{SFRs from CANDELS (y-axis) compared to SFRs derived from CIGALE using CFHT ($u,g,r,i,z$), NEWFIRM ($K$-band), and IRAC (3.6 and 4.5 $\mu$m) photometry for a sample of 38 galaxies with X-ray luminous AGN having $L_X > 10^{43}$ erg s$^{-1}$. Also shown is the one-to-one line (dashed, black).}
    \label{sfr_comp1}
\end{figure}

In Figure \ref{sfr_comp1}, we show the SFRs obtained for the CANDELS sample described above using CIGALE to the published SFR values of the CANDELS catalog described in \cite{2011ApJS..197...35G, 2011ApJS..197...36K}. Only at the highest SFRs does the mean CANDELS SFR deviate from the one-to-one line by a factor of $\sim 0.5$ dex and we note only one source has an SFR that varies by an order of magnitude between the CIGALE SFR and CANDELS reported SFRs.

\subsection{Testing CIGALE SFRs Against SFRs from Mock SEDs}

In this section, we explore whether the SFRs derived by CIGALE for galaxies hosting X-ray luminous AGN are reliable by performing tests on a set of synthetic mock galaxy SEDs. We use CIGALE to generate mock galaxy SEDs with AGN emission and then perform SED fitting on the mock galaxy SED fluxes. We generate two sets of mock galaxy SEDs using the stellar population synthesis models of \cite{2003MNRAS.344.1000B}, assume a \cite{2003PASP..115..763C} IMF, and for one set of mocks we employ the \cite{2006MNRAS.366..767F} AGN emission templates with an updated version of the \cite{2007ApJ...657..810D} dust emission models, and for the other set of mocks we employ the \cite{2014ApJ...784...83D} combined dust and AGN emission models. We use varying dust and AGN emission templates when generating mock galaxy SEDs in order to test whether CIGALE can recover the properties of mock galaxies produced with varying models using the same emission templates discussed in Section \ref{sed_fitting}.

\begin{figure}[H]
\centering
\includegraphics[scale=0.55]{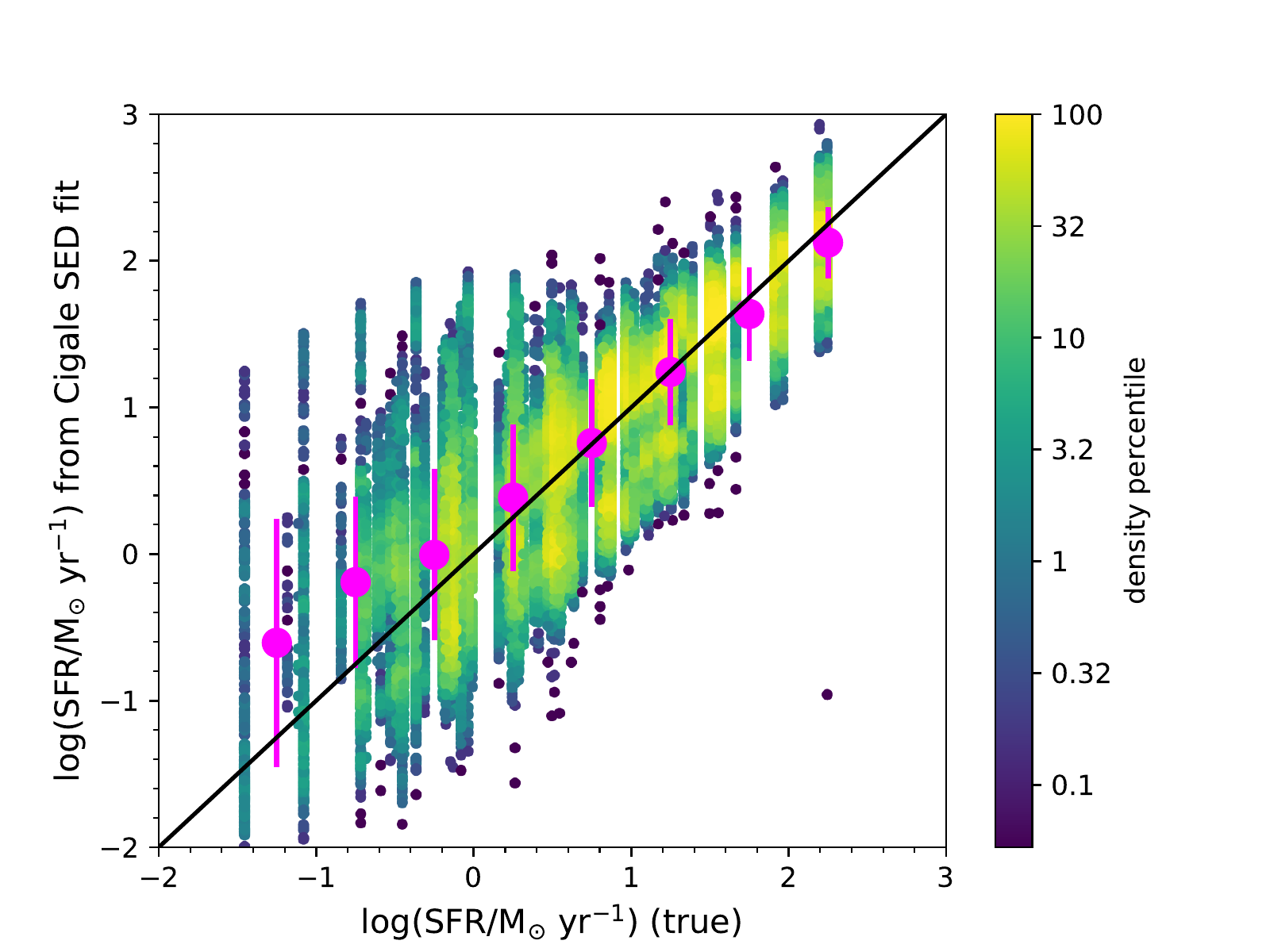}
\caption{Comparison of SFRs from the mock galaxies (x-axis) versus SFRs obtained from the SED fit with CIGALE to the mock galaxy fluxes (y-axis). We do not include WISE photometry in this test in order to resemble the fits we are doing with our S1-Lum-AGN sample. Also shown is the one-to-one line and the mean of the $\log(\rm SFR)$ (magenta circles) with the standard deviation represented by the error bars. All points are color-coded by their density on the x (mock SFR) and y (fit SFR) plane, where density is calculated by counting number of neighbors around each data point inside a circular aperture. We find relatively good agreement between the true SFRs of the mock galaxies and the CIGALE-derived SFRs above $\sim 1 \ M_{\odot} \rm{yr}^{-1}$. At the low end of the true SFRs, some mock galaxies have high CIGALE-derived SFRs, however, only $2 \%$ of the mock galaxies have true SFRs that differ from their CIGALE-derived SFR by a factor of 10 or more and only $10 \%$ of mock galaxies have true SFRs that differ from the CIGALE-derived SFRs by a factor of 4 or more.} 
\label{mock_SED_comp}
\end{figure}

When we generate mock galaxy SEDs in CIGALE, we obtain mock observation fluxes in the DECam $u,g,r,i,z$ bands, NEWFIRM K-band, and IRAC 3.6 and 4.5 $\mu$m bands. In order to assign flux errors for a given photometric band, we calculate the mean S/N of all sources in our data as a function of magnitude. We then use this mean S/N to assign flux errors to mock sources depending on what their magnitude is. This method produces reasonable flux errors in the mock photometry that match our data. At this stage we perturb the fluxes within the errors by drawing from a normal distribution centered around the mock flux with a standard deviation equal to the photometric error. With the new photometric fluxes and flux errors, we perform the same cuts on the mock objects as the real data to obtain a mock photometric sample that resembles our data in terms of magnitude and S/N ratios. That is, we only perform SED fitting on sources with a S/N greater than 5 in the $K_S$-band, a S/N greater than 2 in the two IRAC bands, and require a flux detection in the $u,g,r,i,z$ bands.

 In Figure \ref{mock_SED_comp}, we compare the SFR derived from SED fitting with CIGALE to the true SFR of the mock galaxy. We only show the results for the mock galaxies generated from the \cite{2006MNRAS.366..767F} and \cite{2007ApJ...657..810D} AGN and dust emission templates. However, our results do not change for the mock galaxies where AGN and dust emission is generated from the \cite{2014ApJ...784...83D} templates, suggesting that the choice of AGN or dust emission templates in CIGALE should not have a large impact on the results. 

In Figure \ref{mock_SED_comp}, we see good agreement between the CIGALE-derived SFRs and the true SFRs of mock galaxies above $\sim 1 \  M_{\odot} \rm{yr}^{-1}$ with a scatter of 0.5 dex. At the low SFR end there are some mock galaxies for which the CIGALE-derived SFRs are significantly higher than the true SFRs. However, only $2 \%$ of the mock galaxies have $|\Delta \log(\rm SFR)| > 1$ (where $\Delta \log (\rm SFR) = \log (\rm SFR_{\rm fit}) - \log (\rm SFR_{\rm true})$) and only $10 \%$ of mock objects have $|\Delta \log(\rm SFR)| > 0.6$, meaning $90 \%$ of our mock sample has true SFRs that agree with the CIGALE-derived SFRs to within a factor of 3-4. Most of the large scatter between the true and fit SFRs lie at low SFRs ($< 1 \ M_{\odot} \rm{yr}^{-1}$), however, we note that most ($> 80 \%$) of our sample have SFRs above $1 \ M_{\odot} \rm{yr}^{-1}$. At the low end of the true SFRs, some mock galaxies have high CIGALE-derived SFRs, however, only $2 \%$ of the mock galaxies have true SFRs that differ from their CIGALE-derived SFR by a factor of 10 or more and only $10 \%$ of mock galaxies have true SFRs that differ from the CIGALE-derived SFRs by a factor of 4 or more. We find that CIGALE does not produce a systematic bias towards higher or lower values of SFRs when recovering SFRs from the true mock galaxy SEDs.

\begin{figure*}
\begin{center}
\includegraphics[scale=0.7]{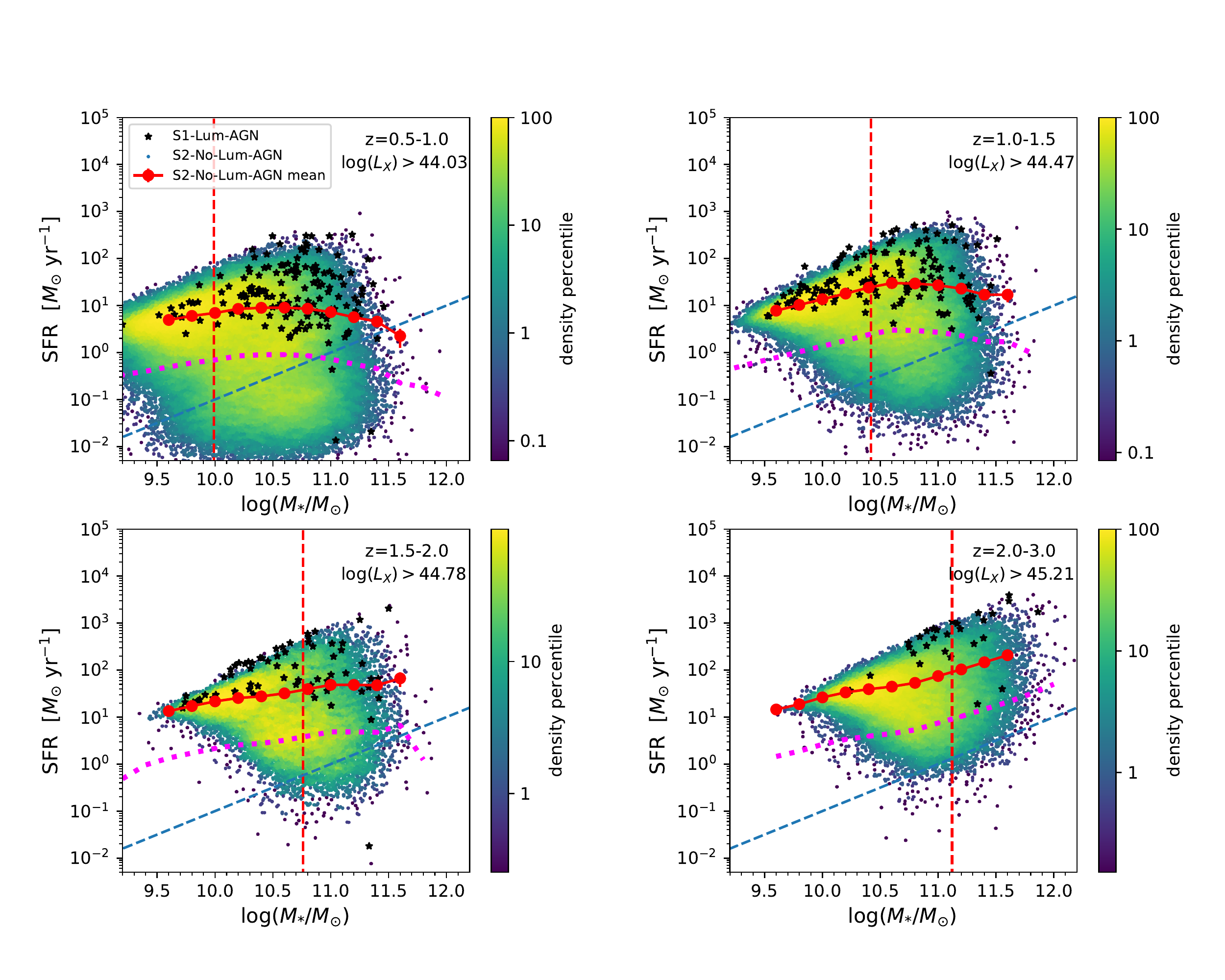}
\caption{SFR vs. stellar mass for our sample of galaxies with (S1-Lum-AGN, black stars) and without (S2-No-Lum-AGN, colored points) X-ray luminous AGN in four different redshift bins. The S2-No-Lum-AGN galaxies are color-coded by their density on the stellar mass-SFR plane (see text).  The dashed vertical line on each panel shows the stellar mass completeness limit in that bin (see Section \ref{mass-distrib}). The X-ray completeness limit for S1-Lum-AGN is shown in each redshift bin in log units of ergs per second. Also shown are the mean SFR of S2-No-Lum-AGN as a function of stellar mass (red circles), which we refer to as the main sequence, the line that falls 1 dex below the main sequence (dotted magenta)  and the line where the specific $\rm{SFR}$ is $10^{-11} \rm{ yr}^{-1}$ (blue dashed). It is striking that galaxies with X-ray luminous AGN have higher mean SFRs than galaxies without X-ray luminous AGN at a given stellar mass (see also Figure \ref{mass_sfr_mean}). Note also that very few galaxies with X-ray luminous AGN have quenched SF if we use the common definition of quenched galaxies as having a specific $\rm{SFR} < 10^{-11} \rm{yr}^{-1}$.}
\label{mass_sfr}
\end{center}

\end{figure*}

\begin{figure*}
\begin{center}
\includegraphics[scale=0.65]{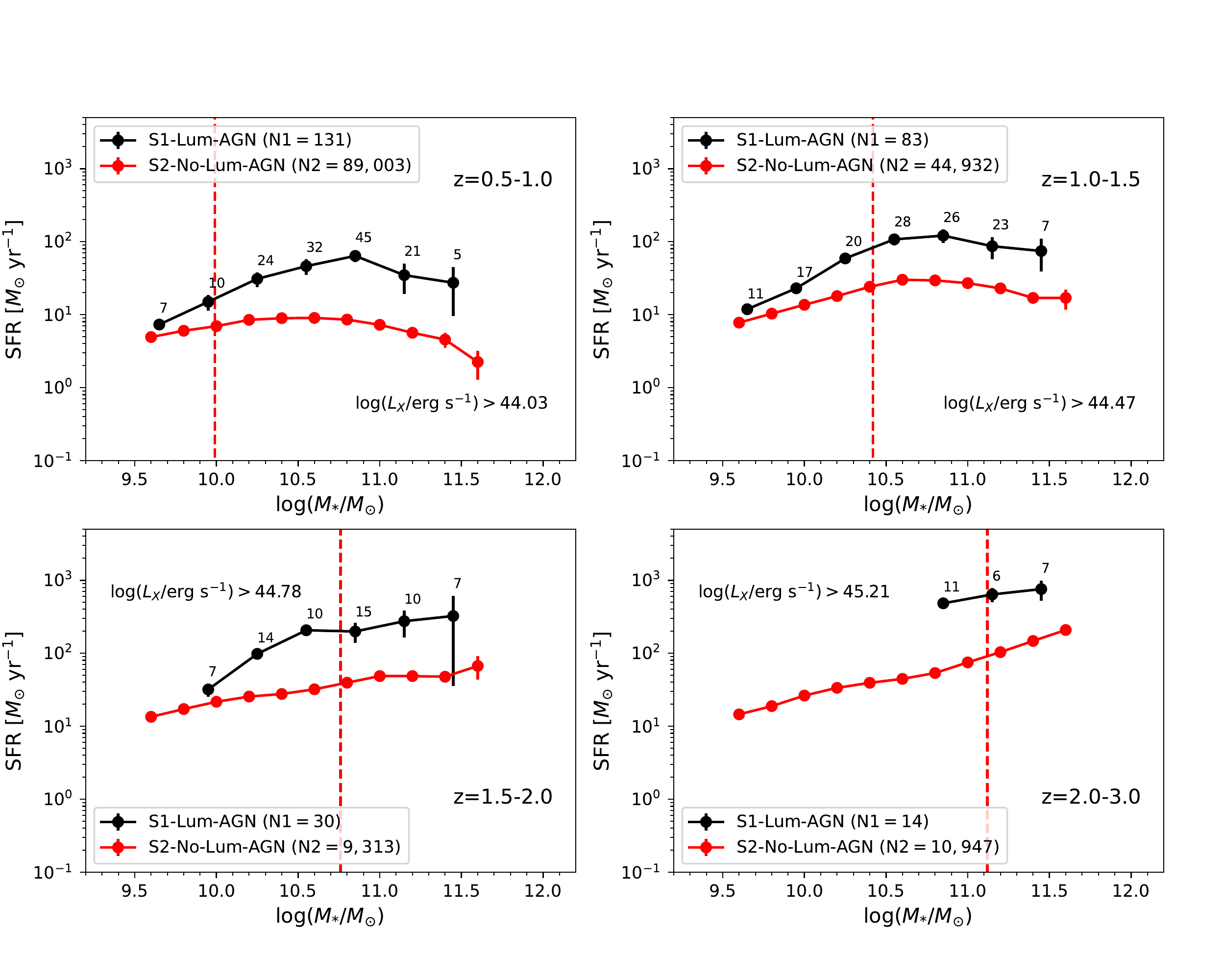}
\caption{The mean SFRs for our sample of galaxies with (S1-Lum-AGN, black circles) and without (S2-No-Lum-AGN, red circles, same as those shown in  Figure \ref{mass_sfr}) X-ray luminous AGN as a function of stellar mass across four redshift bins. We show the total number (N1 and N2) of galaxies in each sample above the stellar mass completeness limit (shown here as vertical dashed lines), as well as the number of galaxies with X-ray luminous AGN in each stellar mass bin above the X-ray luminosity completeness limit. Error bars are 1$\sigma$ values from a bootstrap analysis. The mean SFRs of galaxies with X-ray luminous AGN are higher by a factor of 3 to 10 than those of galaxies without X-ray luminous AGN at a given stellar mass.}
\label{mass_sfr_mean}
\end{center}
\end{figure*}

\section{Results}
In the following section we compare the properties of the sample galaxies with (S1-Lum-AGN)  and without (S2-No-Lum-AGN) X-ray luminous AGN at fixed redshift and stellar mass. Many studies have looked at the evolution of SFR with stellar mass across redshift \citep{2014ApJ...795..104W, 2014ApJS..214...15S, 2007ApJ...670..156D, 2007A&A...468...33E}, however, galaxies with X-ray luminous AGN are usually not included in these studies due to AGN emission contaminating the UV and IR wavelengths.  Although there have been a few studies that look at the relation between stellar mass and SFR of galaxies with luminous AGN \citep{2015MNRAS.452.1841S, 2018A&A...618A..31M, 2012A&A...540A.109S}, they do not derive stellar masses and SFRs for their galaxy samples with  and without luminous AGN in a self-consistent manner,  and their control sample of galaxies without luminous AGN is at least an order of magnitude smaller than the control sample (S2-No-Lum-AGN) in our study.

\subsection{SFR as a Function of Stellar Mass and Redshift} \label{results}

In Figure \ref{mass_sfr} we show the stellar mass and SFR of our sample of galaxies with (S1-Lum-AGN) and without (S2-No-Lum-AGN) X-ray luminous AGN{}.  The galaxies in S2-No-Lum-AGN are color-coded by their density on the stellar mass-SFR plane, where density is calculated by counting number of neighbors around each data point inside a circular aperture with a radius equal to 0.05. The color bar shows the percentile value of the data point's density, where the 100th percentile value corresponds to the highest density and the 0.1 percentile value corresponds to the data points whose density is 1/10th the maximum density. Also shown in Figure \ref{mass_sfr} is the mean SFR in different bins of stellar mass for S2-No-Lum-AGN, which we refer to as the main sequence, with 1$\sigma$ errors calculated through a bootstrap method. We calculate errors in the mean SFR by resampling (i.e., drawing randomly from) the SFR distribution inside a bin of stellar mass $x$ times, where $x$ is the number of objects inside the bin, build a sample from the random draws and calculate the mean for that sample. We draw objects with replacement so the same object in a given bin can be sampled more than once. We do this 1,000 times and make a distribution of mean SFR values from each resampling and take the 16th and 84th percentile of this distribution to calculate the error on the mean of the SFR{}. 

We make note of the four galaxies with X-ray luminous AGN (S1-Lum-AGN) in Figure \ref{mass_sfr} that have SFRs low enough to be considered outliers with respect to the rest of the galaxies in S1-Lum-AGN on the mass-SFR plane. Two of these objects exist at $z=0.5-1.0$, one object exists at $z=1.0-1.5$, and one object exists at $z=1.5-2.0$. We have inspected the DECam and IRAC images, the flags in the source extractor catalogs, as well as the quality of the SED fits of the two outliers at $z=0.5-1$ and the two outliers at $z=1-1.5$ and $z=1.5-2$. The inspection revealed nothing out of the ordinary or anything that could impair the photometry for these objects. Furthermore, the reduced $\chi^2$ of the SED fit to the photometry is less than 5 in all cases, indicating a good fit. As a result, we have no reason to believe that the measured SFRs are erroneous.

In Figure \ref{mass_sfr_mean}, we show the mean SFRs of our sample of galaxies with (S1-Lum-AGN) and without (S2-No-Lum-AGN) X-ray luminous AGN as a function of stellar mass across four redshift bins. Error bars are determined from the bootstrap method described above. In all redshift bins we find that the mean SFRs of galaxies with X-ray luminous AGN are higher by a factor of 3 to 10 than those of galaxies without X-ray luminous AGN at a given stellar mass. This suggests that X-ray luminous AGN tend to coexist in galaxies with enhanced SFRs. In Section \ref{discussion}, we discuss how these results fits into possible evolutionary scenarios connecting AGN and SF activity.

Our results are consistent with those of \cite{2012A&A...540A.109S} who find that the SF activity of galaxies with luminous AGN is enhanced with respect to a mass-matched sample of inactive galaxies (i.e., galaxies without AGN activity) at $z=0.5-2.5$. Furthermore, they find that the level of enhancement in SF activity amongst galaxies with luminous AGN is higher for galaxies with high X-ray luminosities ($L_X > 10^{43.5} \rm{erg s}^{-1}$). Our results are also consistent with those of \cite{2018A&A...618A..31M} who compare the mean SFRs of their sample of galaxies with X-ray luminous AGN to the SFR values of star forming galaxies from \cite{2015A&A...575A..74S} at fixed stellar mass and find that the mean SFRs at fixed mass of galaxies with X-ray luminous AGN are higher than those of galaxies without X-ray luminous AGN{}. We also find that our results are consistent with those of \cite{2019ApJ...879...41K} and \cite{2019arXiv190804795K}. In \cite{2019ApJ...879...41K}, they find that AGN from $z=1-4$ have high SFRs and star formation efficiencies and show no signs of quenching. \cite{2019arXiv190804795K} find that some of the most luminous quasars in Stripe 82 at z=1-2 have the highest SFRs that are a factor of $\sim 3-7$ above the main sequence.

The results in \cite{2015MNRAS.452.1841S} may appear to contradict the results we find here, as they claim their sample of AGN-host galaxies appear to have SFRs that fall off the main sequence of star forming galaxies. However, a subsequent paper \cite{2017MNRAS.466.3161S} claims that when they mass-match their sample of AGN-host galaxies to a control sample, the AGN-host galaxies have higher SFR on average. The erroneous result in \cite{2015MNRAS.452.1841S} is explained by a mismatch between their sample of AGN-host galaxies and their comparative control sample: the mass distribution of their AGN-host galaxies is shifted towards higher values than the mass distribution of their sample of star forming galaxies, and thus, their AGN-host galaxies have lower specific SFR values typical of higher mass ($M_* > 10^{10.5} \ M_{\odot}$) galaxies.

In summary, we find that the mean SFR in galaxies with X-ray luminous AGN is significantly larger than in galaxies without X-ray luminous AGN at a given stellar mass. Our results are consistent with those from several earlier studies described above, but they are significantly more robust because our sample of galaxies without X-ray luminous AGN is 10 to 100 times larger than those of earlier studies, and we analyze both our AGN sample (S1-Lum-AGN) and our mass-matched non-AGN sample (S2-No-Lum-AGN) using the same SED fitting code and methodology. Our results are consistent with a scenario where the high SFR and AGN luminosity are triggered by processes that produce large gas inflow rates into the regions (on scales of a few hundreds to few kpc) typically associated with high SFRs, as well as the sub-pc region associated with the AGN accretion disk. We refer the reader to Section \ref{discussion} for a more detailed discussion of potential evolutionary sequences between AGN and SF activity. 

\begin{figure*}
    \centering
    \includegraphics[scale=0.55]{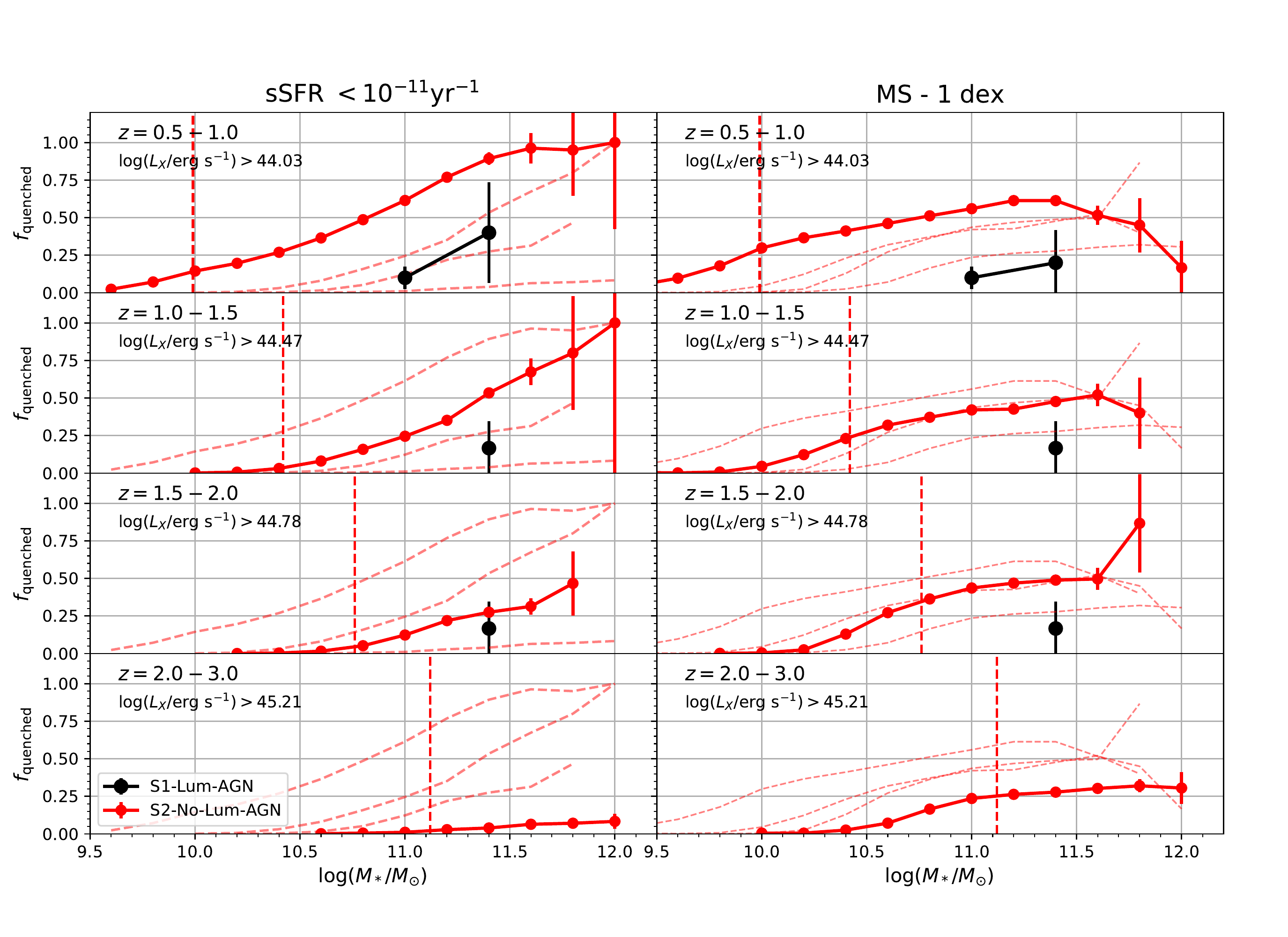}
       \caption{This figure shows the quenched fraction as a function of stellar mass in four different bins of redshift spanning $z=0.5-3$ using two definitions of quiescence, whereby quenched galaxies are defined as having a $\rm{sSFR} < 10^{-11} \rm{yr}^{-1}$ (left panels) or as having a SFR at least 1 dex below the main sequence at a given stellar mass (right panels). In all panels, we plot the fraction of galaxies that are quenched among the sample of galaxies with (S1-Lum-AGN; black circles) and without (S2-No-Lum-AGN; red circles) X-ray luminous AGN{}. The small dashed red lines in each panel represent the quenched fraction of S2-No-Lum-AGN in all the redshift ranges for easy visualization of the evolution of the quenched fraction with redshift. Poison errors are shown on this plot as error bars. The quenched fractions based on the two definitions of quiescence are roughly consistent at $z < 1.5$, but differ significantly at higher redshifts ($z = 1.5-3$) where the quenched fraction based on sSFR is lower by a factor of $\sim 2-3$ at $z=1.5-2$ and by a factor of $\sim 5-10$ at $z=2-3$ for galaxies with $M_* > 10^{11} \ M_{\odot}$. The quenched fractions are a strong function of stellar mass in each redshift bin and generally increase with stellar mass, except possibly at the very highest stellar masses.}
    \label{quenched_fraction} 
\end{figure*}

\subsection{Properties and Fraction of Galaxies with Quenched Star Formation} \label{quenched_galaxies}
In this section, we explore the quenched fraction of our sample of galaxies with (S1-Lum-AGN) and without (S2-No-Lum-AGN) X-ray luminous AGN at fixed stellar mass. The processes that quench SF (i.e., significantly suppress SF) are important for our understanding of galaxy evolution as they intimately regulate the growth of the stellar mass in galaxies. Theorists often invoke AGN feedback in simulations as a way to quench SF \citep{2011ApJ...738...16H, 2012ARA&A..50..455F, 2013MNRAS.436.3031V, 2015MNRAS.449.4105C, 2015ApJ...800...19R, 2016MNRAS.458..816H, 2017MNRAS.464.1854B} in massive galaxies and prevent the overproduction of massive galaxies relative to the observed mass and luminosity functions. By measuring the quenched fraction of massive galaxies with and without X-ray luminous AGN, we aim to shed light on the role of quenching mechanisms, such as AGN feedback. While most earlier studies estimate the mean quenched fraction (averaged over massive galaxies) in different redshift ranges, the unprecedented size of our sample of massive galaxies (e.g., $\sim 30,000$ galaxies with $M_* > 10^{11} \ M_{\odot}$) enables us to study how the quenched fraction ($f_{\rm quenched}$) varies with stellar mass at different redshifts (Figure \ref{quenched_fraction}).

In the literature, multiple methods are used to identify quenched galaxies. Numerous works define quenched galaxies as galaxies having specific SFR (sSFR) less than $10^{-11}\rm{yr}^{-1}$ \citep{2009MNRAS.397.1776F, 2013ApJ...768...92S}. Given the bimodality in color and star formation, it is also common to separate quenched galaxies from star-forming galaxies using a selection based on $U-V$ and $V-J$ colors \citep[UVJ diagram,][]{2007ApJ...655...51W, 2011ApJ...735...86W, 2013ApJ...777...18M,2018ApJ...858..100F}. In \cite{2018ApJ...858..100F} and \cite{2019MNRAS.485.4817D}, a separate method used to define quiescence based on distance from the main sequence is introduced, where galaxies that fall 1 dex or 2$\sigma$ below the main sequence are defined as quenched. 

For this work, we select quenched galaxies based on a galaxy's sSFR ($\rm{sSFR} < 10^{-11} \rm{yr}^{-1}$) and based on distance from the main sequence of star formation. In Figure \ref{mass_sfr}, the blue dashed line represents $\rm{sSFR}$ of $10^{-11} \rm{ yr}^{-1}$. If we follow the approach of \cite{2009MNRAS.397.1776F}, where quenched galaxies are defined as having a $\rm{sSFR} < 10^{-11} \rm{ yr}^{-1}$, then Figure \ref{mass_sfr} shows that very few galaxies with X-ray luminous AGN ($< 5 \%$) have quenched SF{}. For the second definition of quiescence based on distance from the main sequence, we define a quiescent sample using an approach similar to \cite{2019MNRAS.485.4817D} and select our quenched sample in four different redshift ranges spanning $0.5 < z < 3$. The quenched galaxies we select using this method are galaxies whose SFR is at least 1 dex below the main sequence. We rely on this second method of defining quiescence because it allows one to measure the quenched fraction of galaxies in observations without having to worry about the discrepancies that arise in the measured stellar mass and SFR values amongst different SED fitting codes, which could affect the measured quenched fraction if selecting galaxies based on sSFR. In Figure \ref{mass_sfr} the red line represents the main sequence, which is the mean SFR at fixed stellar mass for galaxies without X-ray luminous AGN, and the dotted magenta line represents the region of the mass-SFR plane that falls exactly 1 dex below the main sequence, meaning any sources that exist below this line satisfy the aforementioned definition of quiescence. We do not use the UVJ diagram method to select quenched galaxies as galaxies with X-ray luminous AGN, typically type I AGN-hosts or quasars, may have extremely blue colors due to emission from the accretion disk and thus affect our selection of quenched objects amongst the sample of galaxies with X-ray luminous AGN{}. For galaxies with X-ray luminous AGN with $L_X$ above the $80 \%$ completeness limit ($L_X = 10^{44.03},\ 10^{44.47}, \ 10^{44.78}$ and $10^{45.21}$ erg s$^{-1}$, respectively, at $z=0.5-1.0$, $z=1.0-1.5$, $z=1.5-2.0$, and $z=2.0-3.0$), the vast majority ($97 \%,\ 98 \%,\ 96 \%$ and $100 \%$, respectively, at $z=0.5-1$, $z=1.0-1.5$, $z=1.5-2.0$, and $z=2.0-3.0$) do not show quenched SF{}.

In Figure \ref{quenched_fraction} we show how the quenched fraction ($f_{\rm quenched}$) varies with stellar mass in four different redshift bins for the two definitions of quiescence that we use here. Figure \ref{quenched_fraction} has two solid curves that represent the fraction of galaxies that are quenched among our two samples: the sample of galaxies with (S1-Lum-AGN) and without (S2-No-Lum-AGN) X-ray luminous AGN{}. The three dashed lines in each panel of Figure \ref{quenched_fraction} correspond to the quenched fraction of the S2-No-Lum-AGN sample in all redshift ranges for easy visualization of the evolution of the quenched fraction with redshift. The error bars represent Poisson errors. The sample without (S2-No-Lum-AGN) X-ray luminous AGN is a factor of $\sim 100$ to $\sim 1,000$ larger than the sample of galaxies with (S1-Lum-AGN) X-ray luminous AGN, so we expect its behavior to be representative of the underlying sample of all galaxies. 

We find that the quenched fractions for galaxies without X-ray luminous AGN, based on the two definitions of quenched galaxies ($\rm{sSFR} < 10^{-11} \rm{yr}^{-1}$ versus SFR at least 1 dex below the main sequence) differ at all redshifts, especially at the highest redshift bin ($ z= 2-3$) where the quenched fraction based on sSFR is lower by a factor of $\sim 5-10$ at $z=2-3$ for galaxies with $M_* > 10^{11} \ M_{\odot}$. These results are not surprising, as one definition of quiescence ($\rm{sSFR} < 10^{-11} \rm{yr}^{-1}$) ignores the evolution with time of the SFR-stellar mass relation \citep{2014ApJ...795..104W, 2014ApJS..214...15S}, while the other definition of quiescence (SFR at least 1 dex below the main sequence) is based on a main sequence which is allowed to evolve as it is empirically determined in each redshift bin. 

We also note that the value of the quenched fraction based on selecting quenched galaxies using their distance from the main sequence is more robust across different studies than the quenched fraction based on a specific value of the sSFR{}. This is because the latter is highly sensitive to systematic effects, such as those introduced by the different fitting assumptions (e.g., different IMF, treatment of attenuation, choice of SPS, etc.) used by different SED fitting codes, impacting the absolute value of SFRs. For example, we find after running multiple tests that CIGALE produces systematically higher SFRs than EAZY-py by a factor of $\sim 2-3$, meaning that the measured quenched fractions (given by $\rm{sSFR} < 10^{-11} \rm{yr}^{-1}$) would be higher for our sample if we use EAZY-py instead of CIGALE to derive SFRs. Figure \ref{quenched_fraction} shows that the quenched fractions based on both definitions of quiescence are a strong function of stellar mass in each redshift bin and generally increase with stellar mass, except possibly at the very highest stellar masses.

While many earlier studies with smaller samples measured the mean quenched fraction, averaged over stellar mass as a function of redshift \citep{2006ApJ...649L..71K, 2013ApJ...777...18M, 2015MNRAS.451.2933B, 2018ApJ...858..100F, 2013ApJ...768...92S}, very few studies \citep[e.g.,][]{2013ApJ...777...18M} measure the quenched fraction as a function of stellar mass over different redshift ranges. Our study is the first to reveal, using such a large sample of massive galaxies, how the quenched fraction ($f_{\rm quenched}$) at a given redshift changes as the galaxy stellar mass varies from $10^{10} \ M_{\odot}$ to a few times $10^{11} \ M_{\odot}$ (Figure \ref{quenched_fraction}). The fact that $f_{\rm quenched}$ rises with stellar mass provides important clues on the mechanisms that quench SF in massive galaxies. 

\section{Discussion} \label{discussion}
In Section \ref{results}, we found that the average SFRs of galaxies with X-ray luminous AGN are higher by a factor of $\sim 3$ to $10$ compared to galaxies without X-ray luminous AGN at a given stellar mass and redshift range (see Figure \ref{mass_sfr} and Figure \ref{mass_sfr_mean}). These results are consistent with a scenario where the high SFR and high AGN luminosity are produced by processes that produce large gas inflow rates both into the regions (on scales of a few hundred pc to few kpc) typically associated with high SFRs, as well as the sub-pc region associated with the AGN accretion disk. Assuming a radiative efficiency of $\epsilon=0.1$, and a bolometric to X-ray luminosity ratio of $L_{\rm{bol}} / L_{X} \sim 30$ \citep{2012MNRAS.425..623L}, we estimate that the black hole accretion rates of our sample above the $80 \%$ X-ray completeness limit exceed values of $0.48 \ M_{\odot} \rm{yr}^{-1}$ (for $L_X > 10^{44.03}$ erg s$^{-1}$ at $z=0.5-1.0$), $1.3 \ M_{\odot} \rm{yr}^{-1}$ (for $L_X > 10^{44.47}$ erg s$^{-1}$ at $z=1.0-1.5$), $2.7 \ M_{\odot} \rm{yr}^{-1}$ (for $L_X > 10^{44.78}$ erg s$^{-1}$ at $z=1.5-2.0$), and $7.3 \ M_{\odot} \rm{yr}^{-1}$ (for $L_X > 10^{45.21}$ erg s$^{-1}$ at $z=2.0-3.0$). Given our maximum X-ray luminosity in each redshift bin, we expect the black hole accretion rates to not exceed values of $14, \ 45, \ 50$, and $100 \ M_{\odot} \rm{yr}^{-1}$ at $z=0.5-1.0, \ z=1.0-1.5, \ z=1.5-2.0$, and $z=2.0-3.0$, respectively.

Examples of processes that drive large gas inflow rates through gravitational torques, tidal torques, and dissipative shocks include gas-rich major mergers, gas-rich minor mergers, and strong tidal interactions in gas-rich systems. It should be noted that in order to drive gas from kpc scales down to the scales of the AGN accretion disk, we need mechanisms that effectively reduce the angular momentum of the gas by several orders of magnitude, such as gravitational torques from a primary bar, secondary bars or/and other non-axisymmetric features, shocks, dynamical friction on gas clumps, feedback processes from local SF, run-away self-gravitational instabilities, tidal disruption of clumps, and hydro-magnetic winds \citep[e.g.,][and references therein]{2006LNP...693..143J}. Another mechanism which can effectively reduce angular momentum has to do with the ram-pressure shocks described in \cite{2017MNRAS.465.2643C}, where large-scale (few kpc) shocks affect the entire galactic disc and decouple the dynamics of the gas from the stars and are a complementary trigger to tidal torques \citep[see Figure 2 of][]{2017MNRAS.465.2643C}.

It is instructive to look at numerical high resolution simulations which explore the onset of high SFR and high AGN luminosity \citep{2005MNRAS.361..776S,2008ApJ...676...33D, 2008MNRAS.384..386C, 2015MNRAS.447.2123C, 2017ApJ...845..128P}. In particular, the more recent numerical simulations of merging galaxies \citep[e.g.,][]{2017ApJ...845..128P, 2015MNRAS.447.2123C} show that large gas inflows during major mergers can simultaneously trigger SF and AGN activity in the merging galaxies. In these simulations, the peak of black hole accretion of SF activity appear to coincide and the decline of black hole accretion rate appears to trace the decline in SF with time. \cite{2017ApJ...845..128P} and \cite{2015MNRAS.447.2123C} show that AGN and SF activity are also triggered during minor mergers, but to a much lesser extent. In simulations of isolated galaxies, however, black hole accretion does not necessarily trace SF across time in the same way \citep{2017MNRAS.467.3475N}. This can be due to the fact that in some isolated systems, SF can happen on very large scales (e.g., kpc to tens of kpc) without any associated AGN activity. In isolated barred galaxies, the primary stellar bar can efficiently drive gas down to scales of a few hundred pc to fuel circumnuclear SF activity \citep[e.g.,][]{1994ApJ...425L..73E, 1995ApJ...454..623K, 1999ApJ...516..660H, 2005ApJ...630..837J}, but in many cases the gas stalls on scales of a few hundred pc as there are no effective mechanisms to further rapidly drain its angular momentum and drive it down to the sub-pc scales of the AGN accretion disk \citep[e.g., see][and references therein]{2006LNP...693..143J}.

Figure \ref{mass_sfr} shows that galaxies which have moderate to low SFRs with respect to the average SFR (shown as the red curve) are not associated with X-ray luminous AGN{}. There are several ways to explain these results. In the aforementioned \cite{2017ApJ...845..128P, 2015MNRAS.447.2123C} simulations of gas-rich major mergers, the contemporaneous phase of high SFR and AGN activity is followed by a phase where the black hole accretion rate and SFR both decline with time. In this scenario, the factors responsible for depressing SF (e.g., a declining gas supply, the heating or redistribution of the gas via stellar or AGN feedback) are also effective at depressing any AGN activity. We note that although the results in these simulations are for lower-mass galaxies than the ones we examine here, these simulation are of isolated systems and the results should be "scalable" and thus should hold to higher masses. In Figure \ref{mass_sfr}, one would expect such an evolution in a merging system to cause low-luminosity AGN to lie in the region of low SFRs. Another possibility is that isolated galaxies (which likely make up the bulk of systems shown in Figure \ref{mass_sfr}) exhibit low SFRs and no AGN activity or low-luminosity AGN activity simply because they lack the strong gravitational torques and shocks, which so efficiently drive gas inflows and fuel high central SFRs and AGN luminosity in gas-rich mergers. The X-ray data in our study are not sensitive to these low luminosity AGN, so we cannot directly test their location on Figure \ref{mass_sfr}. 

A more detailed comparison of our quenched fraction results to a wide range of numerical simulations, including hydrodynamical simulations and semi-analytic models is presented in Sherman et al. (submitted) as well as a discussion of the physical mechanisms that contribute to galaxy quenching across different environments, stellar masses, and epochs.

\section{Summary} \label{summary}
We have analyzed the relation between AGN and SF activity at $0.5 < z < 3$ by comparing the stellar masses and SFRs of 898 massive galaxies with X-ray luminous AGN ($L_X > 10^{44}$ erg s$^{-1}$) and a large comparison sample of $\sim 320,000$ galaxies without X-ray luminous AGN (see Figure \ref{sample_flowchart} and Table \ref{tab1}). Our samples are selected from a large (11.8 deg$^2$) area in Stripe82 that has multi-wavelength (X-ray to far-IR) data and corresponds to a very large comoving volume ($\sim 0.3$ Gpc$^3$) at $0.5 < z < 3$, thus minimizing the effects of cosmic variance and captures a large number of rare massive galaxies ($\sim 30,000$ galaxies with $M_* > 10^{11} \ M_{\odot}$) and X-ray luminous AGN{}.  While many galaxy evolution studies discard the hosts of X-ray luminous AGN due to the inability of common SED fitting codes to handle such systems, a strength of our study is that we fit the SED of both galaxies with and without X-ray luminous AGN hosts with the CIGALE SED fitting code, which includes AGN emission templates. We summarize our findings below:

\begin{enumerate}
    \item The stellar mass and SFRs of galaxies with X-ray luminous AGN are likely to be overestimated if AGN emission is not included in the SED fit (Figure \ref{SEDs}). For galaxies with large AGN fraction contamination ($f_{\rm{AGN}} > 0.4$), the stellar mass can be overestimated by factor of up to 5, on average, while SFRs can be overestimated by a factor of up to 10, on average, if AGN emission templates are not included in the SED fit (Figure \ref{SED_comp}). 
    \item The stellar mass function of galaxies with X-ray luminous AGN (Figure \ref{SMF}) shows that the number density of galaxies with X-ray luminous AGN is two to three orders of magnitude lower than galaxies without X-ray luminous AGN for stellar masses in the range of $10^{10}$ to $3 \times 10^{11} \ M_{\odot}$ at redshifts of $0.5 < z < 3$. This suggests that X-ray luminous AGN are a rare and rapid phase in galaxy evolution.
    \item We find that the average SFR of galaxies with X-ray luminous AGN is higher by a factor of $\sim 3$ to $10$ compared to galaxies without X-ray luminous AGN at a given stellar mass and redshift (Figures \ref{mass_sfr} and \ref{mass_sfr_mean}). We remind the reader that these results only hold for X-ray luminous AGN with X-ray luminosities above the $80 \%$ completeness limit at each redshift bin ($L_X > 10^{44.03}, \ 10^{44.47},\ 10^{44.78}$ and $10^{45.21}$ erg s$^{-1}$, respectively, at $z=0.5-1.0, \ z=1.0-1.5, \ z=1.5-2.0$ and $z=2.0-3.0$). These results are consistent with a scenario where the high SFR and high AGN luminosity are produced by processes that produce large gas inflow rates both into the regions (on scales of a few hundred pc to few kpc) typically associated with high SFRs, as well as the sub-pc region associated with the AGN accretion disk. Examples of processes that drive large gas inflow rates through gravitational torques, tidal torques, and dissipative shocks include gas-rich major mergers, gas-rich minor mergers, and strong tidal interactions in gas-rich systems. It should be noted that in order to drive gas from kpc scales down to the scales of the AGN accretion disk, we need mechanisms that effectively reduce the angular momentum of the gas by several orders of magnitude.
    \item Due to the unprecedented size of our sample of massive galaxies, we are able to perform one of the first robust explorations of how the quenched fraction of galaxies varies with stellar mass in each redshift bin. The quenched fraction, based on both definitions of quiescence (galaxies with $\rm{sSFR} < 10^{-11} \  \rm{yr}^{-1}$ or galaxies with SFR at least 1 dex below the main sequence) rises with galaxy stellar mass over the range $10^{10} \  M_{\odot}$ to about $3 \times 10^{11} \ M_{\odot}$ in each of our four redshift bins in the range $0.5 < z < 3$. The vast majority ($> 95 \%$) of galaxies with X-ray luminous AGN at $z=0.5-3$ do not show quenched SF: this suggests that if AGN feedback quenches SF, the associated quenching process takes a significant time to act and the quenched phase sets in after the highly luminous phases of AGN activity.
\end{enumerate}

\noindent

\section*{Data Availability}
The data underlying this work will be shared upon reasonable request to the corresponding author. Email: jflorez06@utexas.edu

\section*{Acknowledgements}
J.F., S.J., and S.S. gratefully acknowledge support from the University of Texas at Austin, as well as NSF grants  AST-1413652 and  AST-1757983. J.F., S.J., S.S., and S.L.F. acknowledge support from NSF grant AST-1614798. J.F. acknowledges support from NSF GRFP grant DGE-1610403. The authors wish to thank Pedro Capelo for useful comments. The authors acknowledge the Texas Advanced Computing Center (TACC) at The University of Texas at Austin for providing High-Performance Computing (HPC) resources that have contributed to the research results reported within this paper. URL: http://www.tacc.utexas.edu.

The Institution for Gravitation and the Cosmos is supported by the Eberly College of Science and the Office of the Senior Vice President for Research at the Pennsylvania State University.




\bibliographystyle{mnras}
\bibliography{agn_draft} 

\begin{thebibliography}{}
\makeatletter
\relax
\def\mn@urlcharsother{\let\do\@makeother \do\$\do\&\do\#\do\^\do\_\do\%\do\~}
\def\mn@doi{\begingroup\mn@urlcharsother \@ifnextchar [ {\mn@doi@}
  {\mn@doi@[]}}
\def\mn@doi@[#1]#2{\def\@tempa{#1}\ifx\@tempa\@empty \href
  {http://dx.doi.org/#2} {doi:#2}\else \href {http://dx.doi.org/#2} {#1}\fi
  \endgroup}
\def\mn@eprint#1#2{\mn@eprint@#1:#2::\@nil}
\def\mn@eprint@arXiv#1{\href {http://arxiv.org/abs/#1} {{\tt arXiv:#1}}}
\def\mn@eprint@dblp#1{\href {http://dblp.uni-trier.de/rec/bibtex/#1.xml}
  {dblp:#1}}
\def\mn@eprint@#1:#2:#3:#4\@nil{\def\@tempa {#1}\def\@tempb {#2}\def\@tempc
  {#3}\ifx \@tempc \@empty \let \@tempc \@tempb \let \@tempb \@tempa \fi \ifx
  \@tempb \@empty \def\@tempb {arXiv}\fi \@ifundefined
  {mn@eprint@\@tempb}{\@tempb:\@tempc}{\expandafter \expandafter \csname
  mn@eprint@\@tempb\endcsname \expandafter{\@tempc}}}

\bibitem[\protect\citeauthoryear{{Acquaviva}, {Gawiser}  \&
  {Guaita}}{{Acquaviva} et~al.}{2011}]{2011ApJ...737...47A}
{Acquaviva} V.,  {Gawiser} E.,   {Guaita} L.,  2011, \mn@doi [\apj]
  {10.1088/0004-637X/737/2/47}, \href
  {http://adsabs.harvard.edu/abs/2011ApJ...737...47A} {737, 47}

\bibitem[\protect\citeauthoryear{{Ananna} et~al.,}{{Ananna}
  et~al.}{2017}]{2017ApJ...850...66A}
{Ananna} T.~T.,  et~al., 2017, \mn@doi [\apj] {10.3847/1538-4357/aa937d}, \href
  {http://adsabs.harvard.edu/abs/2017ApJ...850...66A} {850, 66}

\bibitem[\protect\citeauthoryear{{Ananna} et~al.,}{{Ananna}
  et~al.}{2019}]{2019ApJ...871..240A}
{Ananna} T.~T.,  et~al., 2019, \mn@doi [\apj] {10.3847/1538-4357/aafb77}, \href
  {https://ui.adsabs.harvard.edu/abs/2019ApJ...871..240A} {871, 240}

\bibitem[\protect\citeauthoryear{{Babi{\'c}}, {Miller}, {Jarvis}, {Turner},
  {Alexander}  \& {Croom}}{{Babi{\'c}} et~al.}{2007}]{2007A&A...474..755B}
{Babi{\'c}} A.,  {Miller} L.,  {Jarvis} M.~J.,  {Turner} T.~J.,  {Alexander}
  D.~M.,   {Croom} S.~M.,  2007, \mn@doi [\aap] {10.1051/0004-6361:20078286},
  \href {http://adsabs.harvard.edu/abs/2007A%26A...474..755B} {474, 755}

\bibitem[\protect\citeauthoryear{{Baldry} et~al.,}{{Baldry}
  et~al.}{2012}]{2012MNRAS.421..621B}
{Baldry} I.~K.,  et~al., 2012, \mn@doi [\mnras]
  {10.1111/j.1365-2966.2012.20340.x}, \href
  {https://ui.adsabs.harvard.edu/abs/2012MNRAS.421..621B} {421, 621}

\bibitem[\protect\citeauthoryear{{Berta} et~al.,}{{Berta}
  et~al.}{2013}]{2013A&A...551A.100B}
{Berta} S.,  et~al., 2013, \mn@doi [\aap] {10.1051/0004-6361/201220859}, \href
  {http://adsabs.harvard.edu/abs/2013A%26A...551A.100B} {551, A100}

\bibitem[\protect\citeauthoryear{{Bertin} \& {Arnouts}}{{Bertin} \&
  {Arnouts}}{1996}]{1996A&AS..117..393B}
{Bertin} E.,  {Arnouts} S.,  1996, \mn@doi [\aaps] {10.1051/aas:1996164}, \href
  {http://adsabs.harvard.edu/abs/1996A%26AS..117..393B} {117, 393}

\bibitem[\protect\citeauthoryear{{Bieri}, {Dubois}, {Rosdahl}, {Wagner}, {Silk}
   \& {Mamon}}{{Bieri} et~al.}{2017}]{2017MNRAS.464.1854B}
{Bieri} R.,  {Dubois} Y.,  {Rosdahl} J.,  {Wagner} A.,  {Silk} J.,   {Mamon}
  G.~A.,  2017, \mn@doi [\mnras] {10.1093/mnras/stw2380}, \href
  {https://ui.adsabs.harvard.edu/abs/2017MNRAS.464.1854B} {464, 1854}

\bibitem[\protect\citeauthoryear{{Bongiorno} et~al.,}{{Bongiorno}
  et~al.}{2016}]{2016A&A...588A..78B}
{Bongiorno} A.,  et~al., 2016, \mn@doi [\aap] {10.1051/0004-6361/201527436},
  \href {http://adsabs.harvard.edu/abs/2016A%26A...588A..78B} {588, A78}

\bibitem[\protect\citeauthoryear{{Boquien}, {Burgarella}, {Roehlly}, {Buat},
  {Ciesla}, {Corre}, {Inoue}  \& {Salas}}{{Boquien}
  et~al.}{2019}]{2019A&A...622A.103B}
{Boquien} M.,  {Burgarella} D.,  {Roehlly} Y.,  {Buat} V.,  {Ciesla} L.,
  {Corre} D.,  {Inoue} A.~K.,   {Salas} H.,  2019, \mn@doi [\aap]
  {10.1051/0004-6361/201834156}, \href
  {https://ui.adsabs.harvard.edu/abs/2019A&A...622A.103B} {622, A103}

\bibitem[\protect\citeauthoryear{{Boselli}, {Boissier}, {Cortese}, {Gil de
  Paz}, {Seibert}, {Madore}, {Buat}  \& {Martin}}{{Boselli}
  et~al.}{2006}]{2006ApJ...651..811B}
{Boselli} A.,  {Boissier} S.,  {Cortese} L.,  {Gil de Paz} A.,  {Seibert} M.,
  {Madore} B.~F.,  {Buat} V.,   {Martin} D.~C.,  2006, \mn@doi [\apj]
  {10.1086/507766}, \href {http://adsabs.harvard.edu/abs/2006ApJ...651..811B}
  {651, 811}

\bibitem[\protect\citeauthoryear{{Brammer}, {van Dokkum}  \& {Coppi}}{{Brammer}
  et~al.}{2008}]{2008ApJ...686.1503B}
{Brammer} G.~B.,  {van Dokkum} P.~G.,   {Coppi} P.,  2008, \mn@doi [\apj]
  {10.1086/591786}, \href {http://adsabs.harvard.edu/abs/2008ApJ...686.1503B}
  {686, 1503}

\bibitem[\protect\citeauthoryear{{Brandt} \& {Alexander}}{{Brandt} \&
  {Alexander}}{2015}]{2015A&ARv..23....1B}
{Brandt} W.~N.,  {Alexander} D.~M.,  2015, \mn@doi [\aapr]
  {10.1007/s00159-014-0081-z}, \href
  {http://adsabs.harvard.edu/abs/2015A%26ARv..23....1B} {23, 1}

\bibitem[\protect\citeauthoryear{{Brandt} \& {Hasinger}}{{Brandt} \&
  {Hasinger}}{2005}]{2005ARA&A..43..827B}
{Brandt} W.~N.,  {Hasinger} G.,  2005, \mn@doi [\araa]
  {10.1146/annurev.astro.43.051804.102213}, \href
  {http://adsabs.harvard.edu/abs/2005ARA%26A..43..827B} {43, 827}

\bibitem[\protect\citeauthoryear{{Brennan} et~al.,}{{Brennan}
  et~al.}{2015}]{2015MNRAS.451.2933B}
{Brennan} R.,  et~al., 2015, \mn@doi [\mnras] {10.1093/mnras/stv1007}, \href
  {https://ui.adsabs.harvard.edu/abs/2015MNRAS.451.2933B} {451, 2933}

\bibitem[\protect\citeauthoryear{{Brusa} et~al.,}{{Brusa}
  et~al.}{2010}]{2010ApJ...716..348B}
{Brusa} M.,  et~al., 2010, \mn@doi [\apj] {10.1088/0004-637X/716/1/348}, \href
  {http://adsabs.harvard.edu/abs/2010ApJ...716..348B} {716, 348}

\bibitem[\protect\citeauthoryear{{Bruzual} \& {Charlot}}{{Bruzual} \&
  {Charlot}}{2003}]{2003MNRAS.344.1000B}
{Bruzual} G.,  {Charlot} S.,  2003, \mn@doi [\mnras]
  {10.1046/j.1365-8711.2003.06897.x}, \href
  {http://adsabs.harvard.edu/abs/2003MNRAS.344.1000B} {344, 1000}

\bibitem[\protect\citeauthoryear{{Buat} et~al.,}{{Buat}
  et~al.}{2014}]{2014A&A...561A..39B}
{Buat} V.,  et~al., 2014, \mn@doi [\aap] {10.1051/0004-6361/201322081}, \href
  {https://ui.adsabs.harvard.edu/abs/2014A&A...561A..39B} {561, A39}

\bibitem[\protect\citeauthoryear{{Burlon}, {Ajello}, {Greiner}, {Comastri},
  {Merloni}  \& {Gehrels}}{{Burlon} et~al.}{2011}]{2011ApJ...728...58B}
{Burlon} D.,  {Ajello} M.,  {Greiner} J.,  {Comastri} A.,  {Merloni} A.,
  {Gehrels} N.,  2011, \mn@doi [\apj] {10.1088/0004-637X/728/1/58}, \href
  {https://ui.adsabs.harvard.edu/abs/2011ApJ...728...58B} {728, 58}

\bibitem[\protect\citeauthoryear{{Calistro Rivera}, {Lusso}, {Hennawi}  \&
  {Hogg}}{{Calistro Rivera} et~al.}{2016}]{2016ApJ...833...98C}
{Calistro Rivera} G.,  {Lusso} E.,  {Hennawi} J.~F.,   {Hogg} D.~W.,  2016,
  \mn@doi [\apj] {10.3847/1538-4357/833/1/98}, \href
  {http://adsabs.harvard.edu/abs/2016ApJ...833...98C} {833, 98}

\bibitem[\protect\citeauthoryear{{Calzetti}, {Armus}, {Bohlin}, {Kinney},
  {Koornneef}  \& {Storchi-Bergmann}}{{Calzetti}
  et~al.}{2000}]{2000ApJ...533..682C}
{Calzetti} D.,  {Armus} L.,  {Bohlin} R.~C.,  {Kinney} A.~L.,  {Koornneef} J.,
   {Storchi-Bergmann} T.,  2000, \mn@doi [\apj] {10.1086/308692}, \href
  {http://adsabs.harvard.edu/abs/2000ApJ...533..682C} {533, 682}

\bibitem[\protect\citeauthoryear{{Capelo} \& {Dotti}}{{Capelo} \&
  {Dotti}}{2017}]{2017MNRAS.465.2643C}
{Capelo} P.~R.,  {Dotti} M.,  2017, \mn@doi [\mnras] {10.1093/mnras/stw2872},
  \href {https://ui.adsabs.harvard.edu/abs/2017MNRAS.465.2643C} {465, 2643}

\bibitem[\protect\citeauthoryear{{Capelo}, {Volonteri}, {Dotti}, {Bellovary},
  {Mayer}  \& {Governato}}{{Capelo} et~al.}{2015}]{2015MNRAS.447.2123C}
{Capelo} P.~R.,  {Volonteri} M.,  {Dotti} M.,  {Bellovary} J.~M.,  {Mayer} L.,
   {Governato} F.,  2015, \mn@doi [\mnras] {10.1093/mnras/stu2500}, \href
  {https://ui.adsabs.harvard.edu/abs/2015MNRAS.447.2123C} {447, 2123}

\bibitem[\protect\citeauthoryear{{Cattaneo} et~al.,}{{Cattaneo}
  et~al.}{2009}]{2009Natur.460..213C}
{Cattaneo} A.,  et~al., 2009, \mn@doi [\nat] {10.1038/nature08135}, \href
  {http://adsabs.harvard.edu/abs/2009Natur.460..213C} {460, 213}

\bibitem[\protect\citeauthoryear{{Chabrier}}{{Chabrier}}{2003}]{2003PASP..115..763C}
{Chabrier} G.,  2003, \mn@doi [\pasp] {10.1086/376392}, \href
  {http://adsabs.harvard.edu/abs/2003PASP..115..763C} {115, 763}

\bibitem[\protect\citeauthoryear{{Choi}, {Ostriker}, {Naab}, {Oser}  \&
  {Moster}}{{Choi} et~al.}{2015}]{2015MNRAS.449.4105C}
{Choi} E.,  {Ostriker} J.~P.,  {Naab} T.,  {Oser} L.,   {Moster} B.~P.,  2015,
  \mn@doi [\mnras] {10.1093/mnras/stv575}, \href
  {https://ui.adsabs.harvard.edu/abs/2015MNRAS.449.4105C} {449, 4105}

\bibitem[\protect\citeauthoryear{{Ciesla} et~al.,}{{Ciesla}
  et~al.}{2015}]{2015A&A...576A..10C}
{Ciesla} L.,  et~al., 2015, \mn@doi [\aap] {10.1051/0004-6361/201425252}, \href
  {http://adsabs.harvard.edu/abs/2015A%26A...576A..10C} {576, A10}

\bibitem[\protect\citeauthoryear{{Ciesla} et~al.,}{{Ciesla}
  et~al.}{2016}]{2016A&A...585A..43C}
{Ciesla} L.,  et~al., 2016, \mn@doi [\aap] {10.1051/0004-6361/201527107}, \href
  {http://adsabs.harvard.edu/abs/2016A%26A...585A..43C} {585, A43}

\bibitem[\protect\citeauthoryear{{Ciesla}, {Elbaz}  \& {Fensch}}{{Ciesla}
  et~al.}{2017}]{2017A&A...608A..41C}
{Ciesla} L.,  {Elbaz} D.,   {Fensch} J.,  2017, \mn@doi [\aap]
  {10.1051/0004-6361/201731036}, \href
  {http://adsabs.harvard.edu/abs/2017A%26A...608A..41C} {608, A41}

\bibitem[\protect\citeauthoryear{{Ciesla}, {Elbaz}, {Schreiber}, {Daddi}  \&
  {Wang}}{{Ciesla} et~al.}{2018}]{2018A&A...615A..61C}
{Ciesla} L.,  {Elbaz} D.,  {Schreiber} C.,  {Daddi} E.,   {Wang} T.,  2018,
  \mn@doi [\aap] {10.1051/0004-6361/201832715}, \href
  {http://adsabs.harvard.edu/abs/2018A%26A...615A..61C} {615, A61}

\bibitem[\protect\citeauthoryear{{Conroy} \& {Wechsler}}{{Conroy} \&
  {Wechsler}}{2009}]{2009ApJ...696..620C}
{Conroy} C.,  {Wechsler} R.~H.,  2009, \mn@doi [\apj]
  {10.1088/0004-637X/696/1/620}, \href
  {https://ui.adsabs.harvard.edu/abs/2009ApJ...696..620C} {696, 620}

\bibitem[\protect\citeauthoryear{{Conroy}, {Gunn}  \& {White}}{{Conroy}
  et~al.}{2009}]{2009ApJ...699..486C}
{Conroy} C.,  {Gunn} J.~E.,   {White} M.,  2009, \mn@doi [\apj]
  {10.1088/0004-637X/699/1/486}, \href
  {https://ui.adsabs.harvard.edu/abs/2009ApJ...699..486C} {699, 486}

\bibitem[\protect\citeauthoryear{{Cox}, {Jonsson}, {Somerville}, {Primack}  \&
  {Dekel}}{{Cox} et~al.}{2008}]{2008MNRAS.384..386C}
{Cox} T.~J.,  {Jonsson} P.,  {Somerville} R.~S.,  {Primack} J.~R.,   {Dekel}
  A.,  2008, \mn@doi [\mnras] {10.1111/j.1365-2966.2007.12730.x}, \href
  {https://ui.adsabs.harvard.edu/abs/2008MNRAS.384..386C} {384, 386}

\bibitem[\protect\citeauthoryear{{Cutri} \& {et al.}}{{Cutri} \& {et
  al.}}{2013}]{2013yCat.2328....0C}
{Cutri} R.~M.,  {et al.} 2013, VizieR Online Data Catalog, \href
  {https://ui.adsabs.harvard.edu/abs/2013yCat.2328....0C} {p. II/328}

\bibitem[\protect\citeauthoryear{{Daddi} et~al.,}{{Daddi}
  et~al.}{2007}]{2007ApJ...670..156D}
{Daddi} E.,  et~al., 2007, \mn@doi [\apj] {10.1086/521818}, \href
  {http://adsabs.harvard.edu/abs/2007ApJ...670..156D} {670, 156}

\bibitem[\protect\citeauthoryear{{Dale}, {Helou}, {Magdis}, {Armus},
  {D{\'{\i}}az-Santos}  \& {Shi}}{{Dale} et~al.}{2014}]{2014ApJ...784...83D}
{Dale} D.~A.,  {Helou} G.,  {Magdis} G.~E.,  {Armus} L.,  {D{\'{\i}}az-Santos}
  T.,   {Shi} Y.,  2014, \mn@doi [\apj] {10.1088/0004-637X/784/1/83}, \href
  {http://adsabs.harvard.edu/abs/2014ApJ...784...83D} {784, 83}

\bibitem[\protect\citeauthoryear{{Dav{\'e}}, {Angl{\'e}s-Alc{\'a}zar},
  {Narayanan}, {Li}, {Rafieferantsoa}  \& {Appleby}}{{Dav{\'e}}
  et~al.}{2019}]{2019MNRAS.486.2827D}
{Dav{\'e}} R.,  {Angl{\'e}s-Alc{\'a}zar} D.,  {Narayanan} D.,  {Li} Q.,
  {Rafieferantsoa} M.~H.,   {Appleby} S.,  2019, \mn@doi [\mnras]
  {10.1093/mnras/stz937}, \href
  {https://ui.adsabs.harvard.edu/abs/2019MNRAS.486.2827D} {486, 2827}

\bibitem[\protect\citeauthoryear{{Davidzon} et~al.,}{{Davidzon}
  et~al.}{2013}]{2013A&A...558A..23D}
{Davidzon} I.,  et~al., 2013, \mn@doi [\aap] {10.1051/0004-6361/201321511},
  \href {https://ui.adsabs.harvard.edu/abs/2013A&A...558A..23D} {558, A23}

\bibitem[\protect\citeauthoryear{{Delvecchio} et~al.,}{{Delvecchio}
  et~al.}{2014}]{2014MNRAS.439.2736D}
{Delvecchio} I.,  et~al., 2014, \mn@doi [\mnras] {10.1093/mnras/stu130}, \href
  {https://ui.adsabs.harvard.edu/abs/2014MNRAS.439.2736D} {439, 2736}

\bibitem[\protect\citeauthoryear{{Di Matteo}, {Colberg}, {Springel},
  {Hernquist}  \& {Sijacki}}{{Di Matteo} et~al.}{2008}]{2008ApJ...676...33D}
{Di Matteo} T.,  {Colberg} J.,  {Springel} V.,  {Hernquist} L.,   {Sijacki} D.,
   2008, \mn@doi [\apj] {10.1086/524921}, \href
  {https://ui.adsabs.harvard.edu/abs/2008ApJ...676...33D} {676, 33}

\bibitem[\protect\citeauthoryear{{Dickinson}, {Papovich}, {Ferguson}  \&
  {Budav{\'a}ri}}{{Dickinson} et~al.}{2003}]{2003ApJ...587...25D}
{Dickinson} M.,  {Papovich} C.,  {Ferguson} H.~C.,   {Budav{\'a}ri} T.,  2003,
  \mn@doi [\apj] {10.1086/368111}, \href
  {http://adsabs.harvard.edu/abs/2003ApJ...587...25D} {587, 25}

\bibitem[\protect\citeauthoryear{{Donnari} et~al.,}{{Donnari}
  et~al.}{2019}]{2019MNRAS.485.4817D}
{Donnari} M.,  et~al., 2019, \mn@doi [\mnras] {10.1093/mnras/stz712}, \href
  {https://ui.adsabs.harvard.edu/abs/2019MNRAS.485.4817D} {485, 4817}

\bibitem[\protect\citeauthoryear{{Draine} \& {Li}}{{Draine} \&
  {Li}}{2007}]{2007ApJ...657..810D}
{Draine} B.~T.,  {Li} A.,  2007, \mn@doi [\apj] {10.1086/511055}, \href
  {http://adsabs.harvard.edu/abs/2007ApJ...657..810D} {657, 810}

\bibitem[\protect\citeauthoryear{{Elbaz} et~al.,}{{Elbaz}
  et~al.}{2007}]{2007A&A...468...33E}
{Elbaz} D.,  et~al., 2007, \mn@doi [\aap] {10.1051/0004-6361:20077525}, \href
  {http://adsabs.harvard.edu/abs/2007A%26A...468...33E} {468, 33}

\bibitem[\protect\citeauthoryear{{Elmegreen}}{{Elmegreen}}{1994}]{1994ApJ...425L..73E}
{Elmegreen} B.~G.,  1994, \mn@doi [\apjl] {10.1086/187313}, \href
  {https://ui.adsabs.harvard.edu/abs/1994ApJ...425L..73E} {425, L73}

\bibitem[\protect\citeauthoryear{{Fabian}}{{Fabian}}{2012}]{2012ARA&A..50..455F}
{Fabian} A.~C.,  2012, \mn@doi [\araa] {10.1146/annurev-astro-081811-125521},
  \href {http://adsabs.harvard.edu/abs/2012ARA%26A..50..455F} {50, 455}

\bibitem[\protect\citeauthoryear{{Fang} et~al.,}{{Fang}
  et~al.}{2018}]{2018ApJ...858..100F}
{Fang} J.~J.,  et~al., 2018, \mn@doi [\apj] {10.3847/1538-4357/aabcba}, \href
  {https://ui.adsabs.harvard.edu/abs/2018ApJ...858..100F} {858, 100}

\bibitem[\protect\citeauthoryear{{Feltre}, {Hatziminaoglou}, {Fritz}  \&
  {Franceschini}}{{Feltre} et~al.}{2012}]{2012MNRAS.426..120F}
{Feltre} A.,  {Hatziminaoglou} E.,  {Fritz} J.,   {Franceschini} A.,  2012,
  \mn@doi [\mnras] {10.1111/j.1365-2966.2012.21695.x}, \href
  {https://ui.adsabs.harvard.edu/abs/2012MNRAS.426..120F} {426, 120}

\bibitem[\protect\citeauthoryear{{Ferrarese} \& {Merritt}}{{Ferrarese} \&
  {Merritt}}{2000}]{2000ApJ...539L...9F}
{Ferrarese} L.,  {Merritt} D.,  2000, \mn@doi [\apjl] {10.1086/312838}, \href
  {http://adsabs.harvard.edu/abs/2000ApJ...539L...9F} {539, L9}

\bibitem[\protect\citeauthoryear{{Fliri} \& {Trujillo}}{{Fliri} \&
  {Trujillo}}{2016}]{2016MNRAS.456.1359F}
{Fliri} J.,  {Trujillo} I.,  2016, \mn@doi [\mnras] {10.1093/mnras/stv2686},
  \href {http://adsabs.harvard.edu/abs/2016MNRAS.456.1359F} {456, 1359}

\bibitem[\protect\citeauthoryear{{Fontanot}, {De Lucia}, {Monaco}, {Somerville}
   \& {Santini}}{{Fontanot} et~al.}{2009}]{2009MNRAS.397.1776F}
{Fontanot} F.,  {De Lucia} G.,  {Monaco} P.,  {Somerville} R.~S.,   {Santini}
  P.,  2009, \mn@doi [\mnras] {10.1111/j.1365-2966.2009.15058.x}, \href
  {http://adsabs.harvard.edu/abs/2009MNRAS.397.1776F} {397, 1776}

\bibitem[\protect\citeauthoryear{{Fritz}, {Franceschini}  \&
  {Hatziminaoglou}}{{Fritz} et~al.}{2006}]{2006MNRAS.366..767F}
{Fritz} J.,  {Franceschini} A.,   {Hatziminaoglou} E.,  2006, \mn@doi [\mnras]
  {10.1111/j.1365-2966.2006.09866.x}, \href
  {http://adsabs.harvard.edu/abs/2006MNRAS.366..767F} {366, 767}

\bibitem[\protect\citeauthoryear{{Fumagalli}, {Gavazzi}, {Scaramella}  \&
  {Franzetti}}{{Fumagalli} et~al.}{2011}]{2011A&A...528A..46F}
{Fumagalli} M.,  {Gavazzi} G.,  {Scaramella} R.,   {Franzetti} P.,  2011,
  \mn@doi [\aap] {10.1051/0004-6361/201015463}, \href
  {http://adsabs.harvard.edu/abs/2011A%26A...528A..46F} {528, A46}

\bibitem[\protect\citeauthoryear{{Geach} et~al.,}{{Geach}
  et~al.}{2017}]{2017ApJS..231....7G}
{Geach} J.~E.,  et~al., 2017, \mn@doi [\apjs] {10.3847/1538-4365/aa74b6}, \href
  {http://adsabs.harvard.edu/abs/2017ApJS..231....7G} {231, 7}

\bibitem[\protect\citeauthoryear{{Gebhardt} et~al.,}{{Gebhardt}
  et~al.}{2000}]{2000ApJ...539L..13G}
{Gebhardt} K.,  et~al., 2000, \mn@doi [\apjl] {10.1086/312840}, \href
  {http://adsabs.harvard.edu/abs/2000ApJ...539L..13G} {539, L13}

\bibitem[\protect\citeauthoryear{{Glazebrook} et~al.,}{{Glazebrook}
  et~al.}{2017}]{2017Natur.544...71G}
{Glazebrook} K.,  et~al., 2017, \mn@doi [\nat] {10.1038/nature21680}, \href
  {https://ui.adsabs.harvard.edu/abs/2017Natur.544...71G} {544, 71}

\bibitem[\protect\citeauthoryear{{Grogin} et~al.,}{{Grogin}
  et~al.}{2011}]{2011ApJS..197...35G}
{Grogin} N.~A.,  et~al., 2011, \mn@doi [\apjs] {10.1088/0067-0049/197/2/35},
  \href {http://adsabs.harvard.edu/abs/2011ApJS..197...35G} {197, 35}

\bibitem[\protect\citeauthoryear{{Hambrick}, {Ostriker}, {Naab}  \&
  {Johansson}}{{Hambrick} et~al.}{2011}]{2011ApJ...738...16H}
{Hambrick} D.~C.,  {Ostriker} J.~P.,  {Naab} T.,   {Johansson} P.~H.,  2011,
  \mn@doi [\apj] {10.1088/0004-637X/738/1/16}, \href
  {https://ui.adsabs.harvard.edu/abs/2011ApJ...738...16H} {738, 16}

\bibitem[\protect\citeauthoryear{{Hao}, {Kennicutt}, {Johnson}, {Calzetti},
  {Dale}  \& {Moustakas}}{{Hao} et~al.}{2011}]{2011ApJ...741..124H}
{Hao} C.-N.,  {Kennicutt} R.~C.,  {Johnson} B.~D.,  {Calzetti} D.,  {Dale}
  D.~A.,   {Moustakas} J.,  2011, \mn@doi [\apj] {10.1088/0004-637X/741/2/124},
  \href {https://ui.adsabs.harvard.edu/abs/2011ApJ...741..124H} {741, 124}

\bibitem[\protect\citeauthoryear{{Heckman} \& {Best}}{{Heckman} \&
  {Best}}{2014}]{2014ARA&A..52..589H}
{Heckman} T.~M.,  {Best} P.~N.,  2014, \mn@doi [\araa]
  {10.1146/annurev-astro-081913-035722}, \href
  {http://adsabs.harvard.edu/abs/2014ARA%26A..52..589H} {52, 589}

\bibitem[\protect\citeauthoryear{{Hill} \& {HETDEX Consortium}}{{Hill} \&
  {HETDEX Consortium}}{2016}]{2016ASPC..507..393H}
{Hill} G.~J.,  {HETDEX Consortium} 2016, {HETDEX and VIRUS: Panoramic Integral
  Field Spectroscopy with 35k Fibers}.
p.~393

\bibitem[\protect\citeauthoryear{{Hopkins}, {Richards}  \&
  {Hernquist}}{{Hopkins} et~al.}{2007}]{2007ApJ...654..731H}
{Hopkins} P.~F.,  {Richards} G.~T.,   {Hernquist} L.,  2007, \mn@doi [\apj]
  {10.1086/509629}, \href {http://adsabs.harvard.edu/abs/2007ApJ...654..731H}
  {654, 731}

\bibitem[\protect\citeauthoryear{{Hopkins}, {Hernquist}, {Cox}  \& {Kere{\v
  s}}}{{Hopkins} et~al.}{2008}]{2008ApJS..175..356H}
{Hopkins} P.~F.,  {Hernquist} L.,  {Cox} T.~J.,   {Kere{\v s}} D.,  2008,
  \mn@doi [\apjs] {10.1086/524362}, \href
  {http://adsabs.harvard.edu/abs/2008ApJS..175..356H} {175, 356}

\bibitem[\protect\citeauthoryear{{Hopkins}, {Torrey}, {Faucher-Gigu{\`e}re},
  {Quataert}  \& {Murray}}{{Hopkins} et~al.}{2016}]{2016MNRAS.458..816H}
{Hopkins} P.~F.,  {Torrey} P.,  {Faucher-Gigu{\`e}re} C.-A.,  {Quataert} E.,
  {Murray} N.,  2016, \mn@doi [\mnras] {10.1093/mnras/stw289}, \href
  {https://ui.adsabs.harvard.edu/abs/2016MNRAS.458..816H} {458, 816}

\bibitem[\protect\citeauthoryear{{Hunt} \& {Malkan}}{{Hunt} \&
  {Malkan}}{1999}]{1999ApJ...516..660H}
{Hunt} L.~K.,  {Malkan} M.~A.,  1999, \mn@doi [\apj] {10.1086/307150}, \href
  {https://ui.adsabs.harvard.edu/abs/1999ApJ...516..660H} {516, 660}

\bibitem[\protect\citeauthoryear{{Ibar} \& {Lira}}{{Ibar} \&
  {Lira}}{2007}]{2007A&A...466..531I}
{Ibar} E.,  {Lira} P.,  2007, \mn@doi [\aap] {10.1051/0004-6361:20065350},
  \href {https://ui.adsabs.harvard.edu/abs/2007A&A...466..531I} {466, 531}

\bibitem[\protect\citeauthoryear{{Ilbert} et~al.,}{{Ilbert}
  et~al.}{2013}]{2013A&A...556A..55I}
{Ilbert} O.,  et~al., 2013, \mn@doi [\aap] {10.1051/0004-6361/201321100}, \href
  {https://ui.adsabs.harvard.edu/abs/2013A&A...556A..55I} {556, A55}

\bibitem[\protect\citeauthoryear{{Jahnke} \& {Macci{\`o}}}{{Jahnke} \&
  {Macci{\`o}}}{2011}]{2011ApJ...734...92J}
{Jahnke} K.,  {Macci{\`o}} A.~V.,  2011, \mn@doi [\apj]
  {10.1088/0004-637X/734/2/92}, \href
  {https://ui.adsabs.harvard.edu/abs/2011ApJ...734...92J} {734, 92}

\bibitem[\protect\citeauthoryear{{Jogee}}{{Jogee}}{2006}]{2006LNP...693..143J}
{Jogee} S.,  2006, in {Alloin} D.,  ed.,  Lecture Notes in Physics, Berlin
  Springer Verlag Vol. 693, Physics of Active Galactic Nuclei at all Scales.
  p.~143 (\mn@eprint {} {astro-ph/0408383}), \mn@doi{10.1007/3-540-34621-X_6}

\bibitem[\protect\citeauthoryear{{Jogee}, {Scoville}  \& {Kenney}}{{Jogee}
  et~al.}{2005}]{2005ApJ...630..837J}
{Jogee} S.,  {Scoville} N.,   {Kenney} J. D.~P.,  2005, \mn@doi [\apj]
  {10.1086/432106}, \href
  {https://ui.adsabs.harvard.edu/abs/2005ApJ...630..837J} {630, 837}

\bibitem[\protect\citeauthoryear{{Jogee} et~al.,}{{Jogee}
  et~al.}{2009}]{2009ApJ...697.1971J}
{Jogee} S.,  et~al., 2009, \mn@doi [\apj] {10.1088/0004-637X/697/2/1971}, \href
  {https://ui.adsabs.harvard.edu/abs/2009ApJ...697.1971J} {697, 1971}

\bibitem[\protect\citeauthoryear{{Kawinwanichakij} et~al.,}{{Kawinwanichakij}
  et~al.}{2020}]{2020ApJ...892....7K}
{Kawinwanichakij} L.,  et~al., 2020, \mn@doi [\apj] {10.3847/1538-4357/ab75c4},
  \href {https://ui.adsabs.harvard.edu/abs/2020ApJ...892....7K} {892, 7}

\bibitem[\protect\citeauthoryear{{Kirkpatrick} et~al.,}{{Kirkpatrick}
  et~al.}{2019a}]{2019arXiv190804795K}
{Kirkpatrick} A.,  et~al., 2019a, arXiv e-prints, \href
  {https://ui.adsabs.harvard.edu/abs/2019arXiv190804795K} {p. arXiv:1908.04795}

\bibitem[\protect\citeauthoryear{{Kirkpatrick}, {Sharon}, {Keller}  \&
  {Pope}}{{Kirkpatrick} et~al.}{2019b}]{2019ApJ...879...41K}
{Kirkpatrick} A.,  {Sharon} C.,  {Keller} E.,   {Pope} A.,  2019b, \mn@doi
  [\apj] {10.3847/1538-4357/ab223a}, \href
  {https://ui.adsabs.harvard.edu/abs/2019ApJ...879...41K} {879, 41}

\bibitem[\protect\citeauthoryear{{Knapen}, {Beckman}, {Heller}, {Shlosman}  \&
  {de Jong}}{{Knapen} et~al.}{1995}]{1995ApJ...454..623K}
{Knapen} J.~H.,  {Beckman} J.~E.,  {Heller} C.~H.,  {Shlosman} I.,   {de Jong}
  R.~S.,  1995, \mn@doi [\apj] {10.1086/176516}, \href
  {https://ui.adsabs.harvard.edu/abs/1995ApJ...454..623K} {454, 623}

\bibitem[\protect\citeauthoryear{{Koekemoer} et~al.,}{{Koekemoer}
  et~al.}{2011}]{2011ApJS..197...36K}
{Koekemoer} A.~M.,  et~al., 2011, \mn@doi [\apjs] {10.1088/0067-0049/197/2/36},
  \href {http://adsabs.harvard.edu/abs/2011ApJS..197...36K} {197, 36}

\bibitem[\protect\citeauthoryear{{Kormendy} \& {Ho}}{{Kormendy} \&
  {Ho}}{2013}]{2013ARA&A..51..511K}
{Kormendy} J.,  {Ho} L.~C.,  2013, \mn@doi [\araa]
  {10.1146/annurev-astro-082708-101811}, \href
  {https://ui.adsabs.harvard.edu/abs/2013ARA&A..51..511K} {51, 511}

\bibitem[\protect\citeauthoryear{{Kriek} et~al.,}{{Kriek}
  et~al.}{2006}]{2006ApJ...649L..71K}
{Kriek} M.,  et~al., 2006, \mn@doi [\apjl] {10.1086/508371}, \href
  {https://ui.adsabs.harvard.edu/abs/2006ApJ...649L..71K} {649, L71}

\bibitem[\protect\citeauthoryear{{Kroupa}}{{Kroupa}}{2001}]{2001MNRAS.322..231K}
{Kroupa} P.,  2001, \mn@doi [\mnras] {10.1046/j.1365-8711.2001.04022.x}, \href
  {https://ui.adsabs.harvard.edu/abs/2001MNRAS.322..231K} {322, 231}

\bibitem[\protect\citeauthoryear{{LaMassa} et~al.,}{{LaMassa}
  et~al.}{2013a}]{2013MNRAS.432.1351L}
{LaMassa} S.~M.,  et~al., 2013a, \mn@doi [\mnras] {10.1093/mnras/stt553}, \href
  {http://adsabs.harvard.edu/abs/2013MNRAS.432.1351L} {432, 1351}

\bibitem[\protect\citeauthoryear{{LaMassa} et~al.,}{{LaMassa}
  et~al.}{2013b}]{2013MNRAS.436.3581L}
{LaMassa} S.~M.,  et~al., 2013b, \mn@doi [\mnras] {10.1093/mnras/stt1837},
  \href {http://adsabs.harvard.edu/abs/2013MNRAS.436.3581L} {436, 3581}

\bibitem[\protect\citeauthoryear{{LaMassa} et~al.,}{{LaMassa}
  et~al.}{2016}]{2016ApJ...817..172L}
{LaMassa} S.~M.,  et~al., 2016, \mn@doi [\apj] {10.3847/0004-637X/817/2/172},
  \href {http://adsabs.harvard.edu/abs/2016ApJ...817..172L} {817, 172}

\bibitem[\protect\citeauthoryear{{LaMassa}, {Georgakakis}, {Vivek}, {Salvato},
  {Ananna}, {Urry}, {MacLeod}  \& {Ross}}{{LaMassa}
  et~al.}{2019}]{2019ApJ...876...50L}
{LaMassa} S.~M.,  {Georgakakis} A.,  {Vivek} M.,  {Salvato} M.,  {Ananna}
  T.~T.,  {Urry} C.~M.,  {MacLeod} C.,   {Ross} N.,  2019, \mn@doi [\apj]
  {10.3847/1538-4357/ab108b}, \href
  {https://ui.adsabs.harvard.edu/abs/2019ApJ...876...50L} {876, 50}

\bibitem[\protect\citeauthoryear{{Lang}, {Hogg}  \& {Mykytyn}}{{Lang}
  et~al.}{2016}]{2016ascl.soft04008L}
{Lang} D.,  {Hogg} D.~W.,   {Mykytyn} D.,  2016, {The Tractor: Probabilistic
  astronomical source detection and measurement}, Astrophysics Source Code
  Library (\mn@eprint {ascl} {1604.008})

\bibitem[\protect\citeauthoryear{{Lehmer} et~al.,}{{Lehmer}
  et~al.}{2008}]{2008ApJ...681.1163L}
{Lehmer} B.~D.,  et~al., 2008, \mn@doi [\apj] {10.1086/588459}, \href
  {https://ui.adsabs.harvard.edu/abs/2008ApJ...681.1163L} {681, 1163}

\bibitem[\protect\citeauthoryear{{Leslie}, {Kewley}, {Sanders}  \&
  {Lee}}{{Leslie} et~al.}{2016}]{2016MNRAS.455L..82L}
{Leslie} S.~K.,  {Kewley} L.~J.,  {Sanders} D.~B.,   {Lee} N.,  2016, \mn@doi
  [\mnras] {10.1093/mnrasl/slv135}, \href
  {https://ui.adsabs.harvard.edu/abs/2016MNRAS.455L..82L} {455, L82}

\bibitem[\protect\citeauthoryear{{Liu} et~al.,}{{Liu}
  et~al.}{2016}]{2016MNRAS.459.1602L}
{Liu} Z.,  et~al., 2016, \mn@doi [\mnras] {10.1093/mnras/stw753}, \href
  {https://ui.adsabs.harvard.edu/abs/2016MNRAS.459.1602L} {459, 1602}

\bibitem[\protect\citeauthoryear{{Lusso} et~al.,}{{Lusso}
  et~al.}{2012}]{2012MNRAS.425..623L}
{Lusso} E.,  et~al., 2012, \mn@doi [\mnras] {10.1111/j.1365-2966.2012.21513.x},
  \href {http://adsabs.harvard.edu/abs/2012MNRAS.425..623L} {425, 623}

\bibitem[\protect\citeauthoryear{{Madau} \& {Dickinson}}{{Madau} \&
  {Dickinson}}{2014}]{2014ARA&A..52..415M}
{Madau} P.,  {Dickinson} M.,  2014, \mn@doi [\araa]
  {10.1146/annurev-astro-081811-125615}, \href
  {http://adsabs.harvard.edu/abs/2014ARA%26A..52..415M} {52, 415}

\bibitem[\protect\citeauthoryear{{Magorrian} et~al.,}{{Magorrian}
  et~al.}{1998}]{1998AJ....115.2285M}
{Magorrian} J.,  et~al., 1998, \mn@doi [\aj] {10.1086/300353}, \href
  {http://adsabs.harvard.edu/abs/1998AJ....115.2285M} {115, 2285}

\bibitem[\protect\citeauthoryear{{Mahoro}, {Povi{\'c}}  \&
  {Nkundabakura}}{{Mahoro} et~al.}{2017}]{2017MNRAS.471.3226M}
{Mahoro} A.,  {Povi{\'c}} M.,   {Nkundabakura} P.,  2017, \mn@doi [\mnras]
  {10.1093/mnras/stx1762}, \href
  {https://ui.adsabs.harvard.edu/abs/2017MNRAS.471.3226M} {471, 3226}

\bibitem[\protect\citeauthoryear{{Masoura}, {Mountrichas}, {Georgantopoulos},
  {Ruiz}, {Magdis}  \& {Plionis}}{{Masoura} et~al.}{2018}]{2018A&A...618A..31M}
{Masoura} V.~A.,  {Mountrichas} G.,  {Georgantopoulos} I.,  {Ruiz} A.,
  {Magdis} G.,   {Plionis} M.,  2018, \mn@doi [\aap]
  {10.1051/0004-6361/201833397}, \href
  {http://adsabs.harvard.edu/abs/2018A%26A...618A..31M} {618, A31}

\bibitem[\protect\citeauthoryear{{McLure} \& {Dunlop}}{{McLure} \&
  {Dunlop}}{2002}]{2002MNRAS.331..795M}
{McLure} R.~J.,  {Dunlop} J.~S.,  2002, \mn@doi [\mnras]
  {10.1046/j.1365-8711.2002.05236.x}, \href
  {http://adsabs.harvard.edu/abs/2002MNRAS.331..795M} {331, 795}

\bibitem[\protect\citeauthoryear{{McNamara} \& {Nulsen}}{{McNamara} \&
  {Nulsen}}{2007}]{2007ARA&A..45..117M}
{McNamara} B.~R.,  {Nulsen} P.~E.~J.,  2007, \mn@doi [\araa]
  {10.1146/annurev.astro.45.051806.110625}, \href
  {http://adsabs.harvard.edu/abs/2007ARA%26A..45..117M} {45, 117}

\bibitem[\protect\citeauthoryear{{Mel{\'e}ndez}, {Mushotzky}, {Shimizu},
  {Barger}  \& {Cowie}}{{Mel{\'e}ndez} et~al.}{2014}]{2014ApJ...794..152M}
{Mel{\'e}ndez} M.,  {Mushotzky} R.~F.,  {Shimizu} T.~T.,  {Barger} A.~J.,
  {Cowie} L.~L.,  2014, \mn@doi [\apj] {10.1088/0004-637X/794/2/152}, \href
  {https://ui.adsabs.harvard.edu/abs/2014ApJ...794..152M} {794, 152}

\bibitem[\protect\citeauthoryear{{Moustakas} et~al.,}{{Moustakas}
  et~al.}{2013}]{2013ApJ...767...50M}
{Moustakas} J.,  et~al., 2013, \mn@doi [\apj] {10.1088/0004-637X/767/1/50},
  \href {http://adsabs.harvard.edu/abs/2013ApJ...767...50M} {767, 50}

\bibitem[\protect\citeauthoryear{{Mullaney}, {Alexander}, {Goulding}  \&
  {Hickox}}{{Mullaney} et~al.}{2011}]{2011MNRAS.414.1082M}
{Mullaney} J.~R.,  {Alexander} D.~M.,  {Goulding} A.~D.,   {Hickox} R.~C.,
  2011, \mn@doi [\mnras] {10.1111/j.1365-2966.2011.18448.x}, \href
  {https://ui.adsabs.harvard.edu/abs/2011MNRAS.414.1082M} {414, 1082}

\bibitem[\protect\citeauthoryear{{Muzzin} et~al.,}{{Muzzin}
  et~al.}{2013}]{2013ApJ...777...18M}
{Muzzin} A.,  et~al., 2013, \mn@doi [\apj] {10.1088/0004-637X/777/1/18}, \href
  {https://ui.adsabs.harvard.edu/abs/2013ApJ...777...18M} {777, 18}

\bibitem[\protect\citeauthoryear{{Naab} \& {Ostriker}}{{Naab} \&
  {Ostriker}}{2017}]{2017ARA&A..55...59N}
{Naab} T.,  {Ostriker} J.~P.,  2017, \mn@doi [\araa]
  {10.1146/annurev-astro-081913-040019}, \href
  {http://adsabs.harvard.edu/abs/2017ARA%26A..55...59N} {55, 59}

\bibitem[\protect\citeauthoryear{{Naiman} et~al.,}{{Naiman}
  et~al.}{2018}]{2018MNRAS.477.1206N}
{Naiman} J.~P.,  et~al., 2018, \mn@doi [\mnras] {10.1093/mnras/sty618}, \href
  {http://adsabs.harvard.edu/abs/2018MNRAS.477.1206N} {477, 1206}

\bibitem[\protect\citeauthoryear{{Negri} \& {Volonteri}}{{Negri} \&
  {Volonteri}}{2017}]{2017MNRAS.467.3475N}
{Negri} A.,  {Volonteri} M.,  2017, \mn@doi [\mnras] {10.1093/mnras/stx362},
  \href {https://ui.adsabs.harvard.edu/abs/2017MNRAS.467.3475N} {467, 3475}

\bibitem[\protect\citeauthoryear{{Nelson} et~al.,}{{Nelson}
  et~al.}{2018}]{2018MNRAS.475..624N}
{Nelson} D.,  et~al., 2018, \mn@doi [\mnras] {10.1093/mnras/stx3040}, \href
  {https://ui.adsabs.harvard.edu/abs/2018MNRAS.475..624N} {475, 624}

\bibitem[\protect\citeauthoryear{{Nenkova}, {Sirocky}, {Ivezi{\'c}}  \&
  {Elitzur}}{{Nenkova} et~al.}{2008}]{2008ApJ...685..147N}
{Nenkova} M.,  {Sirocky} M.~M.,  {Ivezi{\'c}} {\v{Z}}.,   {Elitzur} M.,  2008,
  \mn@doi [\apj] {10.1086/590482}, \href
  {https://ui.adsabs.harvard.edu/abs/2008ApJ...685..147N} {685, 147}

\bibitem[\protect\citeauthoryear{{Noll}, {Burgarella}, {Giovannoli}, {Buat},
  {Marcillac}  \& {Mu{\~n}oz-Mateos}}{{Noll}
  et~al.}{2009}]{2009A&A...507.1793N}
{Noll} S.,  {Burgarella} D.,  {Giovannoli} E.,  {Buat} V.,  {Marcillac} D.,
  {Mu{\~n}oz-Mateos} J.~C.,  2009, \mn@doi [\aap]
  {10.1051/0004-6361/200912497}, \href
  {http://adsabs.harvard.edu/abs/2009A%26A...507.1793N} {507, 1793}

\bibitem[\protect\citeauthoryear{{Papovich} et~al.,}{{Papovich}
  et~al.}{2016}]{2016ApJS..224...28P}
{Papovich} C.,  et~al., 2016, \mn@doi [\apjs] {10.3847/0067-0049/224/2/28},
  \href {http://adsabs.harvard.edu/abs/2016ApJS..224...28P} {224, 28}

\bibitem[\protect\citeauthoryear{{Park}, {Smith}  \& {Yi}}{{Park}
  et~al.}{2017}]{2017ApJ...845..128P}
{Park} J.,  {Smith} R.,   {Yi} S.~K.,  2017, \mn@doi [\apj]
  {10.3847/1538-4357/aa81c6}, \href
  {https://ui.adsabs.harvard.edu/abs/2017ApJ...845..128P} {845, 128}

\bibitem[\protect\citeauthoryear{{Peterson} \& {Fabian}}{{Peterson} \&
  {Fabian}}{2006}]{2006PhR...427....1P}
{Peterson} J.~R.,  {Fabian} A.~C.,  2006, \mn@doi [\physrep]
  {10.1016/j.physrep.2005.12.007}, \href
  {http://adsabs.harvard.edu/abs/2006PhR...427....1P} {427, 1}

\bibitem[\protect\citeauthoryear{{Powell}, {Urry}, {Cappelluti}, {Johnson},
  {LaMassa}, {Ananna}  \& {Kollmann}}{{Powell}
  et~al.}{2020}]{2020ApJ...891...41P}
{Powell} M.~C.,  {Urry} C.~M.,  {Cappelluti} N.,  {Johnson} J.~T.,  {LaMassa}
  S.~M.,  {Ananna} T.~T.,   {Kollmann} K.~E.,  2020, \mn@doi [\apj]
  {10.3847/1538-4357/ab6e65}, \href
  {https://ui.adsabs.harvard.edu/abs/2020ApJ...891...41P} {891, 41}

\bibitem[\protect\citeauthoryear{{Pozzetti} et~al.,}{{Pozzetti}
  et~al.}{2010}]{2010A&A...523A..13P}
{Pozzetti} L.,  et~al., 2010, \mn@doi [\aap] {10.1051/0004-6361/200913020},
  \href {https://ui.adsabs.harvard.edu/abs/2010A&A...523A..13P} {523, A13}

\bibitem[\protect\citeauthoryear{{Rigby}, {Best}, {Brookes}, {Peacock},
  {Dunlop}, {R{\"o}ttgering}, {Wall}  \& {Ker}}{{Rigby}
  et~al.}{2011}]{2011MNRAS.416.1900R}
{Rigby} E.~E.,  {Best} P.~N.,  {Brookes} M.~H.,  {Peacock} J.~A.,  {Dunlop}
  J.~S.,  {R{\"o}ttgering} H.~J.~A.,  {Wall} J.~V.,   {Ker} L.,  2011, \mn@doi
  [\mnras] {10.1111/j.1365-2966.2011.19167.x}, \href
  {http://adsabs.harvard.edu/abs/2011MNRAS.416.1900R} {416, 1900}

\bibitem[\protect\citeauthoryear{{Roos}, {Juneau}, {Bournaud}  \&
  {Gabor}}{{Roos} et~al.}{2015}]{2015ApJ...800...19R}
{Roos} O.,  {Juneau} S.,  {Bournaud} F.,   {Gabor} J.~M.,  2015, \mn@doi [\apj]
  {10.1088/0004-637X/800/1/19}, \href
  {https://ui.adsabs.harvard.edu/abs/2015ApJ...800...19R} {800, 19}

\bibitem[\protect\citeauthoryear{{Rosario} et~al.,}{{Rosario}
  et~al.}{2013}]{2013ApJ...771...63R}
{Rosario} D.~J.,  et~al., 2013, \mn@doi [\apj] {10.1088/0004-637X/771/1/63},
  \href {http://adsabs.harvard.edu/abs/2013ApJ...771...63R} {771, 63}

\bibitem[\protect\citeauthoryear{{Sanders}, {Soifer}, {Elias}, {Madore},
  {Matthews}, {Neugebauer}  \& {Scoville}}{{Sanders}
  et~al.}{1988}]{1988ApJ...325...74S}
{Sanders} D.~B.,  {Soifer} B.~T.,  {Elias} J.~H.,  {Madore} B.~F.,  {Matthews}
  K.,  {Neugebauer} G.,   {Scoville} N.~Z.,  1988, \mn@doi [\apj]
  {10.1086/165983}, \href {http://adsabs.harvard.edu/abs/1988ApJ...325...74S}
  {325, 74}

\bibitem[\protect\citeauthoryear{{Santini} et~al.,}{{Santini}
  et~al.}{2012}]{2012A&A...540A.109S}
{Santini} P.,  et~al., 2012, \mn@doi [\aap] {10.1051/0004-6361/201118266},
  \href {http://adsabs.harvard.edu/abs/2012A%26A...540A.109S} {540, A109}

\bibitem[\protect\citeauthoryear{{Schmidt}}{{Schmidt}}{1968}]{1968ApJ...151..393S}
{Schmidt} M.,  1968, \mn@doi [\apj] {10.1086/149446}, \href
  {http://adsabs.harvard.edu/abs/1968ApJ...151..393S} {151, 393}

\bibitem[\protect\citeauthoryear{{Schreiber} et~al.,}{{Schreiber}
  et~al.}{2015}]{2015A&A...575A..74S}
{Schreiber} C.,  et~al., 2015, \mn@doi [\aap] {10.1051/0004-6361/201425017},
  \href {https://ui.adsabs.harvard.edu/abs/2015A&A...575A..74S} {575, A74}

\bibitem[\protect\citeauthoryear{{Sherman} et~al.,}{{Sherman}
  et~al.}{2020}]{2020MNRAS.491.3318S}
{Sherman} S.,  et~al., 2020, \mn@doi [\mnras] {10.1093/mnras/stz3229}, \href
  {https://ui.adsabs.harvard.edu/abs/2020MNRAS.491.3318S} {491, 3318}

\bibitem[\protect\citeauthoryear{{Shimizu}, {Mushotzky}, {Mel{\'e}ndez}, {Koss}
   \& {Rosario}}{{Shimizu} et~al.}{2015}]{2015MNRAS.452.1841S}
{Shimizu} T.~T.,  {Mushotzky} R.~F.,  {Mel{\'e}ndez} M.,  {Koss} M.,
  {Rosario} D.~J.,  2015, \mn@doi [\mnras] {10.1093/mnras/stv1407}, \href
  {http://adsabs.harvard.edu/abs/2015MNRAS.452.1841S} {452, 1841}

\bibitem[\protect\citeauthoryear{{Shimizu}, {Mushotzky}, {Mel{\'e}ndez},
  {Koss}, {Barger}  \& {Cowie}}{{Shimizu} et~al.}{2017}]{2017MNRAS.466.3161S}
{Shimizu} T.~T.,  {Mushotzky} R.~F.,  {Mel{\'e}ndez} M.,  {Koss} M.~J.,
  {Barger} A.~J.,   {Cowie} L.~L.,  2017, \mn@doi [\mnras]
  {10.1093/mnras/stw3268}, \href
  {http://adsabs.harvard.edu/abs/2017MNRAS.466.3161S} {466, 3161}

\bibitem[\protect\citeauthoryear{{Somerville} \& {Dav{\'e}}}{{Somerville} \&
  {Dav{\'e}}}{2015}]{2015ARA&A..53...51S}
{Somerville} R.~S.,  {Dav{\'e}} R.,  2015, \mn@doi [\araa]
  {10.1146/annurev-astro-082812-140951}, \href
  {http://adsabs.harvard.edu/abs/2015ARA%26A..53...51S} {53, 51}

\bibitem[\protect\citeauthoryear{{Somerville}, {Hopkins}, {Cox}, {Robertson}
  \& {Hernquist}}{{Somerville} et~al.}{2008}]{2008MNRAS.391..481S}
{Somerville} R.~S.,  {Hopkins} P.~F.,  {Cox} T.~J.,  {Robertson} B.~E.,
  {Hernquist} L.,  2008, \mn@doi [\mnras] {10.1111/j.1365-2966.2008.13805.x},
  \href {http://adsabs.harvard.edu/abs/2008MNRAS.391..481S} {391, 481}

\bibitem[\protect\citeauthoryear{{Speagle}, {Steinhardt}, {Capak}  \&
  {Silverman}}{{Speagle} et~al.}{2014}]{2014ApJS..214...15S}
{Speagle} J.~S.,  {Steinhardt} C.~L.,  {Capak} P.~L.,   {Silverman} J.~D.,
  2014, \mn@doi [\apjs] {10.1088/0067-0049/214/2/15}, \href
  {http://adsabs.harvard.edu/abs/2014ApJS..214...15S} {214, 15}

\bibitem[\protect\citeauthoryear{{Springel}, {Di Matteo}  \&
  {Hernquist}}{{Springel} et~al.}{2005}]{2005MNRAS.361..776S}
{Springel} V.,  {Di Matteo} T.,   {Hernquist} L.,  2005, \mn@doi [\mnras]
  {10.1111/j.1365-2966.2005.09238.x}, \href
  {https://ui.adsabs.harvard.edu/abs/2005MNRAS.361..776S} {361, 776}

\bibitem[\protect\citeauthoryear{{Springel} et~al.,}{{Springel}
  et~al.}{2018}]{2018MNRAS.475..676S}
{Springel} V.,  et~al., 2018, \mn@doi [\mnras] {10.1093/mnras/stx3304}, \href
  {http://adsabs.harvard.edu/abs/2018MNRAS.475..676S} {475, 676}

\bibitem[\protect\citeauthoryear{{Stefanon}, {Marchesini}, {Rudnick}, {Brammer}
   \& {Whitaker}}{{Stefanon} et~al.}{2013}]{2013ApJ...768...92S}
{Stefanon} M.,  {Marchesini} D.,  {Rudnick} G.~H.,  {Brammer} G.~B.,
  {Whitaker} K.~E.,  2013, \mn@doi [\apj] {10.1088/0004-637X/768/1/92}, \href
  {https://ui.adsabs.harvard.edu/abs/2013ApJ...768...92S} {768, 92}

\bibitem[\protect\citeauthoryear{{Str{\"u}der} et~al.,}{{Str{\"u}der}
  et~al.}{2001}]{2001A&A...365L..18S}
{Str{\"u}der} L.,  et~al., 2001, \mn@doi [\aap] {10.1051/0004-6361:20000066},
  \href {https://ui.adsabs.harvard.edu/abs/2001A&A...365L..18S} {365, L18}

\bibitem[\protect\citeauthoryear{{Sutherland} \& {Saunders}}{{Sutherland} \&
  {Saunders}}{1992}]{1992MNRAS.259..413S}
{Sutherland} W.,  {Saunders} W.,  1992, \mn@doi [\mnras]
  {10.1093/mnras/259.3.413}, \href
  {http://adsabs.harvard.edu/abs/1992MNRAS.259..413S} {259, 413}

\bibitem[\protect\citeauthoryear{{Timlin} et~al.,}{{Timlin}
  et~al.}{2016}]{2016ApJS..225....1T}
{Timlin} J.~D.,  et~al., 2016, \mn@doi [\apjs] {10.3847/0067-0049/225/1/1},
  \href {http://adsabs.harvard.edu/abs/2016ApJS..225....1T} {225, 1}

\bibitem[\protect\citeauthoryear{{Tristram} et~al.,}{{Tristram}
  et~al.}{2007}]{2007A&A...474..837T}
{Tristram} K.~R.~W.,  et~al., 2007, \mn@doi [\aap]
  {10.1051/0004-6361:20078369}, \href
  {https://ui.adsabs.harvard.edu/abs/2007A&A...474..837T} {474, 837}

\bibitem[\protect\citeauthoryear{{Viero} et~al.,}{{Viero}
  et~al.}{2014}]{2014ApJS..210...22V}
{Viero} M.~P.,  et~al., 2014, \mn@doi [\apjs] {10.1088/0067-0049/210/2/22},
  \href {http://adsabs.harvard.edu/abs/2014ApJS..210...22V} {210, 22}

\bibitem[\protect\citeauthoryear{{Vogelsberger}, {Genel}, {Sijacki}, {Torrey},
  {Springel}  \& {Hernquist}}{{Vogelsberger}
  et~al.}{2013}]{2013MNRAS.436.3031V}
{Vogelsberger} M.,  {Genel} S.,  {Sijacki} D.,  {Torrey} P.,  {Springel} V.,
  {Hernquist} L.,  2013, \mn@doi [\mnras] {10.1093/mnras/stt1789}, \href
  {https://ui.adsabs.harvard.edu/abs/2013MNRAS.436.3031V} {436, 3031}

\bibitem[\protect\citeauthoryear{{Wellons} et~al.,}{{Wellons}
  et~al.}{2015}]{2015MNRAS.449..361W}
{Wellons} S.,  et~al., 2015, \mn@doi [\mnras] {10.1093/mnras/stv303}, \href
  {https://ui.adsabs.harvard.edu/abs/2015MNRAS.449..361W} {449, 361}

\bibitem[\protect\citeauthoryear{{Whitaker} et~al.,}{{Whitaker}
  et~al.}{2011}]{2011ApJ...735...86W}
{Whitaker} K.~E.,  et~al., 2011, \mn@doi [\apj] {10.1088/0004-637X/735/2/86},
  \href {https://ui.adsabs.harvard.edu/abs/2011ApJ...735...86W} {735, 86}

\bibitem[\protect\citeauthoryear{{Whitaker} et~al.,}{{Whitaker}
  et~al.}{2014}]{2014ApJ...795..104W}
{Whitaker} K.~E.,  et~al., 2014, \mn@doi [\apj] {10.1088/0004-637X/795/2/104},
  \href {http://adsabs.harvard.edu/abs/2014ApJ...795..104W} {795, 104}

\bibitem[\protect\citeauthoryear{{White} \& {Rees}}{{White} \&
  {Rees}}{1978}]{1978MNRAS.183..341W}
{White} S.~D.~M.,  {Rees} M.~J.,  1978, \mn@doi [\mnras]
  {10.1093/mnras/183.3.341}, \href
  {http://adsabs.harvard.edu/abs/1978MNRAS.183..341W} {183, 341}

\bibitem[\protect\citeauthoryear{{Wilkins}, {Trentham}  \& {Hopkins}}{{Wilkins}
  et~al.}{2008}]{2008MNRAS.385..687W}
{Wilkins} S.~M.,  {Trentham} N.,   {Hopkins} A.~M.,  2008, \mn@doi [\mnras]
  {10.1111/j.1365-2966.2008.12885.x}, \href
  {http://adsabs.harvard.edu/abs/2008MNRAS.385..687W} {385, 687}

\bibitem[\protect\citeauthoryear{{Wold} et~al.,}{{Wold}
  et~al.}{2019}]{2019ApJS..240....5W}
{Wold} I.~G.~B.,  et~al., 2019, \mn@doi [\apjs] {10.3847/1538-4365/aaee85},
  \href {http://adsabs.harvard.edu/abs/2019ApJS..240....5W} {240, 5}

\bibitem[\protect\citeauthoryear{{Wright} et~al.,}{{Wright}
  et~al.}{2010}]{2010AJ....140.1868W}
{Wright} E.~L.,  et~al., 2010, \mn@doi [\aj] {10.1088/0004-6256/140/6/1868},
  \href {http://adsabs.harvard.edu/abs/2010AJ....140.1868W} {140, 1868}

\bibitem[\protect\citeauthoryear{{Wuyts} et~al.,}{{Wuyts}
  et~al.}{2007}]{2007ApJ...655...51W}
{Wuyts} S.,  et~al., 2007, \mn@doi [\apj] {10.1086/509708}, \href
  {https://ui.adsabs.harvard.edu/abs/2007ApJ...655...51W} {655, 51}

\bibitem[\protect\citeauthoryear{{Yang} et~al.,}{{Yang}
  et~al.}{2017}]{2017ApJ...842...72Y}
{Yang} G.,  et~al., 2017, \mn@doi [\apj] {10.3847/1538-4357/aa7564}, \href
  {http://adsabs.harvard.edu/abs/2017ApJ...842...72Y} {842, 72}

\bibitem[\protect\citeauthoryear{{Yang}, {Brandt}, {Alexander}, {Chen}, {Ni},
  {Vito}  \& {Zhu}}{{Yang} et~al.}{2019}]{2019MNRAS.485.3721Y}
{Yang} G.,  {Brandt} W.~N.,  {Alexander} D.~M.,  {Chen} C.-T.~J.,  {Ni} Q.,
  {Vito} F.,   {Zhu} F.-F.,  2019, \mn@doi [\mnras] {10.1093/mnras/stz611},
  \href {http://adsabs.harvard.edu/abs/2019MNRAS.485.3721Y} {485, 3721}

\bibitem[\protect\citeauthoryear{{da Cunha}, {Charlot}  \& {Elbaz}}{{da Cunha}
  et~al.}{2008}]{2008MNRAS.388.1595D}
{da Cunha} E.,  {Charlot} S.,   {Elbaz} D.,  2008, \mn@doi [\mnras]
  {10.1111/j.1365-2966.2008.13535.x}, \href
  {http://adsabs.harvard.edu/abs/2008MNRAS.388.1595D} {388, 1595}

\bibitem[\protect\citeauthoryear{{van Dokkum} et~al.,}{{van Dokkum}
  et~al.}{2009}]{2009PASP..121....2V}
{van Dokkum} P.~G.,  et~al., 2009, \mn@doi [\pasp] {10.1086/597138}, \href
  {http://adsabs.harvard.edu/abs/2009PASP..121....2V} {121, 2}

\makeatother
\end{thebibliography}



\appendix
\section{Assessing the Impact of  WISE Data on the Analysis}
Figure \ref{sample_flowchart} and Table \ref{tab1} show that if we require WISE detection, our sample of galaxies with X-ray luminous AGN would be reduced from 932 sources (in sample S1-Lum-AGN) to 356 sources (in sample S1-Lum-AGN-WISE), while our sample of galaxies without with X-ray luminous AGN be reduced from 318,904 sources (in sample S2-No-Lum-AGN) to 4,695 sources (in sample S2-No-Lum-AGN-WISE). In order to better sample the X-ray luminosity function and prevent a drastic reduction in sample size, we use the samples without WISE data (S1-Lum-AGN and S2-No-Lum-AGN) in our main analysis. However, here in the Appendix, we perform some tests where we compare the SFR and stellar mass derived with and without WISE mid-IR data, and demonstrate that the inclusion of WISE data would not change the results of this work.

\begin{figure*}
\begin{center}
\includegraphics[scale=0.6]{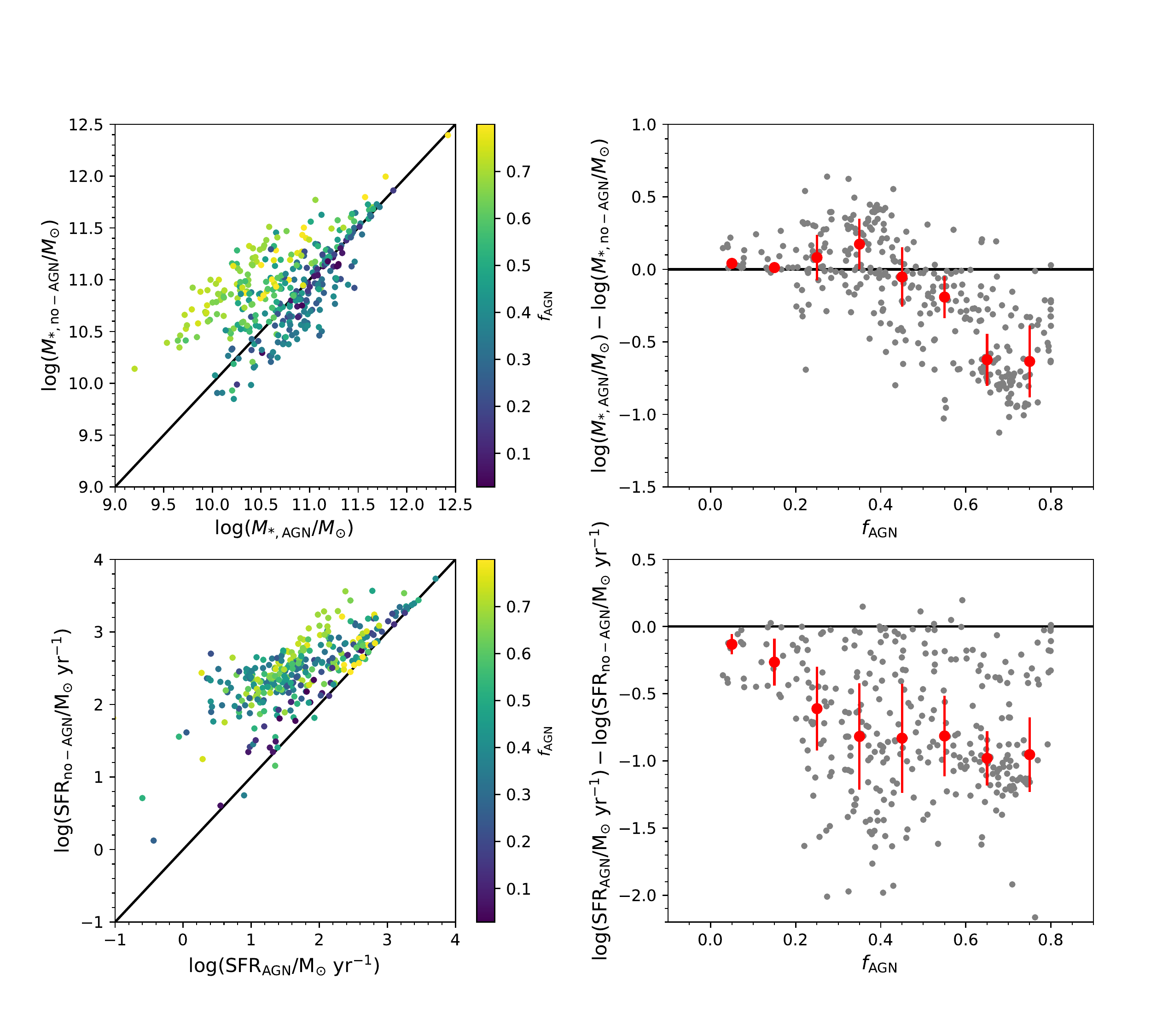}
\caption{Similar to Figure \ref{SED_comp} but for sources in S1-Lum-AGN-WISE, that is, galaxies with X-ray luminous AGN and WISE photometry available. Left: Stellar mass and SFR estimates for our sample of galaxies with X-ray luminous AGN (S1-Lum-AGN-WISE) when AGN emission is included in the SED fit (x-axis) versus when AGN emission is not included (y-axis). Points are colored according to their fractional AGN contamination ($f_{\rm{AGN}}$), defined as the fraction of light in the $8-1000$ $\mu$m wavelength range that is contributed by the AGN{}. Right: Difference in log stellar mass and SFR as a function of the fractional AGN contamination. Also shown is the median (red circles) log difference of stellar mass and SFR with and without the AGN emission in the SED fit in four bins of $f_{\rm{AGN}}$ with the median absolute deviation shown as error bars. We note that the results of this test are qualitatively similar to those shown in Figure \ref{SED_comp}, where WISE photometry is not used in the sample.}
\label{SED_comp-WISE}
\end{center}
\end{figure*}

In Figure \ref{SED_comp-WISE}, we show the results of a test similar to that shown in Figure \ref{SED_comp} for galaxies with X-ray luminous AGN and with a WISE detection. We find that when WISE photometry is included in the SED fitting, our results remain largely the same for this test. That is, we still find that stellar masses and SFRs can be overestimated, on average, by a factor of up to $\sim 5$ and $\sim 10$ for $f_{\rm AGN} > 0.4$, respectively, if AGN emission templates are not included in the SED fit.

\begin{figure*}
\centering
\subfigure{\includegraphics[scale=0.5]{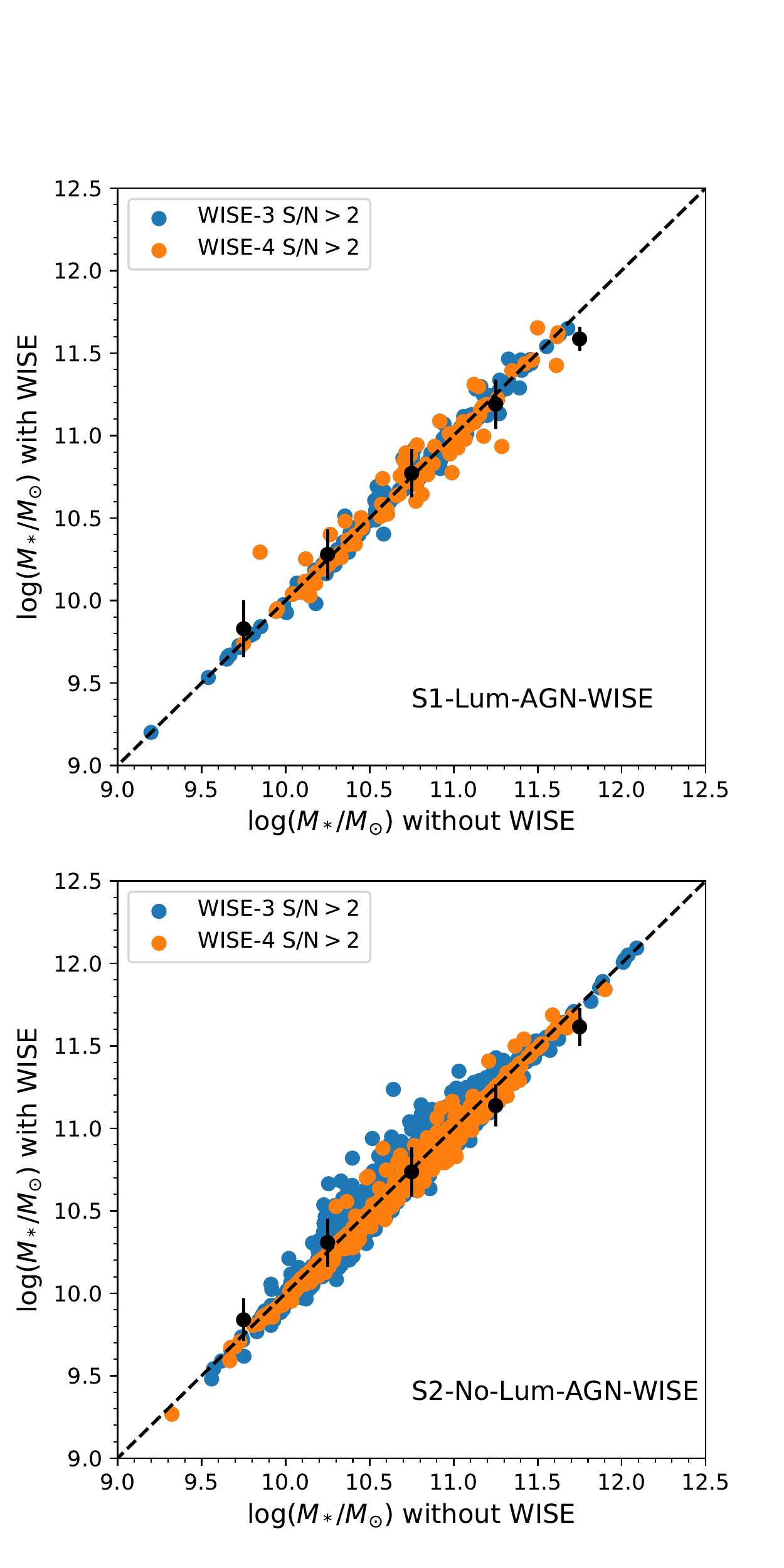}}
\subfigure{\includegraphics[scale=0.5]{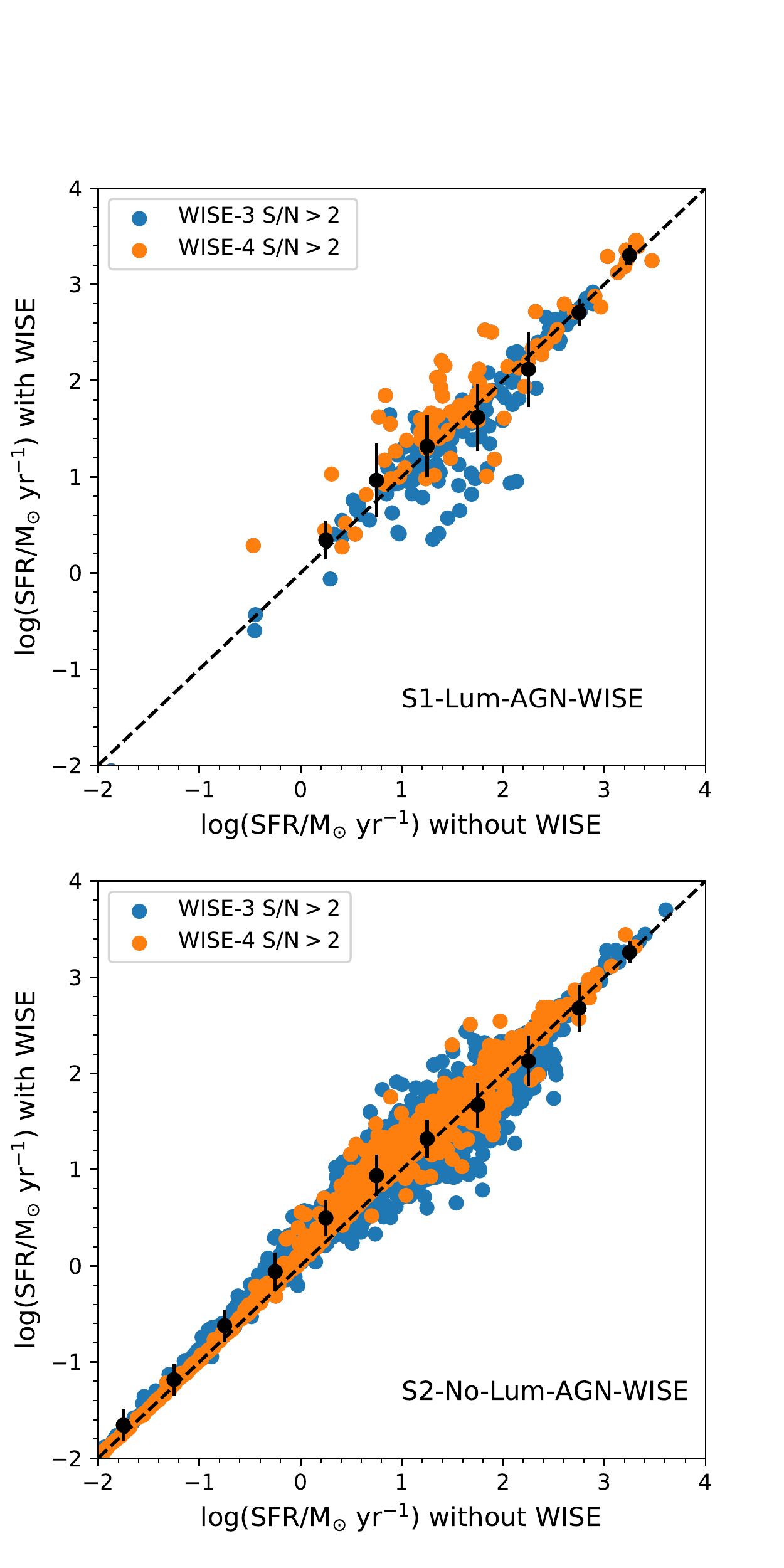}}
\caption{Left: The stellar mass estimate of galaxies with (S1-Lum-AGN-WISE, top) and without (S2-No-Lum-AGN, bottom) X-ray luminous AGN{}. Right: The The SFR estimate of galaxies with (S1-Lum-AGN-WISE, top) and without (S2-No-Lum-AGN, bottom) X-ray luminous AGN{}. The y-axis on all panels shows the stellar mass (or SFR) value obtained when WISE-3 or WISE-4 photometry is included in the SED fit, while the x-axis shows the value that is obtained when WISE photometry is not included in the SED fit. We find that stellar masses do not vary by more than 0.5 dex when WISE data is excluded from the photometry in either sample. SFRs, on the other hand, can vary by a factor of $\sim 1$ dex when WISE photometry is not included in the SED fit, however, we find no systematic bias. } 
\label{WISE-no_WISE_comparison}
\end{figure*}

The top panels of Figure \ref{WISE-no_WISE_comparison} show the stellar mass estimate using WISE data in the SED fit versus the stellar mass estimate without using WISE data for galaxies with and without X-ray luminous AGN{}. The bottom panels of Figure \ref{WISE-no_WISE_comparison} show the SFR estimate using WISE data in the SED fit versus the SFR estimate without using WISE data for both samples of galaxies with and without X-ray luminous AGN{}. The stellar mass estimates do not change by much when WISE data is added to the SED fit. The SFR estimates, however, change by quite a bit on a case to case basis and have a large scatter of $\sim 0.3$ dex. There is no systematic offset, however, in SFRs derived with and without WISE data.

\begin{figure*}
\centering
\includegraphics[scale=0.6]{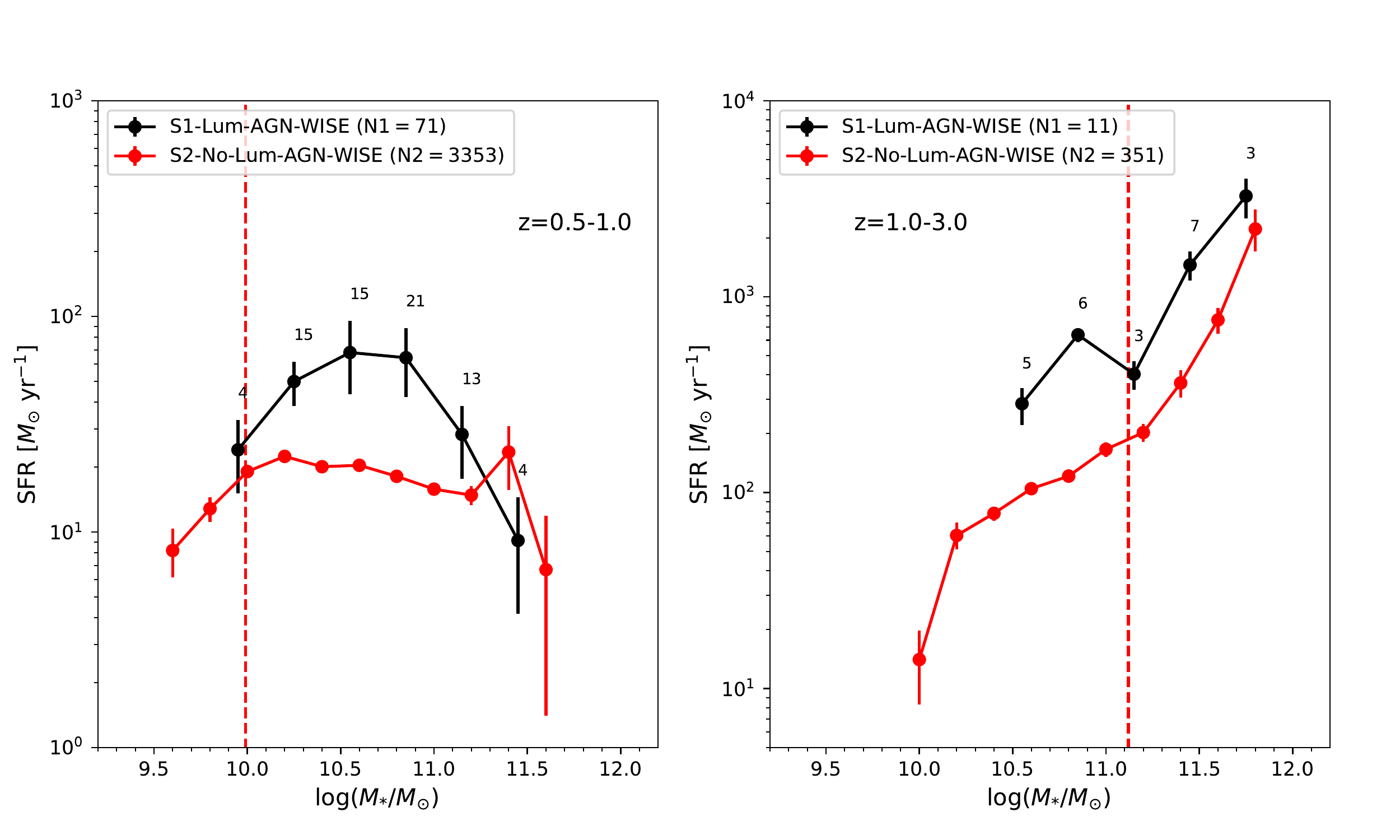}
\caption{The mean SFR of galaxies with (S1-Lum-AGN-WISE) and without (S2-No-Lum-AGN-WISE) X-ray luminous AGN as a function of stellar mass in two different bins of redshift. The dashed vertical line shows the mass completeness limit discussed in Section \ref{mass-distrib}. Our results here for the two samples with WISE photometry do not change qualitatively from those of Figure \ref{mass_sfr_mean}.}
\label{mass_vs_sfr_mean-WISE}
\end{figure*}

In Figure \ref{mass_vs_sfr_mean-WISE}, we show the mean SFR of galaxies with (S1-Lum-AGN-WISE) and without X-ray luminous AGN (S2-No-Lum-AGN-WISE) and with photometry out to 22 $\mu$m. Although we have poor number statistics, we find that the trends are consistent with the trends we see in Figure \ref{mass_sfr_mean} for both samples of galaxies with and without X-ray luminous AGN as the sample of galaxies with X-ray luminous AGN has higher SFRs, on average, at a given stellar mass.

\bsp	
\label{lastpage}
\end{document}